\documentclass[final,3p,times]{elsarticle}
\usepackage{tikz}
\usepackage{amssymb}
\usepackage{amsthm}
\usepackage[]{amsmath}
\usepackage{graphicx}
\usepackage{graphics}
\usepackage{subfigure}
\usepackage{latexsym}
\usepackage[]{amsmath}
\usepackage{amssymb}
\usepackage{indentfirst}
\usepackage{epsfig}
\usepackage{epstopdf}
\usepackage{float}
\usepackage{setspace}
\usepackage{booktabs}
\usepackage{times}
\usepackage{anysize}
\usepackage{endnotes}
\usepackage{amsmath}
\usepackage{mathtools}
\usepackage{mathrsfs}
\usepackage{enumitem}
\usepackage{color}
\usepackage{empheq}

\numberwithin{equation}{section}

\biboptions{numbers,sort&compress}
\usepackage[colorlinks, citecolor=red]{hyperref}
\usepackage{amsmath}
\usepackage{times}
\usepackage{anysize}
\marginsize{1.5cm}{1.5cm}{0.7cm}{1.5cm}
\selectfont
\journal {}

\makeatletter
\def\ps@pprintTitle{%
	\let\@oddhead\@empty
	\let\@evenhead\@empty
	\let\@oddfoot\@empty
	\let\@evenfoot\@oddfoot}
\makeatother

\begin{document}

\begin{frontmatter}

\title{Inverse scattering transform for the $N$-component Gerdjikov-Ivanov equation with nonzero boundary conditions}

\author{ Feng Zhang\fnref{label1} }
\author{Kun Zhu\fnref{label2}}
\author{ Yi Zhang\fnref{label1}$^*$ }
\cortext[cor1]{Corresponding author}
\ead{zy2836@163.com}
\address[label1]{ School of Mathematical Sciences, Zhejiang Normal University, Jinhua 321004, PR China}
\address[label2]{ School of Mathematics and Information Science, Nanchang Normal University, Nanchang 330032, PR China}

\begin{abstract}
This paper investigates the inverse scattering theory of the $N$-component Gerdjikov-Ivanov equation with nonzero boundary conditions. The analyticity of the Jost eigenfunctions is analyzed, and auxiliary eigenfunctions are constructed using tensor products and generalized cross products, thereby yielding a set of analytic eigenfunctions required to formulate the inverse problem. Three fundamental symmetry relations satisfied by the eigenfunctions and scattering data are established, and the asymptotic behavior of both is examined. We provide a complete characterization of the discrete spectrum and derive the corresponding residue conditions by analyzing the ranks of the relevant matrices. The inverse problem is then formulated as a matrix Riemann-Hilbert problem whose solution is used to reconstruct the potential. In the reflectionless case, an explicit solution formula is obtained. Finally, taking the four-component case as an example, we illustrate the structural features of solitons, breathers, and their interactions.
\end{abstract}

\end{frontmatter}

\section{Introduction}
\label{sec:1}

Solitary waves are nonlinear waves that maintain their shapes during propagation and exhibit spatial localization. They arise in a wide range of physical contexts, including fluid mechanics, nonlinear optics, and plasma physics \cite{A1,A2,A3}. Nonlinear integrable evolution equations play a significant role in the mathematical modeling of such wave phenomena. These equations possess rich mathematical structures, such as infinitely many conservation laws and Lax pairs, which provide a theoretical framework for investigating soliton propagation and interactions \cite{A4,A5}. Various methods have been developed to analyze and solve these equations, notably the inverse scattering transform (IST) \cite{A8}, the Darboux transformation \cite{A7}, and the Hirota bilinear method \cite{A6}. Among these methods, the IST relates the potential to its scattering data through an analysis of the associated linear spectral problem, and then reconstructs the potential from these data by solving the inverse problem. This method originated in research on the initial value problem of the KdV equation \cite{A9} and was subsequently generalized to the nonlinear Schr\"odinger (NLS) equation \cite{A10}. It has become an effective tool for analyzing the spectral properties of integrable systems, constructing exact solutions, and investigating the long-time asymptotics of solutions. 

Among various integrable models, the NLS equation serves as a classical model for the evolution of dispersive wave trains in the weakly nonlinear regime \cite{A11}. To further capture the nonlinear effects associated with spatial variations in wave-packet envelopes, a class of derivative NLS (DNLS) equations has attracted considerable attention \cite{A12}. The Gerdjikov-Ivanov (GI) equation belongs to this family and is also known as the DNLS III equation \cite{A13}. Its scalar form is given by
\begin{equation}\label{eq0}
\mathrm{i}u_t+u_{xx}-\mathrm{i}u^2u^*_x+\frac{1}{2}|u|^4u=0,
\end{equation}
where the superscript $*$ denotes complex conjugate. This equation contains both a derivative nonlinear term and a quintic nonlinear term and is completely integrable. Other classical members of the DNLS family include the Kaup-Newell equation \cite{A14} and the Chen-Lee-Liu equation \cite{A15}, known as the DNLS I and DNLS II equations, respectively. The GI equation can be used to study the propagation of nonlinear optical pulses in optical fibers and nonlinear Alfv\'en waves in magnetized plasmas \cite{A16}. And extensive research has focused on its Hamiltonian structure and various types of exact solutions, including solitons, breathers, and rogue waves \cite{A17,A18,A19,A20,A21}.

The GI equation models the nonlinear evolution of a single complex-valued field. However, for wave systems with multiple interacting components, scalar models are often insufficient to fully capture the coupling effects between different components \cite{A22,A23}. To investigate such coupling phenomena, many classical scalar integrable equations have been generalized to multi-component integrable systems that describe both the evolution of individual components and their nonlinear interactions \cite{A24,A25,A26}. Compared with their scalar counterparts, these systems incorporate couplings between components and can exhibit a broader range of localized wave structures and interaction dynamics. The GI equation likewise admits an integrable multi-component generalization that provides a mathematical model for studying localized waves and their interactions in the presence of derivative nonlinear couplings \cite{A27,A28,A29}. More generally, the $N$-component GI equation provides a unified mathematical framework for investigating coupled structures, spectral properties, and localized wave solutions for systems with different numbers of components.

In the present study, we investigate the following $N$-component GI equation $(N \ge 2)$ with nonzero boundary conditions (NZBCs):
\begin{subequations}\label{eq2}
\begin{empheq}[left=\empheqlbrace]{align}
&\mathrm{i}\boldsymbol{u}_t+\boldsymbol{u}_{xx}
+2\mathrm{i}u_0^2\boldsymbol{u}_x
-\mathrm{i}(\boldsymbol{u}_x^\dagger\boldsymbol{u})\boldsymbol{u}
-u_0^2(\boldsymbol{u}^\dagger\boldsymbol{u})\boldsymbol{u}
+\frac12u_0^4\boldsymbol{u}
+\frac12(\boldsymbol{u}^\dagger\boldsymbol{u})^2\boldsymbol{u}
=\boldsymbol{0}, \label{eq2a} \\
&\lim_{x\rightarrow\pm\infty}\boldsymbol{u}(x,t)
=\boldsymbol{u}_\pm=\boldsymbol{u}_0\mathrm{e}^{\mathrm{i}\theta_\pm}, \label{eq2b}
\end{empheq}
\end{subequations}
where ${\boldsymbol{u}}=({u}_1,{u}_2,\dots,{u}_N)^T,$ $\theta_\pm$ are real numbers, and $\boldsymbol{0}$ denotes the zero vector in $\mathbb{C}^{N}$. The superscripts $T$ and $\dagger$ represent transpose and conjugate transpose, respectively. By exploiting the unitary invariance of Eq. (\ref{eq2a}), we may choose, without loss of generality, that the background vector takes the form $\boldsymbol{u}_0={( {0, \ldots ,0,}{u_0})^T} $, where $u_0>0$. Under the gauge transformation ${\boldsymbol{u}}=\tilde{\boldsymbol{u}} \mathrm{e}^{\frac{3}{2}\mathrm{i}u_0^4t-\mathrm{i}u_0^2x}$, Eq. (\ref{eq2a}) transforms into the standard form of the $N$-component GI equation
\begin{equation}\label{eq1}
\mathrm{i}\tilde{\boldsymbol{u}}_t+\tilde{\boldsymbol{u}}_{xx}
-\mathrm{i}(\tilde{\boldsymbol{u}}^\dagger_x\tilde{\boldsymbol{u}})\tilde{\boldsymbol{u}}
+\frac12\bigl(\tilde{\boldsymbol{u}}^\dagger\tilde{\boldsymbol{u}}\bigr)^2\tilde{\boldsymbol{u}}=\boldsymbol{0}.
\end{equation}

For the single-component case $\tilde{\boldsymbol{u}}=(0,\dots,0,u)^T$, Eq. (\ref{eq1}) recovers to the scalar GI equation (\ref{eq0}), while setting $N = 2$ yields the coupled GI equation. In recent years, the integrable structures, spectral theory, and localized wave solutions of Eq. (\ref{eq1}) have attracted considerable attention. Gerdjikov et al. investigated its spectral properties using quadratic bundle Lax pairs associated with symmetric spaces and constructed soliton solutions via the dressing method \cite{A27}. The Hamiltonian structures of this system with zero boundary condition (ZBC) were studied in \cite{A30,A31,A32}, and multi-soliton solutions were constructed using the Riemann-Hilbert approach. Dong et al. further extended the Riemann-Hilbert formulation and the construction of multi-soliton solutions to the nonlocal $N$-component GI equation through a reverse-time nonlocal reduction \cite{A33}. In addition, various nonlinear wave solutions arising in the coupled GI equation, including solitons, breathers, and rogue waves, along with their interactions, have been studied in \cite{A29,A34,A35}. For NZBCs, Ma and Zhu examined the IST of the two-component GI equation, constructed analytic eigenfunctions using the adjoint spectral problem, and discussed the classification of breather and soliton solutions \cite{A36}. Despite these advances, the IST for the general $N$-component GI equation with NZBCs remains to be developed.

With the advancement of the inverse scattering theory, this method has been generalized from scalar equations to multi-component systems \cite{A37,A38}. Although such extensions are relatively straightforward for ZBC, they become considerably more challenging for NZBCs. These difficulties arise mainly because the associated spectral problems typically involve a two-sheeted Riemann surface with a nontrivial branch structure, and not all Jost eigenfunctions admit analytic continuation into the corresponding spectral regions  \cite{A39,A40,A41}. In the two-component setting with NZBCs, the IST for the Manakov system was first developed in \cite{A42}. Building on this work, subsequent studies further refined this theory, providing a detailed analysis and justification of the analytic properties of the Jost eigenfunctions, as well as a complete characterization of the discrete spectrum \cite{A43,A44}. The related techniques were later extended to the coupled mKdV equation \cite{A45}, the two-component Hirota equation \cite{A46,A47}, and other integrable systems. For the more general multi-component case with NZBCs, Prinari et al. investigated the IST for the multi-component NLS equation and constructed the auxiliary eigenfunctions using the tensor product method \cite{A48}. However, the symmetry relations for the eigenfunctions and scattering data were not established, and the discrete spectrum was not characterized. Subsequently, Biondini et al. addressed these issues for the three-component NLS equation by combining the triangular decomposition of the scattering matrix with the generalized cross products and the tensor products \cite{A49}. Liu and Geng further investigated the IST for the $N$-component NLS equation and extended the tensor product and generalized cross product methods to construct auxiliary eigenfunctions without requiring triangular decompositions of the scattering matrix \cite{A50,A51}. Motivated by these developments, this paper investigates the IST for the $N$-component GI equation with NZBCs. Compared with the two-component case, the construction of the auxiliary eigenfunctions requires combining tensor products with generalized cross products. Moreover, at discrete eigenvalues, the linear dependencies among the columns of the Jost eigenfunctions cannot be determined directly but are instead identified by analyzing the ranks of the the matrices formed by these columns, which complicates the derivation of the residue conditions. Compared with the $N$-component NLS equation, the Eq. (\ref{eq2}) admits three fundamental symmetry relations of its eigenfunctions and scattering data, resulting in a more intricate symmetric distribution of the discrete eigenvalues. Consequently, more residue conditions need to be considered, complicating both the formulation of the Riemann-Hilbert problem and the recovery of the potential. This work provides an effective approach for analyzing derivative-type nonlinear couplings, localized waves, and their interactions for an arbitrary number of components with a nonzero background, and offers a generalizable theoretical framework for studying the IST of other $N$-component DNLS equations with NZBCs.

The subsequent sections are organized as follows. Section \ref{sec:2} addresses the direct scattering problem, beginning with the associated spectral problem. We first introduce the Jost eigenfunctions and their modified forms, establish the corresponding Volterra integral equations, and then analyze their analytic properties. The scattering relations are then established, and the analytic properties of the scattering coefficients are investigated. Using generalized cross products and tensor products, we construct the auxiliary Jost eigenfunctions, which yields a set of analytic eigenfunctions required to formulate the inverse problem. We further investigate the three fundamental symmetry relations satisfied by the eigenfunctions and scattering coefficients, and derive several combined symmetry relations from them. In addition, the asymptotic behavior of the Jost eigenfunctions is analyzed, and the product representations of the auxiliary Jost eigenfunctions in terms of the scattering coefficients and Jost eigenfunctions are derived. Section \ref{sec:3} focuses on the inverse problem. We construct sectionally meromorphic matrices and classify the discrete eigenvalues into two classes according to whether they lie on or off the circle. The corresponding residue conditions are then derived by analyzing the ranks of the relevant matrices. Subsequently, the inverse problem is formulated as an $(N+1)\times(N+1)$ matrix Riemann-Hilbert problem, whose solution is used to reconstruct the potential. In the reflectionless case, an explicit solution formula is derived. Taking the four-component GI equation as an example, we further discuss the solution structures associated with the two classes of discrete eigenvalues. Finally, Section \ref{sec:4} presents the main conclusions of this work along with some related discussions.

\section{Direct Scattering}
\label{sec:2}

\subsection{Riemann surface and uniformization}
\label{subsec:2.1}

The Lax pair associated with the $N$-component GI equation (\ref{eq2a}) takes the following form:
\begin{equation}\label{eq3}
\begin{aligned}
\boldsymbol{\varphi}_x&=\boldsymbol{X}\boldsymbol{\varphi},\\
\boldsymbol{\varphi}_t&=\boldsymbol{Y}\boldsymbol{\varphi},
\end{aligned}
\end{equation}
where
$$\begin{aligned}
\boldsymbol{X}
&=-\mathrm{i}k^2\boldsymbol{\sigma}_3+\frac{\mathrm{i}}{2}\boldsymbol{U}^2\boldsymbol{\sigma}_3-k\boldsymbol{U}\boldsymbol{\sigma}_3-\frac{\mathrm{i}}{2}u_0^2\boldsymbol{\sigma}_3,\\
\boldsymbol{Y}&=\frac{3\mathrm{i}}{4}u_0^4\boldsymbol{\sigma}_3-2\mathrm{i}k^4\boldsymbol{\sigma}_3-2k^3\boldsymbol{U}\boldsymbol{\sigma}_3+\mathrm{i}k^2\boldsymbol{\sigma}_3
\boldsymbol{U}^2+\frac{\mathrm{i}}{4}\boldsymbol{U}^4\boldsymbol{\sigma}_3+\mathrm{i}k\boldsymbol{U}_x+\frac12[\boldsymbol{U},\boldsymbol{U}_x]-\frac{k}{2}u_0^2[\boldsymbol{\sigma}_3,
\boldsymbol{U}]-\mathrm{i}u_0^2\boldsymbol{\sigma}_3\boldsymbol{U}^2,\\
\end{aligned}$$\vspace{-1.5ex}
$$\begin{aligned}
&\begin{gathered}
\boldsymbol{\sigma}_3=\begin{pmatrix}1&\boldsymbol{0}^T\\
\boldsymbol{0}&-\boldsymbol{I}_{N\times N}\end{pmatrix},\quad\boldsymbol{U}(x,t)=\begin{pmatrix}0&\boldsymbol{u}^T\\
\boldsymbol{u}^*&\boldsymbol{0}_{N\times N}
\end{pmatrix}.
\end{gathered}
\end{aligned}$$

\begin{figure}[htbp]
\centering
\begin{tikzpicture}[scale=1.2, every node/.style={font=\small}]

\begin{scope}
\fill[blue!20] (0,0) rectangle (2.5,2.5);
\fill[blue!20] (-2.5,-2.5) rectangle (0,0);
\clip (0,0) circle (1.4);
\end{scope}

\draw[black, thin] (-2.5,2.5) rectangle (2.5,0);
\draw[black, thin] (-2.5,0) rectangle (2.5,-2.5);
\draw[black, thin] (0,0) circle (1.4);

\draw[->, thick] (-3,0) -- (3,0) node[below] {Re $z$};
\draw[->, thick] (0,-3) -- (0,3) node[left] {Im $z$};
\draw[->, thick] (-1.9,0) -- (-2,0);
\draw[->, thick] (-0.6,0) -- (-0.8,0);
\draw[->, thick] (0.7,0) -- (0.8,0);
\draw[->, thick] (1.9,0) -- (2,0);
\draw[->, thick] (0,0.8) -- (0,0.7);
\draw[->, thick] (0,2) -- (0,1.9);
\draw[->, thick] (0,-0.8) -- (0,-0.7);
\draw[->, thick] (0,-2) -- (0,-1.9);

\fill[black] (0,0) circle (1.2pt);

\node at (-2,0.2) {$\mathbb{R}$};

\node at (-0.5,-1.5) {$\boldsymbol{C}_0$};

\fill[red!60] (0.99,0.99) circle (1.2pt);\node at (1.2,0.99) {$\omega_l$};
\fill[red!60] (0.6,0.6) circle (1.2pt);\node at (0.8,0.6) {$z_m$};
\fill[red!60] (1.6,1.6) circle (1.2pt);\node at (1.85,1.6) {${\dfrac{u_0^2}{z_m^*}}$};
\fill[red!60] (-0.6,-0.6) circle (1.2pt);\node at (-0.85,-0.6) {$-z_m$};
\fill[red!60] (-0.99,-0.99) circle (1.2pt);\node at (-1.25,-0.99) {$-\omega_l$};
\fill[red!60] (-1.6,-1.6) circle (1.2pt);\node at (-1.9,-1.6) {$-{\dfrac{u_0^2}{z_m^*}}$};
\fill[red!60] (0.99,-0.99) circle (1.2pt);\node at (1.2,-0.99) {${\omega^*_l}$};
\fill[red!60] (1.6,-1.6) circle (1.2pt);\node at (1.85,-1.6) {${\dfrac{u_0^2}{z_m^*}}$};
\fill[red!60] (0.6,-0.6) circle (1.2pt);\node at (0.8,-0.6) {$z^*_m$};
\fill[red!60] (-0.6,0.6) circle (1.2pt);\node at (-0.85,0.6) {$-{z^*_m}$};
\fill[red!60] (-0.99,0.99) circle (1.2pt);\node at (-1.25,1.05) {$-\omega_l^*$};
\fill[red!60] (-1.6,1.6) circle (1.2pt);\node at (-1.9,1.6) {$-{\dfrac{u_0^2}{z_m}}$};
\end{tikzpicture}
\caption{Analytic regions $\Xi_+$ (blue) and $\Xi_-$ (white) in the complex $z$-plane, the distribution of discrete spectrum, and the orientation of the jump contour $\Sigma$.}
\label{fig:0}
\end{figure}
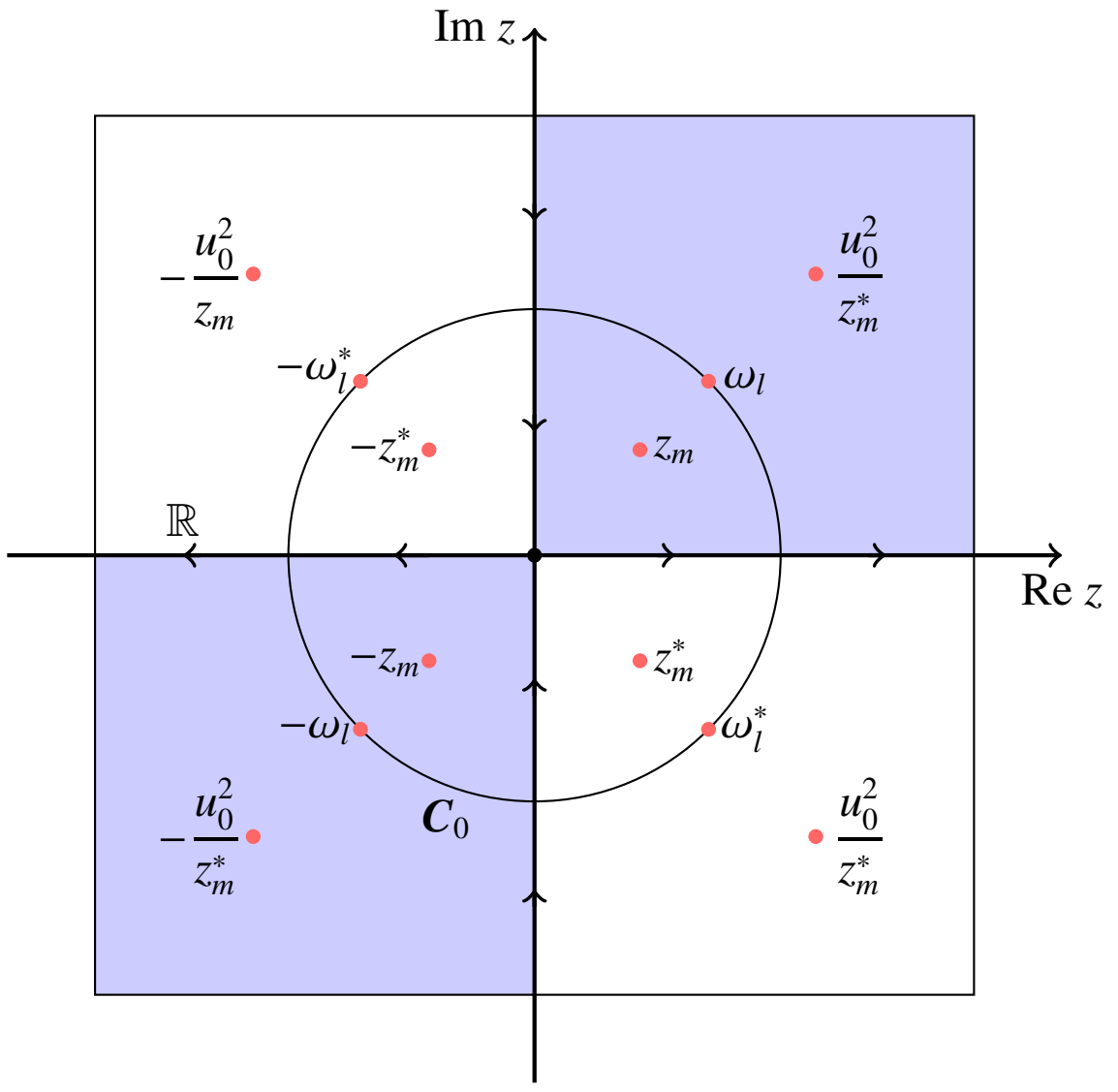

To construct the Jost eigenfunctions and analyze their analytic properties, we consider the limiting spectral problem as $x\to\pm\infty$:
\begin{equation}\label{eq4}
\begin{aligned}
\boldsymbol{\varphi}_x&=\boldsymbol{X}_{\pm}\boldsymbol{\varphi},\\
\boldsymbol{\varphi}_t&=\boldsymbol{Y}_{\pm}\boldsymbol{\varphi},
\end{aligned}
\end{equation}
where
$$\begin{aligned}\boldsymbol{X}_{\pm}
&=-\mathrm{i}k^2{\boldsymbol{\sigma}}_3+\frac{\mathrm{i}}{2}\boldsymbol{U}_{\pm}^2{\boldsymbol{\sigma}}_3-k\boldsymbol{U}_{\pm}{\boldsymbol{\sigma}}_3-\frac{\mathrm{i}}{2}u_0^{2}{\boldsymbol{\sigma}}_3,\\
\boldsymbol{Y}_{\pm}
&=\frac{3\mathrm{i}}{4}u_0^{4}{\boldsymbol{\sigma}}_3-2\mathrm{i}k^{4}{\boldsymbol{\sigma}}_3-2k^{3}\boldsymbol{U}_{\pm}{\boldsymbol{\sigma}}_3+\mathrm{i}k^{2}{\boldsymbol{\sigma}}_3\boldsymbol{U}_{\pm}^{2}
+\frac{\mathrm{i}}{4}\boldsymbol{U}_{\pm}^{4}{\boldsymbol{\sigma}}_3-\frac{k}{2}u_0^2[{\boldsymbol{\sigma}}_3,\boldsymbol{U}_{\pm}]-\mathrm{i}u_0^{2}{\boldsymbol{\sigma}}_3\boldsymbol{U}_{\pm}^{2},\end{aligned}$$
$$\begin{aligned}
\boldsymbol{U}_{\pm}
&=
\begin{pmatrix}
0 & \boldsymbol{u}_{\pm}^{T}\\
\boldsymbol{u}_{\pm}^{*} & \boldsymbol{0}_{N\times N}
\end{pmatrix}.
\end{aligned}$$
The eigenvalues of $\boldsymbol{X}_{\pm}$ are $\mathrm{i}\left(k^2+\frac{1}{2}u_0^2\right)$ with multiplicity  $N-1$, and $\pm \mathrm{i}k\lambda$, where $\lambda^2=k^2+u_0^2$. The corresponding eigenvalues of $\boldsymbol{Y}_{\pm}$ are $\mathrm{i}\left(2k^4-\frac{3}{4}u_0^4\right)$ with multiplicity $N-1$, and $\pm \mathrm{i}k\lambda(2k^2-u_0^2).$ To make $\lambda$ single-valued, we introduce the two-sheeted Riemann surface defined by $\lambda^2=k^2+u_0^2$, with branch points at $k=\pm\mathrm{i}u_0$. This Riemann surface is obtained by taking two copies of the complex $k$-plane slit along the branch cut $[-\mathrm{i}u_0,\mathrm{i}u_0]$ and gluing the opposite banks of the cut crosswise. We label the sheets $S_1$ and $S_2$ so that $\lambda\sim k$ on $S_1$ and $\lambda\sim-k$ on $S_2$ as $k$ tends to infinity. In particular, for $k\in\mathbb{R}\setminus\{0\}$,
$$\lambda(k)=
\begin{cases}
\operatorname{sgn}(k)\sqrt{k^2+u_0^2},
& \text{on } S_1,\\[2mm]
-\operatorname{sgn}(k)\sqrt{k^2+u_0^2},
& \text{on } S_2,
\end{cases}$$
where $\sqrt{k^2+u_0^2}$ denotes the positive square root. The spectral problem on the Riemann surface can be reformulated on a single complex plane using $z=\lambda+k$, which is known as the uniformization variable, with the inverse mapping
\begin{equation}\label{eq5}
k=\frac12\left(z-\frac{u_0^2}{z}\right),\quad\lambda=\frac12\left(z+\frac{u_0^2}{z}\right).
\end{equation}
This transformation maps the sheets $S_1$ and $S_2$ onto the regions $|z|>u_0$ and $0<|z|<u_0$ of the $z$-plane, respectively. As $|k|\to\infty$, the corresponding limits are $|z|\to\infty$ on $S_1$ and $z\to0$ on $S_2$. The two sides of the branch cut on each sheet correspond to the right and left semicircles of $\boldsymbol{C}_0 = \{ z\in\mathbb{C} \mid |z| = u_0 \}$, while the branch points $k=\pm\mathrm{i}u_0$ are mapped to $z=\pm\mathrm{i}u_0$. Furthermore, the regions $\operatorname{Im}(\lambda k)<0$ and $\operatorname{Im}(\lambda k)>0$ are mapped onto $\Xi_-$ and $\Xi_+$ in the $z$-plane, respectively, while $\operatorname{Im}(\lambda k)=0$ corresponds to $z\in\Sigma=(\mathbb{R}\cup\mathrm{i}\mathbb{R})\setminus\{0\}$, where
\begin{equation}\label{eq6}
\Xi_\pm=\{z\in\mathbb{C}:\operatorname{Re}z\,\operatorname{Im}z\gtrless0\}.
\end{equation}
The relevant details are shown in Fig. \ref{fig:0}.

\subsection{Jost eigenfunctions}
\label{subsec:2.2}

Since the asymptotic matrices $\boldsymbol{X}_\pm$ and $\boldsymbol{Y}_\pm$ commute, they can be simultaneously diagonalized by a common eigenvector matrix $\boldsymbol{A}_\pm(z)$ as follows:
\begin{equation}\label{eq7}
\begin{aligned}
\boldsymbol{A}_\pm^{-1}\boldsymbol{X}_\pm\boldsymbol{A}_\pm
&= \mathrm{i}\boldsymbol{\Lambda},\quad\boldsymbol{\Lambda}=\operatorname{diag}\Big(-k\lambda,\,\underbrace{\tfrac12u_0^2+k^2,\ldots,\tfrac12u_0^2+k^2}_{N-1},\,k\lambda\Big),\\
\boldsymbol{A}_\pm^{-1}\boldsymbol{Y}_\pm\boldsymbol{A}_\pm
&= \mathrm{i}\tilde{\boldsymbol{\Lambda}},\quad\tilde{\boldsymbol{\Lambda}}=\operatorname{diag}\!\Big(-k\lambda(2k^2-u_0^2),\,\underbrace{2k^4-\tfrac34u_0^4,\ldots,2k^4-\tfrac34u_0^4}_{N-1},\,k\lambda(2k^2-u_0^2)\Big),
\end{aligned}
\end{equation}
where
\[\boldsymbol{A}_\pm(z)=
\begin{pmatrix}
1 & \boldsymbol{0}_{1\times(N-1)} & -\dfrac{\mathrm{i} u_\pm}{z}\\
\boldsymbol{0}_{(N-1)\times1} & \boldsymbol{I}_{(N-1)\times(N-1)} & \boldsymbol{0}_{(N-1)\times1}\\
-\dfrac{\mathrm{i} u_\pm^*}{z} & \boldsymbol{0}_{1\times(N-1)} & 1
\end{pmatrix}.\]
A direct calculation gives
\begin{equation}\label{eq8}
\det\boldsymbol{A}_\pm=1+\frac{u_0^2}{z^2}=:\alpha,\quad\boldsymbol{A}_\pm^{-1}=
\begin{pmatrix}
\dfrac{1}{\alpha} & \boldsymbol{0}_{1\times(N-1)} & \dfrac{\mathrm{i} u_\pm}{z\alpha}\\[5pt]
\boldsymbol{0}_{(N-1)\times1} & \boldsymbol{I}_{(N-1)\times(N-1)} & \boldsymbol{0}_{(N-1)\times1}\\[3pt]
\dfrac{\mathrm{i} u_\pm^*}{z\alpha} & \boldsymbol{0}_{1\times(N-1)} & \dfrac{1}{\alpha}
\end{pmatrix}.
\end{equation}

The Jost eigenfunctions $\boldsymbol{\varphi}_\pm$ are defined as simultaneous solutions of the Lax pair (\ref{eq3}) satisfying
\begin{equation}\label{eq9}
\boldsymbol{\varphi}_\pm(x,t,z)=\boldsymbol{A}_\pm(z)\mathrm{e}^{\mathrm{i}\boldsymbol{\Theta}(x,t,z)}+o(1),\quad x\to\pm\infty,\quad z\in\Sigma,
\end{equation}
where
$$\boldsymbol{\Theta}(x,t,z)=\operatorname{diag}\left(\theta_1,\theta_2,\ldots,\theta_2,-\theta_1\right)(x,t,z),$$
and
$$\theta_1=-k\lambda x-k\lambda\left(2k^2-u_0^2\right)t,\quad\theta_2=\left(k^2+\frac12u_0^2\right)x+\left(2k^4-\frac34u_0^4\right)t.$$

To eliminate the oscillatory exponential factors, we define the modified Jost eigenfunctions
\begin{equation}\label{eq10}
\boldsymbol{\upsilon}_\pm(x,t,z)=\boldsymbol{\varphi}_\pm(x,t,z)\mathrm{e}^{-\mathrm{i}\boldsymbol{\Theta}(x,t,z)},\quad\boldsymbol{\upsilon}_\pm(x,t,z)=\boldsymbol{A}_\pm(z)+o(1),\quad x\to\pm\infty,
\end{equation}
which satisfy the Lax pair in two equivalent forms:
\begin{subequations}\label{eq11}
\begin{flalign}
\left(\mathrm{e}^{-\mathrm{i}x\boldsymbol{\Lambda}}\boldsymbol{A}_\pm^{-1}\boldsymbol{\upsilon}_\pm\mathrm{e}^{\mathrm{i}x\boldsymbol{\Lambda}}\right)_x
&=\mathrm{e}^{-\mathrm{i}x\boldsymbol{\Lambda}}\boldsymbol{A}_\pm^{-1}\Delta\boldsymbol{X}_\pm\boldsymbol{\upsilon}_\pm\mathrm{e}^{\mathrm{i}x\boldsymbol{\Lambda}},\label{eq11a}\\
\left(\mathrm{e}^{-\mathrm{i}x\boldsymbol{\Lambda}}\boldsymbol{A}_\mp^{-1}\boldsymbol{\upsilon}_\pm\mathrm{e}^{\mathrm{i}x\boldsymbol{\Lambda}}\right)_x
&=\mathrm{e}^{-\mathrm{i}x\boldsymbol{\Lambda}}\boldsymbol{A}_\mp^{-1}\Delta\boldsymbol{X}_\mp\boldsymbol{\upsilon}_\pm\mathrm{e}^{\mathrm{i}x\boldsymbol{\Lambda}},\label{eq11b}
\end{flalign}
\end{subequations}
where $\Delta\boldsymbol X_\pm=\boldsymbol X-\boldsymbol X_\pm$.  For brevity, the variables $x$, $t$, and $z$ are omitted here. Using the asymptotic conditions in Eqs. (\ref{eq10}), we integrate Eq. (\ref{eq11a}) from $\pm\infty$ to $x$ and Eq. (\ref{eq11b}) from $0$ to $x$, yielding the Volterra integral equations
\begin{subequations}\label{eq12}
\begin{flalign}
\boldsymbol{\upsilon}_\pm
={}&
\boldsymbol{A}_\pm+\int_{\pm\infty}^{x}\boldsymbol{A}_\pm\mathrm{e}^{\mathrm{i}(x-\zeta)\boldsymbol{\Lambda}}\boldsymbol{A}_\pm^{-1}\Delta\boldsymbol{X}_\pm(\zeta,t,z)\boldsymbol{\upsilon}_\pm(\zeta,t,z)
\mathrm{e}^{-\mathrm{i}(x-\zeta)\boldsymbol{\Lambda}}\,\mathrm{d}\zeta,\quad x\in\mathbb{R}_\pm,\label{eq12a}
\\
\boldsymbol{\upsilon}_\pm
={}&
\boldsymbol{A}_\mp\mathrm{e}^{\mathrm{i}x\boldsymbol{\Lambda}}\boldsymbol{A}_\mp^{-1}\boldsymbol{\upsilon}_\pm(0,t,z)\mathrm{e}^{-\mathrm{i}x\boldsymbol{\Lambda}}+\int_{0}^{x}\boldsymbol{A}_\mp
\mathrm{e}^{\mathrm{i}(x-\zeta)\boldsymbol{\Lambda}}\boldsymbol{A}_\mp^{-1}\Delta\boldsymbol{X}_\mp(\zeta,t,z)\boldsymbol{\upsilon}_\pm(\zeta,t,z)\mathrm{e}^{-\mathrm{i}(x-\zeta)\boldsymbol{\Lambda}}\,\mathrm{d}\zeta,
\quad x\in\mathbb{R}_\mp,\label{eq12b}
\end{flalign}
\end{subequations}
where $\boldsymbol{\upsilon}_\pm(0,t,z)$ is determined by Eq. (\ref{eq12a}).

A direct calculation yields the following identity
$$\boldsymbol{A}_\pm\mathrm{e}^{\mathrm{i}x\boldsymbol{\Lambda}}\boldsymbol{A}_\pm^{-1}
=\begin{pmatrix}
\dfrac{z^2\mathrm{e}^{-\mathrm{i}k\lambda x} + u_0^2\mathrm{e}^{\mathrm{i}k\lambda x}}{2z\lambda} & \boldsymbol{0}_{1\times(N-1)} & \dfrac{\mathrm{i}u_\pm\bigl(\mathrm{e}^{-\mathrm{i}k\lambda x}-\mathrm{e}^{\mathrm{i}k\lambda x}\bigr)}{2\lambda} \\[10pt]
\boldsymbol{0}_{(N-1)\times 1} & \mathrm{e}^{\mathrm{i}(k^2+\frac{u_0^2}{2})x}\boldsymbol{I}_{(N-1)\times(N-1)} & \boldsymbol{0}_{(N-1)\times 1} \\[10pt]
\dfrac{\mathrm{i}u_\pm^*\bigl(\mathrm{e}^{\mathrm{i}k\lambda x}-\mathrm{e}^{-\mathrm{i}k\lambda x}\bigr)}{2\lambda} & \boldsymbol{0}_{1\times(N-1)} & \dfrac{z^2\mathrm{e}^{\mathrm{i}k\lambda x} + u_0^2\mathrm{e}^{-\mathrm{i}k\lambda x}}{2z\lambda}
\end{pmatrix}.$$
Although $\boldsymbol{A}_\pm$ is singular at the branch points $z=\pm\mathrm{i}u_0$, the above matrix product has removable singularities at these points, with
$$\lim_{z\to\varsigma  \mathrm{i}u_0}\boldsymbol A_\pm\mathrm{e}^{\mathrm{i}x\boldsymbol{\Lambda}}\boldsymbol A_\pm^{-1}
=
\begin{pmatrix}
1+\mathrm{i}u_0^2x & \boldsymbol0_{1\times(N-1)} & \varsigma  \mathrm{i}u_0u_\pm x\\[6pt]
\boldsymbol0_{(N-1)\times1} & \mathrm{e}^{-\frac{\mathrm{i}}{2}u_0^2x}\boldsymbol I_{(N-1)\times(N-1)} & \boldsymbol0_{(N-1)\times1}\\[6pt]
-\varsigma  \mathrm{i}u_0u_\pm^*x & \boldsymbol0_{1\times(N-1)} & 1-\mathrm{i}u_0^2x
\end{pmatrix},
\quad \varsigma =\pm1.$$

We partition the modified Jost eigenfunctions into column blocks, where $\boldsymbol\upsilon_{\pm1}$ and $\boldsymbol\upsilon_{\pm3}$ are the first and last columns of $\boldsymbol\upsilon_{\pm}(x,t,z)$, respectively, and $\boldsymbol{\upsilon}_{\pm2}$ consists of columns $2$ through $N$. Using the above Eqs. (\ref{eq12}), we establish that $\boldsymbol\upsilon_{\pm j}$ are well defined and have the following analyticity and continuity properties.
\newtheorem{proposition}{Proposition}
\begin{proposition}\label{Pro:1}
Suppose that $\boldsymbol u(\cdot,t)-\boldsymbol u_\pm\in L^{1,1}(\mathbb R_\pm)$. Then, as functions of $z$, $\boldsymbol\upsilon_{-1}$ and $\boldsymbol\upsilon_{+3}$ are analytic in $\Xi_+$ and extend continuously to $\Sigma$; $\boldsymbol\upsilon_{+1}$ and $\boldsymbol\upsilon_{-3}$ are analytic in $\Xi_-$ and extend continuously to $\Sigma$; and $\boldsymbol\upsilon_{\pm2}$ is continuous on $\Sigma$.
\end{proposition}
\begin{proof}
We prove that $\boldsymbol{\upsilon}_{-1}$ is analytic in $\Xi_+$ and extends continuously to $\Sigma$, while the same argument yields the corresponding properties for the remaining columns.

For $x\in\mathbb R_-$, the Volterra integral equation (\ref{eq12a}) yields the following relation for its first column:
\begin{equation}\label{eq13}
\boldsymbol\upsilon_{-1}=\begin{pmatrix}
1\\
\boldsymbol0_{(N-1)\times1}\\
-\dfrac{\mathrm{i}u_-^*}{z}
\end{pmatrix}
+
\int_{-\infty}^{x}\boldsymbol S_-(x-\zeta,z)\Delta\boldsymbol X_-(\zeta,t,z)\boldsymbol\upsilon_{-1}(\zeta,t,z)\,\mathrm{d}\zeta,
\end{equation}
where
$$\boldsymbol S_\pm(x-\zeta,z)
=
\boldsymbol A_\pm\mathrm{e}^{\mathrm{i}(x-\zeta)\boldsymbol \Lambda}
\boldsymbol A_\pm^{-1}\mathrm{e}^{\mathrm{i}(x-\zeta)k\lambda}
=
\begin{pmatrix}
\dfrac{z^2+u_0^2\mathrm{e}^{2\mathrm{i}k\lambda(x-\zeta)}}{2z\lambda} & \boldsymbol0_{1\times(N-1)} & \dfrac{\mathrm{i}u_\pm\bigl(1-\mathrm{e}^{2\mathrm{i}k\lambda(x-\zeta)}\bigr)}{2\lambda}\\[6pt]
\boldsymbol0_{(N-1)\times1} & \mathrm{e}^{\frac{\mathrm{i}}2z^2(x-\zeta)}\boldsymbol I_{(N-1)\times(N-1)} & \boldsymbol0_{(N-1)\times1}\\[6pt]
\dfrac{\mathrm{i}u_\pm^*\bigl(\mathrm{e}^{2\mathrm{i}k\lambda(x-\zeta)}-1\bigr)}{2\lambda} & \boldsymbol0_{1\times(N-1)} & \dfrac{z^2\mathrm{e}^{2\mathrm{i}k\lambda(x-\zeta)}+u_0^2}{2z\lambda}
\end{pmatrix}.$$
For the integral equation (\ref{eq13}), we construct the Neumann series $\boldsymbol\upsilon_{-1}=\sum_{n=0}^{\infty}\boldsymbol\upsilon_n(x,t,z)$, whose terms are defined by
$$\boldsymbol\upsilon_0=
\begin{pmatrix}
1\\
\boldsymbol0_{(N-1)\times1}\\
-\dfrac{\mathrm{i}u_-^*}{z}
\end{pmatrix},\quad
\boldsymbol\upsilon_{n+1}
=
\int_{-\infty}^{x}\boldsymbol E(x,\zeta,t,z)\boldsymbol\upsilon_n(\zeta,t,z)\,\mathrm{d}\zeta,\quad\boldsymbol E
=
\boldsymbol S_-(x-\zeta,z)\Delta\boldsymbol X_-(\zeta,t,z).$$
Let $ K$ be an arbitrary nonempty compact subset of $\Xi_+\cup\Sigma$. For $z\in K$, we have $0<\delta_\varepsilon\leq|z|\leq\delta_\infty<\infty,$ and hence
$$|k|=\frac12\left|z-\frac{u_0^2}{z}\right|\leq c_K,\quad c_K=\frac12\left(\delta_\infty+\frac{u_0^2}{\delta_\varepsilon}\right)<\infty.$$
The matrix $\boldsymbol S_-(x-\zeta,z)$ is analytic in $\Xi_+$ and extends continuously to $\Sigma$. Furthermore, for $z\in K$ and $\zeta\leq x$, its explicit expression yields
\begin{equation}\label{eq14}
\|\boldsymbol S_-(x-\zeta,z)\|\leq c_1\bigl(1+|x-\zeta|\bigr)\leq c_1\bigl(1+|\zeta|\bigr),
\end{equation}
where $c_1>0$ depends only on $ K$ and $u_0$, and $\|\cdot\|$ denotes the vector $1$-norm and its induced matrix norm. Since $\boldsymbol u$ is continuous with respect to $x$ and satisfies the boundary conditions (\ref{eq2b}), it is bounded on $\mathbb R$. Set $c_u=\mathop {\sup }\limits_{x \in \mathbb{R}}\|\boldsymbol u(x,t)\|<\infty.$ For all $x\in\mathbb R_-$ and $z\in K$, we have
\begin{equation}\label{eq15}
\begin{aligned}
\|\Delta\boldsymbol X_-\|
&\leq |k|\,\|\boldsymbol U-\boldsymbol U_-\|+\frac12\|\boldsymbol U^2-\boldsymbol U_-^2\|=|k|\,\|\boldsymbol U-\boldsymbol U_-\|+\frac12\left\|(\boldsymbol U-\boldsymbol U_-)\boldsymbol U+\boldsymbol U_-(\boldsymbol U-\boldsymbol U_-)\right\|\\
&\leq\left(|k|+\frac12\bigl(\|\boldsymbol U\|+u_0\bigr)\right)\|\boldsymbol U-\boldsymbol U_-\|\leq\left(c_K+\frac12(c_u+u_0)\right)\|\boldsymbol u-\boldsymbol u_-\|.
\end{aligned}
\end{equation}
From the two equations above, we get
\begin{equation}\label{eq16}
\|\boldsymbol E(x,\zeta,t,z)\|\leq f(\zeta),\quad f(\zeta)=c_1(1+|\zeta|)\left(c_K+\frac12(c_u+u_0)\right)\|\boldsymbol u(\zeta,t)-\boldsymbol u_-\|,
\end{equation}
where $f\in L^1(\mathbb R_-)$ by the assumption that $\boldsymbol u(\cdot,t)-\boldsymbol u_\pm\in L^{1,1}(\mathbb R_\pm)$.

By induction, we get
\begin{equation}\label{eq17}
\|\boldsymbol\upsilon_n(x,t,z)\|\leq\frac{c_3}{n!}F(x)^n\leq\frac{c_3}{n!}\tilde F_-^{\,n},\quad n\geq0,
\end{equation}
where
$$c_3=1+\frac{u_0}{\delta_\varepsilon},\quad F(x)=\int_{-\infty}^{x}f(\zeta)\,\mathrm{d}\zeta, \quad \tilde F_-=\int_{-\infty}^{0}f(\zeta)\,\mathrm{d}\zeta<\infty.$$
Consequently,
\begin{equation}\label{eq17a}
\sum_{n=0}^{\infty}\|\boldsymbol\upsilon_n(x,t,z)\|\leq c_3\sum_{n=0}^{\infty}\frac{\tilde F_-^{\,n}}{n!}=c_3\,\mathrm{e}^{\tilde F_-}<\infty.
\end{equation}
The Weierstrass M-test then implies that the Neumann series converges absolutely and uniformly for $x\in\mathbb R_-$ and $z\in K$. Together with the integrable kernel bound, this uniform convergence ensures that the $\boldsymbol\upsilon_{-1}$ satisfies the Eq. (\ref{eq13}). Moreover, the same estimates show that a sufficiently high power of the integral operator is a contraction, establishing uniqueness of the bounded solution.

We next prove the continuity and analyticity of each term in the Neumann series with respect to $z$, starting with continuity. The initial term $\boldsymbol\upsilon_0$ is analytic in $\Xi_+$ and admits a continuous extension to $\Sigma$. Assume that $\boldsymbol\upsilon_{n-1}$ is continuous on $\Xi_+\cup\Sigma$. Take any $z_0\in\Xi_+\cup\Sigma$ and a compact relative neighborhood ${K_{{z_0}}} \subset {\Xi _ + } \cup \Sigma $ of $z_0$. By Eqs. (\ref{eq16}) and (\ref{eq17}), we derive
\begin{equation}\label{eq18}
\|\boldsymbol E(x,\zeta,t,z)\boldsymbol\upsilon_{n-1}(\zeta,t,z)\|\leq\frac{c_3}{(n-1)!}f(\zeta)\tilde F_-^{\,n-1},
\end{equation}
where the right-hand side is integrable with respect to $\zeta$ and independent of $z$. Applying the dominated convergence theorem, we obtain
$$\begin{aligned}
\lim_{\substack{z\to z_0\\z\in\Xi_+\cup\Sigma}}\boldsymbol\upsilon_n(x,t,z)=\int_{-\infty}^{x}\boldsymbol E(x,\zeta,t,z_0)\boldsymbol\upsilon_{n-1}(\zeta,t,z_0)\,\mathrm{d}\zeta=\boldsymbol\upsilon_n(x,t,z_0).
\end{aligned}$$
Therefore, $\boldsymbol\upsilon_n$ is continuous on $\Xi_+\cup\Sigma$. Next, we prove that $\boldsymbol\upsilon_n$ is analytic in $\Xi_+$. Suppose that $\boldsymbol\upsilon_{n-1}$ is analytic in $\Xi_+$, and let $C\subset\Xi_+$ be an arbitrary piecewise smooth closed curve. The above estimates remain valid on a compact set containing $C$. Hence, from the recurrence relation, we get
$$\oint_C\boldsymbol\upsilon_n(x,t,z)\mathrm{d}z = \oint_C\int_{-\infty}^{x} \boldsymbol E(x,\zeta,t,z)\boldsymbol\upsilon_{n-1}(\zeta,t,z) \,\mathrm{d}\zeta\,\mathrm{d}z.$$
Moreover, Eq. (\ref{eq18}) implies that
$$\int_C\int_{-\infty}^{x}\|\boldsymbol E(x,\zeta,t,z)\boldsymbol\upsilon_{n-1}(\zeta,t,z)\|\,\mathrm{d}\zeta\,|\mathrm{d}z|<\infty.$$
Therefore, Fubini's theorem allows us to interchange the order of integration. Since $\boldsymbol E$ is analytic as a function of $z$, and $\boldsymbol\upsilon_{n-1}$ is analytic in $\Xi_+$ by the induction hypothesis, Cauchy's theorem yields
$$\begin{aligned}
\oint_C\boldsymbol\upsilon_n(x,t,z)\mathrm{d}z
&=
\int_{-\infty}^{x}\oint_C\boldsymbol E(x,\zeta,t,z)\boldsymbol\upsilon_{n-1}(\zeta,t,z)\,\mathrm{d}z\,\mathrm{d}\zeta=0.
\end{aligned}$$
Together with the continuity established above, Morera's theorem implies that $\boldsymbol\upsilon_n$ is analytic in $\Xi_+$. Since the Neumann series converges uniformly on every compact subset of $\Xi_+\cup\Sigma$, and each term is continuous on this set and analytic in $\Xi_+$, its sum $\boldsymbol\upsilon_{-1}$ is analytic in $\Xi_+$ and extends continuously to $\Sigma$.

For $x\in\mathbb R_+$, the Volterra integral equation (\ref{eq12b}) yields
$$\boldsymbol{\upsilon}_{-1}=\boldsymbol S_+(x,z)\boldsymbol{\upsilon}_{-1}(0,t,z)+\int_0^x\boldsymbol S_+(x-\zeta,z)\Delta\boldsymbol X_+(\zeta,t,z)\boldsymbol{\upsilon}_{-1}(\zeta,t,z)\,\mathrm{d}\zeta .$$
To control the growth of the integral kernel on the positive half-axis and obtain a uniform estimate, we introduce the weight function $1+|x|$ and divide both sides of the above integral equation by $1+|x|$, which gives
$$\frac{\boldsymbol{\upsilon}_{-1}}{1+|x|}=\frac{\boldsymbol S_+(x,z)}{1+|x|}\boldsymbol{\upsilon}_{-1}(0,t,z)+\int_0^x\boldsymbol S_+(x-\zeta,z)\frac{1+|\zeta|}{1+|x|}\Delta\boldsymbol X_+(\zeta,t,z)\frac{\boldsymbol{\upsilon}_{-1}(\zeta,t,z)}
{1+|\zeta|}\,\mathrm{d}\zeta .$$
We seek a Neumann series solution in the form $\frac{\boldsymbol{\upsilon}_{-1}}{1+|x|}=\sum_{n=0}^{\infty}\tilde{\boldsymbol{\upsilon}}_n(x,t,z),$ where
$$ \tilde{\boldsymbol{\upsilon}}_0=\frac{\boldsymbol S_+(x,z)}{1+|x|}\boldsymbol{\upsilon}_{-1}(0,t,z),\quad\tilde{\boldsymbol{\upsilon}}_{n+1}=\int_0^x\tilde{\boldsymbol E}(x,\zeta,t,z)\tilde{\boldsymbol{\upsilon}}_n(\zeta,t,z)\,d\zeta ,\quad\tilde{\boldsymbol E}=\frac{1+|\zeta|}{1+|x|}\boldsymbol S_+(x-\zeta,z)\Delta\boldsymbol X_+.$$
As in Eq. (\ref{eq14}), let $ K$ be an arbitrary compact set in $\Xi_+\cup\Sigma$. For $0\le\zeta\le x$ and $z \in  K$, we have
$$\|\boldsymbol{S}_+ (x - \zeta, z)\| \le c_1\bigl(1 + |x - \zeta|\bigr) \le c_1\bigl(1 + |x|\bigr).$$
Consequently,
$$\|\tilde{\boldsymbol{\upsilon}}_0\|\leq c_1c_3\mathrm{e}^{\tilde F_-}:= c_4 ,\quad\|\tilde{\boldsymbol E}(x,\zeta,t,z)\|\leq\tilde f(\zeta),$$
where
$$\tilde f(\zeta)=c_1(1+|\zeta|)\left(c_K+\frac12(c_u+u_0)\right)\|\boldsymbol u(\zeta,t)-\boldsymbol u_+\|,\quad\tilde F_+=\int_0^\infty\tilde f(\zeta)\mathrm{d}\zeta<+\infty .$$
Using the same inductive estimate as in Eq. (\ref{eq17}), we obtain
$$\|\tilde{\boldsymbol\upsilon}_n(x,t,z)\|\le \frac{c_4}{n!}\left(\int_0^x\tilde f(\zeta)\,\mathrm d\zeta\right)^n\le \frac{c_4}{n!}\tilde F_+^{\,n}.$$
Therefore, the Neumann series exhibits absolute and uniform convergence for $x\in\mathbb R_+$ and $z\in K$, with
\begin{equation}\label{eq19}
\left\|\frac{\boldsymbol{\upsilon}_{-1}}{1+|x|}\right\|=\left\|\sum_{n=0}^{\infty}\tilde{\boldsymbol{\upsilon}}_n\right\|\leq c_4\mathrm{e}^{\tilde F_+}.
\end{equation}
The analyticity in $\Xi_+$ and the continuous extension to $\Sigma$ then follow by the same argument as for $x\in\mathbb R_-$.\end{proof}

From the above proposition and Eq. (\ref{eq10}), it follows that the columns of $\boldsymbol{\varphi}_\pm$ have the same properties as those of $\boldsymbol\upsilon_\pm$.

\subsection{Scattering matrix}
\label{subsec:2.3}

A direct calculation using Eq. (\ref{eq4}) yields $\operatorname{Tr} \boldsymbol{X}_\pm = \mathrm{i}(N-1)(k^2+\frac{1}{2} u_0^2)$, and $\operatorname{Tr} \boldsymbol{Y}_\pm = \mathrm{i}(N-1)(2k^4-\frac{1}{4} u_0^4)$. By the Abel's formula, ${\partial _x}\left( {\det {\boldsymbol{\upsilon} _ \pm }} \right) = {\partial _t}\left( {\det {\boldsymbol{\upsilon} _ \pm }} \right) = 0$, so that the determinant of ${\boldsymbol{\upsilon} _ \pm }$ is independent of $x$ and $t$. Hence
\begin{equation}\label{eq20}
\det {\boldsymbol{\upsilon} _ \pm }=\det \boldsymbol A_\pm=\alpha,\quad z\in\mathring{\Sigma},
\end{equation}
where $\mathring{\Sigma}=\Sigma\setminus\{\pm \mathrm{i} u_0\}$. Combining this with Eq. (\ref{eq10}), we conclude that
\begin{equation}\label{eq20a}
\det\boldsymbol\varphi_\pm=\alpha\mathrm{e}^{\mathrm{i}(N-1)\theta_2},\quad z\in\mathring{\Sigma}.
\end{equation}

For notational simplicity, the eigenfunctions are written as functions of $z$ alone in what follows, with their dependence on $(x,t)$ understood. Since $\boldsymbol\varphi_+$ and $\boldsymbol\varphi_-$ are fundamental matrix solutions of the Lax pair (\ref{eq3}), they are related by $(x,t)$-independent matrices $\boldsymbol S(z)$ and $\boldsymbol G(z)$ for $ z\in\mathring{\Sigma}$, such that
\begin{subequations}\label{eq21}
\begin{align}
\boldsymbol\varphi_-(z) &= \boldsymbol\varphi_+(z) \boldsymbol G(z),\label{eq21b}\\
\boldsymbol\varphi_+(z) &= \boldsymbol\varphi_-(z) \boldsymbol B(z),\label{eq21a}
\end{align}
\end{subequations}
where
$$ \boldsymbol B(z)=
\begin{pmatrix}
b_{11} &  b_{12} & b_{13} \\
b_{21} &  b_{22} & b_{23} \\
b_{31} &  b_{32} & b_{33}
\end{pmatrix},
\quad
\boldsymbol G(z)=\boldsymbol B^{-1}(z)=
\begin{pmatrix}
g_{11} & g_{12} & g_{13} \\
g_{21} & g_{22} & g_{23} \\
g_{31} & g_{32} & g_{33}
\end{pmatrix}.$$
The matrices $\boldsymbol B(z)$ and $\boldsymbol G(z)$ are called scattering matrices, and their entries $b_{ij}$ and $g_{ij}$ are called scattering coefficients. With respect to the block structure displayed above,  $b_{11}$, $b_{13}$, $b_{31}$ and $b_{33}$ are scalars; $b_{12}$ and $b_{32}$ are $1\times(N-1)$ row vectors; $b_{21}$ and $b_{23}$ are $(N-1)\times1$ column vectors; and $b_{22}$ is an $(N-1) \times (N-1)$ matrix. The blocks $g_{ij}$ have the same dimensions as $b_{ij}$. Furthermore, taking determinants of the scattering relations (\ref{eq21}) and using Eq. (\ref{eq20a}), we arrive at $\det \boldsymbol S(z) = \det \boldsymbol G(z) = 1.$

Combining Eqs. (\ref{eq12b}) and (\ref{eq21}), and using the asymptotic relation $\mathop {\lim }\limits_{x \to  \pm \infty } \boldsymbol A_ \pm ^{ - 1}{\boldsymbol\upsilon _ \pm } = {\boldsymbol I_{(N + 1) \times (N + 1)}}$, we obtain
\begin{subequations}\label{eq22}
\begin{align}
\boldsymbol G(z) &= \boldsymbol A_+^{-1} \boldsymbol \upsilon_-(0,0,z) + \int_{0}^{+\infty} \mathrm{e}^{-\mathrm{i}\zeta \boldsymbol \Lambda} \boldsymbol A_+^{-1} \Delta \boldsymbol X_+(\zeta,0,z) \boldsymbol \upsilon_-(\zeta,0,z) \mathrm{e}^{\mathrm{i}\zeta \boldsymbol \Lambda} \mathrm{d}\zeta,\label{eq22a} \\
\boldsymbol B(z) &= \boldsymbol A_-^{-1} \boldsymbol \upsilon_+(0,0,z) + \int_{0}^{-\infty} \mathrm{e}^{-\mathrm{i}\zeta \boldsymbol \Lambda} \boldsymbol A_-^{-1} \Delta \boldsymbol X_-(\zeta,0,z) \boldsymbol \upsilon_+(\zeta,0,z) \mathrm{e}^{\mathrm{i}\zeta \boldsymbol \Lambda} \mathrm{d}\zeta.\label{eq22b}
\end{align}
\end{subequations}

The integral equations (\ref{eq22}) together with Proposition \ref{Pro:1} yield the following properties of the scattering coefficients.

\begin{proposition}\label{Pro:2}
Under the assumptions of Proposition \ref{Pro:1}, the scattering coefficients $g_{11}$, $g_{33}$, $b_{11}$, and $b_{33}$ are well-defined for $ z\in\mathring{\Sigma}$. Moreover, $g_{11}$ and $b_{33}$ admit analytic extensions to $\Xi_+$, whereas $g_{33}$ and $b_{11}$ admit analytic extensions to $\Xi_-$.
\end{proposition}
\begin{proof}
We show only that $g_{11}$ can be analytically continued to $\Xi_+$, while the corresponding properties of the other scattering coefficients follow by the same method.

The $(1, 1)$ entry of Eq. (\ref{eq22a}) yields
\begin{equation}\label{eq23}
\begin{aligned}
g_{11}=\left[\boldsymbol A_+^{-1} \boldsymbol\upsilon_{-1}(0,0,z)\right]_1 + \int_{0}^{+\infty} \left[\boldsymbol A_+^{-1} \Delta \boldsymbol X_+(\zeta,0,z) \boldsymbol \upsilon_{-1}(\zeta,0,z)\right]_1 \mathrm{d}\zeta,
\end{aligned}
\end{equation}
where the subscript $1$ denotes the first component of the vector.
From Eq. (\ref{eq19}), it follows that for any compact set $K\subset\Xi_+$ and all $z\in K$, $\frac{\boldsymbol\upsilon_{-1}(x,0,z)}{1+\left|x\right|}$ is uniformly bounded for $x\in\mathbb{R}_+$; therefore, there exists a constant $\tilde{c}_K>0$ such that
$$\begin{aligned}
\left| \left[\boldsymbol A_+^{-1}\Delta \boldsymbol X_+(\zeta,0,z) \boldsymbol\upsilon_{-1}(\zeta,0,z)\right]_1 \right|
&\le \left\| \boldsymbol  A_+^{-1}\Delta \boldsymbol  X_+(\zeta,0,z) \boldsymbol\upsilon_{-1}(\zeta,0,z) \right\|
\le \tilde{c}_K (1+|\zeta|) \left\| \Delta \boldsymbol  X_+(\zeta,0,z) \right\| \\
&\le \tilde{c}_K \left(c_K + \frac12\big(c_u + u_0\big)\right)(1+|\zeta|) \left\| \boldsymbol u(\zeta,0)-\boldsymbol  u_+ \right\|.
\end{aligned}$$
Under the assumption that $\boldsymbol u(\cdot,0)-\boldsymbol u_+\in L^{1,1}(\mathbb R_+)$, we have $(1+|\zeta|) \left\| \boldsymbol u(\zeta,0)-\boldsymbol  u_+ \right\|\in L^1(\mathbb R_+)$, which establishes the required $L^1$-bound for the integrand. Moreover, Proposition \ref{Pro:1} ensures that $\boldsymbol\upsilon_{-1}$ is analytic in $z\in\Xi_+$. Hence, the integrand in Eq. (\ref{eq23}) is analytic in $\Xi_+$. Therefore, by the analytic theorem for parameter integrals, the integral term in Eq. (\ref{eq23}) is analytic in $\Xi_+$. Since $[\boldsymbol A_+^{-1} \boldsymbol\upsilon_{-1}(0,0,z)]_1$ is also analytic in $\Xi_+$, the right-hand side of Eq. (\ref{eq23}) provides an analytic extension of $g_{11}$ to $\Xi_+$.\end{proof}

\subsection{Auxiliary eigenfunctions}
\label{subsec:2.4}

Formulating the inverse problem as a Riemann-Hilbert problem requires a complete set of eigenfunctions that are analytic in $\Xi_\pm$. However, as shown in Proposition \ref{Pro:1}, the middle blocks $\boldsymbol{\varphi}_{\pm2}$ are not analytic in $\Xi_\pm$. To address this issue, following \cite{A50,A51}, we employ tensor products and generalized cross products of vectors in $\mathbb{C}^{N+1}$ to construct auxiliary Jost eigenfunctions.  For clarity, we use $\boldsymbol{\upsilon}_{\pm,j}$ to represent the $j$-th column of $\boldsymbol{\upsilon}_{\pm}$. Accordingly, the column-block decomposition of $\boldsymbol{\upsilon}_{\pm}$ is given by
$${\boldsymbol\upsilon _{ \pm 1}} = {\boldsymbol\upsilon _{ \pm ,1}},\quad{\boldsymbol\upsilon _{ \pm 2}} = \left( {{\boldsymbol\upsilon _{ \pm ,2}}, \dots ,{\boldsymbol\upsilon _{ \pm ,N}}} \right),\quad{\boldsymbol\upsilon _{ \pm 3}} = {\boldsymbol\upsilon _{ \pm ,N+1}}.$$
The same notation is adopted for the corresponding column blocks of $\boldsymbol\varphi_\pm(x,t,z)$.

To construct the auxiliary eigenfunctions required below, for $n = 2, \ldots, N$, we define
\begin{equation}\label{eq24}
\begin{aligned}
\boldsymbol p_{n+} (x,t;z) &= \left( \boldsymbol\upsilon_{+,2} \wedge \dots \wedge \widehat{\boldsymbol\upsilon}_{+,n} \wedge \dots \wedge \boldsymbol\upsilon_{+,N+1} \right)(x,t,z),\\
\boldsymbol p_{n-} (x,t;z) &= \left( \boldsymbol\upsilon_{+,1} \wedge \dots \wedge \widehat{\boldsymbol\upsilon}_{+,n}  \wedge \dots \wedge \boldsymbol\upsilon_{+,N} \right)(x,t;z),\\
\boldsymbol P_{n+} (x,t;z) &= \left( \boldsymbol\varphi_{+,2} \wedge \dots \wedge \widehat{\boldsymbol\varphi}_{+,n}  \wedge \dots \wedge \boldsymbol\varphi_{+,N+1} \right)(x,t;z),\\
\boldsymbol P_{n-} (x,t;z) &= \left( \boldsymbol\varphi_{+,1} \wedge \dots \wedge \widehat{\boldsymbol\varphi}_{+,n}  \wedge \dots \wedge \boldsymbol\varphi_{+,N} \right)(x,t;z),
\end{aligned}
\end{equation}
where $\wedge$ denotes the wedge product and $\widehat{\scalebox{0.65}{$\bullet$}}$ indicates that the corresponding factor is omitted from the product. Let $\breve{\boldsymbol{p}}_{n\pm}(x,t,z) = (\boldsymbol A_+^{-1})^{(N-1)} \boldsymbol p_{n\pm}(x,t,z)$. To specify the operators appearing below, we introduce some notation on the exterior power $\Lambda^{N-1}(\mathbb{C}^{N+1})$. For a linear map $\boldsymbol A$ on $\mathbb C^{N+1}$, we let $\boldsymbol A_{(N-1)}$ denote the associated linear map on $\Lambda^{N-1}(\mathbb C^{N+1})$, defined by ${\boldsymbol{A} _{(N - 1)}}\left( {{\boldsymbol{v}_1} \wedge  \dots  \wedge {\boldsymbol{v}_{N - 1}}} \right) = \frac{1}{{N - 1}}\sum\nolimits_{j = 1}^{N - 1} {{\boldsymbol{v}_1} \wedge  \dots  \wedge \boldsymbol{A} {\boldsymbol{v}_j} \wedge  \dots  \wedge {\boldsymbol{v}_{N - 1}}} $. Analogously, $\boldsymbol A^{(N-1)}$ denotes the map induced by $\boldsymbol A$ on the same exterior power, given by ${\boldsymbol{A}^{(N - 1)}}\left( {{\boldsymbol{v}_1} \wedge  \dots  \wedge {\boldsymbol{v}_{N - 1}}} \right) = \boldsymbol{A}{\boldsymbol{v}_1} \wedge  \dots  \wedge \boldsymbol{A}{\boldsymbol{v}_{N - 1}}$. The identity operator on $\Lambda^{N-1}(\mathbb C^{N+1})$ is denoted by ${{I}_{\Lambda^{N-1}(\mathbb{C}^{N+1})}}$. For brevity, the variables $x$, $t$, and $z$ are omitted in what follows. Using the Lax pair (\ref{eq11a}), we obtain the following differential equations and asymptotic conditions for $\breve{\boldsymbol{p}}_{n\pm}$:
\begin{equation}\label{eq25}
\begin{aligned}
&\partial_x \breve{\boldsymbol{p}}_{n\pm} = \breve{\boldsymbol{A}}_\pm\breve{\boldsymbol{p}}_{n\pm} + \breve{\boldsymbol{X}}\breve{\boldsymbol{p}}_{n\pm},\\
&\lim_{x\to+\infty} \breve{\boldsymbol{p}}_{n+} = \boldsymbol{e}_2 \wedge \dots \wedge \widehat{\boldsymbol{e}}_n \wedge \dots \wedge \boldsymbol{e}_{N+1},\\
&\lim_{x\to+\infty} \breve{\boldsymbol{p}}_{n-} = \boldsymbol{e}_1 \wedge \dots \wedge \widehat{\boldsymbol{e}}_n \wedge \dots \wedge \boldsymbol{e}_{N},
\end{aligned}
\end{equation}
where
$$\begin{array}{l}
{{\breve {\boldsymbol{A}}}_ \pm }(z) =    \mathrm{i}\left( {(2-N)\left( {{k^2} + \frac{u_0^2}{2}} \right) \pm k\lambda } \right){{I}_{\Lambda^{N-1}(\mathbb{C}^{N+1})}}+\mathrm{i}(N - 1)\boldsymbol \Lambda_{(N-1)},\\
\breve {\boldsymbol{X}}(x,t,z) = (N - 1){(\boldsymbol{A}_ + ^{ - 1})^{(N - 1)}}\left(\boldsymbol{X}_{(N-1)} - {(\boldsymbol{X}_ {+})_{(N-1)} }\right)\boldsymbol{A}_ +^{(N - 1)},
\end{array}$$
and $\{ {\boldsymbol{e}_1}, \dots ,{\boldsymbol{e}_{N+1}}\} $ is the standard basis of $\mathbb{C}^{N+1}$. To represent the differential equations in Eqs. (\ref{eq25}) as a finite-dimensional matrix system, we choose the standard basis $\{ {\boldsymbol{\Gamma} _1}, \dots ,{\boldsymbol{\Gamma} _{N(N + 1)/2}}\} $ of the exterior power $\Lambda^{N-1}(\mathbb{C}^{N+1})$, which is defined explicitly as follows:
\begin{equation}\label{eq26}
{\boldsymbol{\Gamma} _\iota} = \left\{ {\begin{array}{*{20}{l}}
{{\boldsymbol{e}_1} \wedge  \dots  \wedge {\widehat{\boldsymbol{e}}_{\iota + 1}} \wedge  \dots  \wedge {\boldsymbol{e}_N},}&{1\le\iota\le N-1,}\\
{{\boldsymbol{e}_2} \wedge  \dots  \wedge {\widehat{\boldsymbol{e}}_{\iota - N + 2}} \wedge  \dots  \wedge {\boldsymbol{e}_{N + 1}},}&{ N\le\iota\le 2N - 2,}\\
{{\boldsymbol{e}_2} \wedge {\boldsymbol{e}_3} \wedge  \dots  \wedge {\boldsymbol{e}_N},}&{\iota = 2N - 1,}
\end{array}} \right.
\end{equation}
and
$${\boldsymbol{\Gamma} _{2N}}, \dots ,{\boldsymbol{\Gamma} _{\frac{{N(N + 1)}}{2}}} \in \left\{ {{\boldsymbol{e}_1} \wedge {\boldsymbol{e}_{{j_1}}} \wedge  \dots  \wedge {\boldsymbol{e}_{{j_{N - 3}}}} \wedge {\boldsymbol{e}_{N + 1}}\left| {2 \le {j_1} <  \dots  < {j_{N - 3}} \le N} \right.} \right\}.$$
The definition of ${{\breve {\boldsymbol{A}}}_ \pm }(z)$ and the construction of this standard basis imply that each basis vector $\boldsymbol{\Gamma}_j$ is an eigenvector of ${{\breve {\boldsymbol{A}}}_ \pm }(z)$. Let $\boldsymbol\Gamma=\left( {\boldsymbol{\Gamma} _1}, \dots ,{\boldsymbol{\Gamma} _{N(N + 1)/2}}\right) $ denote the matrix whose columns are these basis vectors. Consequently,
\begin{equation}\label{eq27}
{\breve {\boldsymbol{A}}_ \pm }(z)\boldsymbol{\Gamma}  = \boldsymbol{\Gamma}\check{\boldsymbol{A}}_\pm(z),
\end{equation}
where $\check{\boldsymbol{A}}_\pm(z)$ are diagonal matrices whose diagonal entries are corresponding eigenvalues of ${\breve {\boldsymbol{A}}_ \pm }(z)$, given explicitly by
\begin{equation}\label{eq28}
\begin{aligned}
\check{\boldsymbol{A}}_+(z) &= \operatorname{diag}\Big[
\underbrace{-2\mathrm{i}k\lambda,\,\dots,\,-2\mathrm{i}k\lambda}_{N - 1},
\underbrace{0,\,\dots,\,0}_{N - 1},
\mathrm{i}\Big(\frac{u_0^2}{2}+k^2- k\lambda\Big),
\underbrace{-\mathrm{i}\Big(\frac{u_0^2}{2}+k^2 + k\lambda\Big),\,\dots,\,-\mathrm{i}\Big( \frac{u_0^2}{2}+k^2  + k\lambda\Big)}_{{(N - 1)(N - 2)}/{2}}
\Big],\\
\check{\boldsymbol{A}}_-(z) &= \operatorname{diag}\Big[
\underbrace{0,\,\dots,\,0}_{N - 1},
\underbrace{2\mathrm{i}k\lambda,\,\dots,\,2\mathrm{i}k\lambda}_{N - 1},
\mathrm{i}\Big( \frac{u_0^2}{2}+k^2 + k\lambda\Big),
\underbrace{-\mathrm{i}\Big( \frac{u_0^2}{2}+k^2 - k\lambda\Big),\,\dots,\,-\mathrm{i}\Big( \frac{u_0^2}{2}+k^2 - k\lambda\Big)}_{{(N - 1)(N - 2)}/{2}}
\Big].
\end{aligned}
\end{equation}

For the linear map $\breve {\boldsymbol{X}}$ on the exterior power $\Lambda^{N-1}(\mathbb C^{N+1})$, the image of each basis vector admits a unique expansion in the standard basis. Consequently, there exists a square matrix $\check{\boldsymbol{X}}(x,t,z)$ of order $N(N+1)/2$, whose entries are determined by the action of $\breve {\boldsymbol{X}}$ on the standard basis $\boldsymbol{\Gamma}$ and which satisfies ${\breve {\boldsymbol{X}} }\boldsymbol{\Gamma}  = \boldsymbol{\Gamma}\check{\boldsymbol{X}}$. Similarly, $\breve{\boldsymbol{p}}_{n\pm}$ can be expressed as $\breve{\boldsymbol{p}}_{n\pm}  = \boldsymbol{\Gamma}\,\check{\boldsymbol{p}}_{n\pm}$, where $\check{\boldsymbol{p}}_{n\pm}(x,t,z)\in\mathbb C^{\frac{N(N+1)}2}$ is the corresponding coordinate vector with respect to $\boldsymbol{\Gamma}$. In these coordinates, Eqs. (\ref{eq25}) can be recast as
\begin{equation}\label{eq29}
\begin{aligned}
&{\partial _x}\check{\boldsymbol{p}}_{n\pm} = \check{\boldsymbol{A}}_\pm\check{\boldsymbol{p}}_{n\pm} + \check{\boldsymbol{X}}\check{\boldsymbol{p}}_{n\pm},\\
&\lim_{x\to+\infty} \check{\boldsymbol{p}}_{n+} = \check{\boldsymbol{e}}_{n+N-2},\\
&\lim_{x\to+\infty} \check{\boldsymbol{p}}_{n-} = \check{\boldsymbol{e}}_{n-1},
\end{aligned}
\end{equation}
where $\{ {\check{\boldsymbol{e}} _1}, \dots ,{\check{\boldsymbol{e}} _{N(N + 1)/2}}\} $ is the standard basis of $\mathbb{C}^{\frac{{N(N + 1)}}{2}}$. Together with the corresponding boundary conditions, the differential equations (\ref{eq29}) can be written equivalently as the following Volterra integral equations:
\begin{equation}\label{eq30}
\begin{aligned}
\check{\boldsymbol{p}}_{n_+} &= \check{\boldsymbol{e}}_{n+N-2} - \int_{x}^{+\infty} \mathrm{e}^{(x-\zeta)\check{\boldsymbol{A}}_+} \check{\boldsymbol{X}}(\zeta,t,z)\,\check{\boldsymbol{p}}_{n_+}(\zeta,t,z)\,\mathrm{d}\zeta,\\
\check{\boldsymbol{p}}_{n_-} &= \check{\boldsymbol{e}}_{n-1} - \int_{x}^{+\infty} \mathrm{e}^{(x-\zeta)\check{\boldsymbol{A}}_-} \check{\boldsymbol{X}}(\zeta,t,z)\,\check{\boldsymbol{p}}_{n_-}(\zeta,t,z)\,\mathrm{d}\zeta.
\end{aligned}
\end{equation}
Based on the definition of the induced operators on the exterior algebra, the diagonal structure of $\check{\boldsymbol{A}}_\pm$, and the linear dependence of $\check{\boldsymbol{X}}$ on $\Delta\boldsymbol X_\pm$, we follow the argument of Proposition \ref{Pro:1} to construct that the corresponding Neumann series converge locally uniformly in $\Xi_\pm$. Their sums are analytic in the respective regions and admit continuous extensions to $\mathring{\Sigma}$. By Eqs. (\ref{eq29}) and the corresponding boundary conditions, the coordinate vectors $\check{\boldsymbol p}_{n\pm}$ satisfy the associated Volterra integral equations on $\mathring{\Sigma}$. The boundary values of the Neumann series sums satisfy the same equations. Since both these boundary values and $\check{\boldsymbol p}_{n\pm}$ are bounded on the right half-axis, uniqueness of solutions to the Volterra integral equations implies that they coincide on $\mathring{\Sigma}$. Hence, for each $n = 2, \dots, N$, $\check{\boldsymbol p}_{n+}$ admits an analytic extension to $\Xi_+$, while $\check{\boldsymbol p}_{n-} $ admits an analytic extension to $\Xi_-$, with both extensions continuous up to $\mathring{\Sigma}$. From the definition of coordinate vectors, we have $ \boldsymbol p_{n\pm} = \boldsymbol A_+^{(N-1)} \boldsymbol\Gamma\, \check{\boldsymbol p}_{n\pm}$. Furthermore, Eqs. (\ref{eq10}) and the definition of the wedge product yield $ \boldsymbol P_{n+} =\mathrm{e}^{\mathrm{i}((N-2)\theta_2-\theta_1)} \boldsymbol p_{n+}$ and $\boldsymbol P_{n-} =\mathrm{e}^{\mathrm{i}(\theta_1+(N-2)\theta_2)} \boldsymbol p_{n-}$. These relations imply that $\boldsymbol p_{n+}$ and $\boldsymbol P_{n+}$ admit analytic extensions to $\Xi_+$, while $\boldsymbol p_{n-}$ and $\boldsymbol P_{n-}$ admit analytic extensions to $\Xi_-$. All these extensions are continuous up to $\mathring{\Sigma}$.

Next, we introduce the generalized cross product \cite{A50} to construct new solutions of Eqs. (\ref{eq3}) from known vector solutions. For any ${\boldsymbol{v}_1}, \dots, {\boldsymbol{v}_N} \in {\mathbb{C}^{N + 1}}$, we define
\begin{equation}\label{eq31}
\boldsymbol{\mathcal{H}}\left[ {{\boldsymbol{v}_1}, \dots ,{\boldsymbol{v}_N}} \right] = \sum\limits_{l=1}^{N + 1} {\det \left( {{\boldsymbol{v}_1}, \dots ,{\boldsymbol{v}_N},{\boldsymbol{e}_l}} \right){\boldsymbol{e}_l}},
\end{equation}
which generalizes the usual cross product to the $(N+1)$-dimensional complex vector space. Equivalently, $\boldsymbol{\mathcal{H}}[{\boldsymbol{v}_1}, \dots ,{\boldsymbol{v}_N}] = \boldsymbol{*}\left( {\boldsymbol{v}_1}\right.  \wedge  \cdots \left. \wedge {\boldsymbol{v}_N} \right)$, where $\boldsymbol{*}(\scalebox{0.65}{$\bullet$})$ denotes the Hodge star operator \cite{A49}.

\begin{proposition}\label{Pro:3}
Let ${\boldsymbol{\varphi} _1}, \dots ,{\boldsymbol{\varphi} _N}$ be $N$ column vector solutions to the Lax pair (\ref{eq3}). Then
\begin{equation}\label{eq32}
\boldsymbol{\Psi} \left( {x,t,z} \right) = {\mathrm{e}^{\mathrm{i}(N - 1){\theta _2}\left( {x,t,z} \right)}}\boldsymbol{\mathcal{H}}\left[ {\boldsymbol{\varphi} _1^*, \dots ,\boldsymbol{\varphi} _N^*} \right]\left( {x,t,{z^*}} \right)
\end{equation}
also satisfies the same Lax pair.
\end{proposition}
\begin{proof}
Equations (\ref{eq5}) imply that $k^*(z^*)=k(z)$ and $\lambda^*(z^*)=\lambda(z)$. Consequently, ${\boldsymbol{X}^\dag }({z^*}) = -\boldsymbol{X}(z)$ and ${\boldsymbol{Y}^\dag }({z^*}) = -\boldsymbol{Y}(z)$. A direct calculation using these identities yields
$$\begin{aligned}
{\boldsymbol{\Psi} _x}\left( {z} \right)
&= {\mathrm{e}^{\mathrm{i}(N - 1){\theta _2}}}\left( {\left(\mathrm{i}(N - 1){\theta _{2,x}} + \mathrm{Tr}({\boldsymbol{X}^*}({z^*}))\right)\boldsymbol{I}_{(N+1)\times(N+1)} - {\boldsymbol{X}^\dag }({z^*})} \right)\boldsymbol{\mathcal{H}}\left[ {\boldsymbol{\varphi} _1^*, \dots ,\boldsymbol{\varphi} _N^*} \right]\left( {{z^*}} \right)\\
&= - {\mathrm{e}^{\mathrm{i}(N - 1){\theta _2}}}{\boldsymbol{X}^\dag }\left( {{z^*}} \right)\boldsymbol{\mathcal{H}}\left[ {\boldsymbol{\varphi} _1^*, \dots ,\boldsymbol{\varphi} _N^*} \right]\left( {{z^*}} \right) = \boldsymbol{X}\left( {z} \right)\boldsymbol{\Psi} \left( {z} \right).
\end{aligned}$$
The temporal part of the Lax pair, given by ${\boldsymbol{\Psi} _t}=\boldsymbol{Y}\boldsymbol{\Psi} $, can be verified similarly. These calculations rely on the identity $\sum\nolimits_{l = 1}^N \boldsymbol{\mathcal{H}}  \left[ {\boldsymbol{v}_1}, \dots ,{\boldsymbol{v}_{l - 1}},\right. $  $ \left.{\boldsymbol{Av}_l},{\boldsymbol{v}_{l + 1}}, \dots ,{\boldsymbol{v}_N} \right]  = ( \mathrm{Tr}\hspace{2pt}(\boldsymbol{A}){\boldsymbol{I}_{(N + 1) \times (N + 1)}} - {\boldsymbol{A}^T} )\boldsymbol{\mathcal{H}}\left[ {{\boldsymbol{v}_1}, \dots ,{\boldsymbol{v}_N}} \right]$, where $\boldsymbol{A} \in {\mathbb{C}^{(N + 1) \times (N + 1)}}$ and ${\boldsymbol{v}_l} \in {\mathbb{C}^{N + 1}}$.
\end{proof}

The functions ${\boldsymbol{\varphi} ^*}\left( {{z^*}} \right)$ are known as the adjoint Jost eigenfunctions. The Schwarz reflection principle allows us to determine the analytic properties of their columns $\boldsymbol{\varphi} _{ \pm ,l}^*\left( {{z^*}} \right)$ for $l = 1, \dots ,N + 1$. The relations between the Jost eigenfunctions and their adjoint forms are given by
\begin{equation}\label{eq33}
\begin{aligned}
{\alpha _{{l_{N + 1}}}}{\boldsymbol{\varphi} _{ \pm ,{l_{N + 1}}}}(z) &= {\mathrm{e}^{\mathrm{i}(N - 1){\theta _2}(z)}}\boldsymbol{\mathcal{H}}\left[ {\boldsymbol{\varphi} _{ \pm ,{l_1}}^*\left( {{z^*}} \right), \dots ,\boldsymbol{\varphi} _{ \pm ,{l_N}}^*\left( {{z^*}} \right)} \right],\\
{\alpha _{{l_{N + 1}}}}\boldsymbol{\varphi} _{ \pm ,{l_{N + 1}}}^*({z^*}) &= {\mathrm{e}^{\mathrm{i}(1 - N){\theta _2}(z)}}\boldsymbol{\mathcal{H}}\left[ {{\boldsymbol{\varphi} _{ \pm ,{l_1}}}\left( {z} \right), \dots ,{\boldsymbol{\varphi} _{ \pm ,{l_N}}}\left( {z} \right)} \right],
\end{aligned}
\end{equation}
where $\left( {{l_1}, \dots,  {l_{N + 1}}} \right)$ is an even permutation of $\left\{ {1, \dots ,N + 1} \right\}$, and
$${\alpha _l} = \left\{ {\begin{array}{*{20}{l}}
{1,}&{l = 1,}\\
{\alpha,}&{ 2\le l \le N,}\\
{1,}&{l = N + 1.}
\end{array}} \right.$$
We introduce the auxiliary Jost eigenfunctions ${\boldsymbol{\psi} _ \pm }(x,t,z) = \left( {{\boldsymbol{\psi} _{2 \pm }}, \dots ,{\boldsymbol{\psi} _{N \pm }}} \right)(x,t,z)$, where
\begin{equation}\label{eq34}
\begin{aligned}
\alpha {\boldsymbol{\psi} _{n + }}(x,t,z) &= {( - 1)^{N + 1 - n}}{\mathrm{e}^{\mathrm{i}(N - 1){\theta _2}(x,t,z)}}\boldsymbol{\ast}\left( {\boldsymbol{P}_{n - }^*\left( {x,t,{z^*}} \right)\wedge\boldsymbol{\varphi} _{ - ,N + 1}^*\left( {x,t,{z^*}} \right)} \right),\\
\alpha {\boldsymbol{\psi} _{n - }}(x,t,z) &= {( - 1)^{N + 1 - n}}{\mathrm{e}^{\mathrm{i}(N - 1){\theta _2}(x,t,z)}}\boldsymbol{\ast}\left( {\boldsymbol{\varphi} _{ - ,1}^*\left( {x,t,{z^*}} \right)\wedge\boldsymbol{P}_{n + }^*\left( {x,t,{z^*}} \right)} \right).
\end{aligned}
\end{equation}
Their modified forms are ${\boldsymbol{\mu} _ \pm }(x,t,z) = {\boldsymbol{\psi} _ \pm }(x,t,z){\mathrm{e}^{ - \mathrm{i}{\theta _2}(x,t,z)}}$, where
\begin{equation}\label{eq35}
\begin{aligned}
\alpha {\boldsymbol{\mu} _{n + }}(x,t,z) &= {( - 1)^{N + 1 - n}}\boldsymbol{\ast}\left( {\boldsymbol{p}_{n - }^*\left( {x,t,{z^*}} \right)\wedge\boldsymbol\upsilon _{ - ,N + 1}^*\left( {x,t,{z^*}} \right)} \right),\\
\alpha {\boldsymbol{\mu} _{n - }}(x,t,z) &= {( - 1)^{N + 1 - n}}\boldsymbol{\ast}\left( {\boldsymbol\upsilon _{ - ,1}^*\left( {x,t,{z^*}} \right)\wedge\boldsymbol{p}_{n + }^*\left( {x,t,{z^*}} \right)} \right).
\end{aligned}
\end{equation}
The analyticity of the Jost eigenfunctions and tensors established earlier, together with the analyticity of the conjugate reflection transform, imply that ${\boldsymbol{\psi} _{n + }}$ and ${\boldsymbol{\mu} _{n + }}$ are analytic in $\Xi_+$, while ${\boldsymbol{\psi} _{n - }}$ and ${\boldsymbol{\mu} _{n - }}$ are analytic in $\Xi_-$. All these functions are continuous up to $\mathring{\Sigma}$. It then follows from Eqs. (\ref{eq35}) that
\begin{equation}\label{eq36}
\begin{array}{l}
\boldsymbol{\upsilon} _{ + ,1}^\dag ({z^*}){\boldsymbol{\mu} _ + }(z) = {\boldsymbol{0}_{1 \times (N - 1)}},\quad\boldsymbol{\upsilon} _{ - ,N + 1}^\dag ({z^*}){\boldsymbol{\mu} _ + }(z) = {\boldsymbol{0}_{1 \times (N - 1)}},\\
\boldsymbol{\upsilon} _{ - ,1}^\dag ({z^*}){\boldsymbol{\mu} _ - }(z) = {\boldsymbol{0}_{1 \times (N - 1)}},\quad\boldsymbol{\upsilon} _{ + ,N + 1}^\dag ({z^*}){\boldsymbol{\mu} _ - }(z) = {\boldsymbol{0}_{1 \times (N - 1)}}.
\end{array}
\end{equation}

\subsection{Symmetries}
\label{subsec:2.5}

The scattering problems of integrable systems with NZBCs generally exhibit more intricate symmetries than those with ZBC. For the system (\ref{eq2}) considered here, these symmetries are closely linked to the structure of the Riemann surface, the transformations of the spectral parameters, and the boundary conditions (\ref{eq9}). Using the uniformization variable $z$, the relevant spectral transformations can be expressed as mappings of the complex $z$-plane. In particular, we consider three fundamental maps: $z \to {z^*}$, $z \to  - z$, and $z \to  - u_0^2/z$. These correspond to complex conjugation, simultaneous sign reversal of $k$ and $\lambda$, and interchange of the two sheets of the Riemann surface, respectively. 

\subsubsection{The first symmetry}

The map $z \to {z^*}$ corresponds to $(\lambda,k ) \to ({\lambda^*},{ k^*})$. If ${\boldsymbol{\varphi} (z)}$ is a fundamental matrix solution of the Lax pair (\ref{eq3}), a direct calculation produces
\begin{equation}\label{eq37}
{\partial _x}\left( {{\boldsymbol{\varphi} ^\dag }({z^*})\boldsymbol{\varphi} (z)} \right) = {\partial _t}\left( {{\boldsymbol{\varphi} ^\dag }({z^*})\boldsymbol{\varphi} (z)} \right) = {\boldsymbol{0}_{(N + 1) \times (N + 1)}}.
\end{equation}
Together with the boundary conditions (\ref{eq9}), this result yields
\begin{equation}\label{eq38}
\boldsymbol{\varphi} _ \pm ^\dag ({z^*}){\boldsymbol{\varphi} _ \pm }(z) = \boldsymbol{\upsilon} _ \pm ^\dag ({z^*}){\boldsymbol{\upsilon} _ \pm }(z) = \boldsymbol{Q}(z),\quad z \in\Sigma,
\end{equation}
where
$$\boldsymbol{Q}(z) = \mathrm{diag}(\alpha (z),\underbrace {1, \dots ,1}_{N - 1},\alpha (z)).$$
The Schwarz reflection principle then yields the orthogonality relations for the columns of the Jost eigenfunctions and their modified forms:
\begin{equation}\label{eq39}
\boldsymbol{\varphi} _{ \pm 1}^\dag ({z^*}){\boldsymbol{\varphi} _{ \pm 3}}(z) = 0,\quad\boldsymbol{\upsilon} _{ \pm 1}^\dag ({z^*}){\boldsymbol{\upsilon} _{ \pm 3}}(z) = 0,\quad z\in\Xi_{\pm}.
\end{equation}
Using the relations between the Jost eigenfunctions and the scattering matrices given in Eqs. (\ref{eq21}), we obtain the explicit expressions for $\boldsymbol G$ and $\boldsymbol B$:
\begin{equation}\label{eq39a}
\begin{array}{l}
\boldsymbol{B}(z) = {\boldsymbol{Q}^{ - 1}}(z)\boldsymbol{\varphi} _ - ^\dag ({z^*}){\boldsymbol{\varphi} _ + }(z),\quad
\boldsymbol{G}(z) = {\boldsymbol{Q}^{ - 1}}(z)\boldsymbol{\varphi} _ + ^\dag ({z^*}){\boldsymbol{\varphi} _ - }(z).
\end{array}
\end{equation}
These expressions imply the following symmetry relation between $\boldsymbol B$ and $\boldsymbol G$:
$$\boldsymbol{G}(z) = {\boldsymbol{Q}^{ - 1}}(z){\boldsymbol{B}^\dag }({z^*})\boldsymbol{Q}(z).$$
Comparing the entries on both sides of this matrix identity yields the symmetry relations among the scattering coefficients:
\begin{equation}\label{eq40}
\begin{aligned}
b_{11}^*(z^*) &= g_{11}(z) = \frac{1}{\alpha} \boldsymbol{\varphi}_{+1}^\dag(z^*) \boldsymbol{\varphi}_{-1}(z)
= \frac{1}{\alpha} \boldsymbol{\upsilon}_{+1}^\dag(z^*) \boldsymbol{\upsilon}_{-1}(z), & z &\in \Xi_+ \cup \Sigma,\\
b_{33}^*(z^*) &= g_{33}(z) = \frac{1}{\alpha} \boldsymbol{\varphi}_{+3}^\dag(z^*) \boldsymbol{\varphi}_{-3}(z)
= \frac{1}{\alpha} \boldsymbol{\upsilon}_{+3}^\dag(z^*) \boldsymbol{\upsilon}_{-3}(z), & z &\in \Xi_- \cup \Sigma.
\end{aligned}
\end{equation}
and
\begin{equation}\label{eq41}
\begin{array}{l}
{g_{12}}^\dag(z^*) = \dfrac{1}{{\alpha }}b_{21} ({z}),\quad{g_{13}}^*(z^*) = b_{31}({z}),\quad{g_{21}}^\dag(z^*) = \alpha b_{12} ({z}),\quad{g_{22}}^\dag(z^*) = b_{22} ({z}),\\
{g_{23}}^\dag(z^*) = \alpha b_{32} ({z}),\quad{\kern 2pt} {g_{31}}^*(z^*) = b_{13}({z}),\quad{g_{32}}^\dag(z^*) = \dfrac{1}{{\alpha }}b_{23} ({z}),\quad z\in\Sigma.
\end{array}
\end{equation}

\subsubsection{The second symmetry}

The map $z \to  - z$ corresponds to $(\lambda,k ) \to ( - \lambda,-k )$. Together with the symmetry ${\boldsymbol{\sigma} _3}\boldsymbol{U}(x,t){\boldsymbol{\sigma} _3} =  - \boldsymbol{U}(x,t)$ of the potential matrix, this gives ${\boldsymbol{\sigma} _3}\boldsymbol{X}( - z){\boldsymbol{\sigma} _3} = \boldsymbol{X}(z)$ and ${\boldsymbol{\sigma} _3}\boldsymbol{Y}( - z){\boldsymbol{\sigma} _3} = \boldsymbol{Y}(z)$. Consequently, both $\boldsymbol{\varphi} (z)$ and ${\boldsymbol{\sigma} _3}\boldsymbol{\varphi} ( - z){\boldsymbol{\sigma} _3}$ satisfy the Lax pair (\ref{eq3}). Moreover, the identity ${\boldsymbol{\sigma} _3}{\boldsymbol{A}_ \pm }( - z){\boldsymbol{\sigma} _3} = {\boldsymbol{A}_ \pm }(z)$ implies that $\boldsymbol{\sigma}_3\boldsymbol{\varphi}_\pm(-z)\boldsymbol{\sigma}_3$ and $\boldsymbol{\varphi}_\pm(z)$ satisfy the same boundary conditions for $x \to \pm \infty$. By uniqueness of the Jost eigenfunctions, we obtain their second symmetry relation
\begin{equation}\label{eq42}
{\boldsymbol{\varphi} _ \pm }(z) = {\boldsymbol{\sigma} _3}{\boldsymbol{\varphi} _ \pm }( - z){\boldsymbol{\sigma} _3}.
\end{equation}
The analytic properties of $\boldsymbol{\varphi}_{\pm}(z)$ yield the following relations for the columns of the Jost eigenfunctions and their modified forms:
\begin{equation}\label{eq43}
\begin{aligned}
\boldsymbol{\varphi}_{\pm 1}(z) &= \boldsymbol{\sigma}_3 \boldsymbol{\varphi}_{\pm 1}(-z), &
\boldsymbol{\upsilon}_{\pm 1}(z) &= \boldsymbol{\sigma}_3 \boldsymbol{\upsilon}_{\pm 1}(-z), & z&\in\Xi_{\mp},\\
\boldsymbol{\varphi}_{\pm 2}(z) &= -\boldsymbol{\sigma}_3 \boldsymbol{\varphi}_{\pm 2}(-z), &
\boldsymbol{\upsilon}_{\pm 2}(z) &= -\boldsymbol{\sigma}_3 \boldsymbol{\upsilon}_{\pm 2}(-z), & z&\in\Sigma,\\
\boldsymbol{\varphi}_{\pm 3}(z) &= -\boldsymbol{\sigma}_3 \boldsymbol{\varphi}_{\pm 3}(-z), &
\boldsymbol{\upsilon}_{\pm 3}(z) &= -\boldsymbol{\sigma}_3 \boldsymbol{\upsilon}_{\pm 3}(-z), & z&\in\Xi_{\pm}.
\end{aligned}
\end{equation}
Substituting Eqs. (\ref{eq42}) into the scattering relations (\ref{eq21}) yields
\begin{equation}\label{eq44}
\boldsymbol{B}(z) = {\boldsymbol{\sigma} _3}\boldsymbol{B}( - z){\boldsymbol{\sigma} _3},\quad\boldsymbol{G}(z) = {\boldsymbol{\sigma} _3}\boldsymbol{G}( - z){\boldsymbol{\sigma} _3},\quad z \in \Sigma.
\end{equation}
The corresponding symmetry relations for the scattering coefficients follow as
\begin{equation}\label{eq45}
\begin{aligned}
b_{11}(z) &= b_{11}(-z), & g_{33}(z) &= g_{33}(-z), & z &\in \Xi_{-},\\
b_{33}(z) &= b_{33}(-z), & g_{11}(z) &= g_{11}(-z), & z &\in \Xi_{+}.
\end{aligned}
\end{equation}
Combining the symmetries (\ref{eq43}) with the tensors in Eqs. (\ref{eq24}) produces
\begin{equation}\label{eq46}
\begin{array}{l}
\boldsymbol{P}_{n - }^*({z^*}) = {( - 1)^{N - 2}}{\boldsymbol{\sigma} _3}^{(N - 1)}\boldsymbol{P}_{n - }^*( - {z^*}),\quad z \in {\Xi _ - },\\
\boldsymbol{P}_{n + }^*({z^*}) = {( - 1)^{N - 1}}{\boldsymbol{\sigma} _3}^{(N - 1)}\boldsymbol{P}_{n + }^*( - {z^*}),\quad z \in {\Xi _ + }.
\end{array}
\end{equation}
Finally, the preceding relations together with Eqs. (\ref{eq34}) and (\ref{eq35}) yield the following symmetry relations:
\begin{equation}\label{eq47}
{\boldsymbol{\psi} _ \pm }(z) =  - {\boldsymbol{\sigma} _3}{\boldsymbol{\psi} _ \pm }( - z),\quad{\boldsymbol{\mu} _ \pm }(z) =  - {\boldsymbol{\sigma} _3}{\boldsymbol{\mu} _ \pm }( - z), \quad z \in {\Xi _ \pm }.
\end{equation}
Here we have used the identity $\boldsymbol{\mathcal{H}}\left[ {{\boldsymbol{\sigma} _3}{\boldsymbol{v}_1}, \dots ,{\boldsymbol{\sigma} _3}{\boldsymbol{v}_N}} \right] = {( - 1)^N}{\boldsymbol{\sigma} _3}\boldsymbol{\mathcal{H}}\left[ {{\boldsymbol{v}_1}, \dots ,{\boldsymbol{v}_N}} \right]$, where ${\boldsymbol{v}_l} \in {\mathbb{C}^{N + 1}}$.

\subsubsection{The third symmetry}

The transformation $z \to  - u_0^2/z$ is equivalent to $(\lambda,k ) \to ( - \lambda,k )$ and establishes a correspondence between the two sheets of the Riemann surface. Since the spectral parameter $k$ remains fixed under this transformation, both $\boldsymbol{\varphi} \left( {z} \right)$ and $\boldsymbol{\varphi} \left( { - u_0^2/z} \right)$ satisfy the same Lax pair (\ref{eq3}). It follows that there exists an invertible matrix $\boldsymbol{L}(z)$ such that $\boldsymbol{\varphi} \left( {z} \right) = \boldsymbol{\varphi} \left( { - u_0^2/z} \right)\boldsymbol{L}(z)$. Comparing their asymptotic behavior as $x \to \pm \infty$ with the boundary conditions (\ref{eq9}) then determines the third symmetry relation, which is
\begin{equation}\label{eq48}
\boldsymbol{\varphi}_\pm(z) = \boldsymbol{\varphi}_\pm\left(-\dfrac{u_0^2}{z}\right)\boldsymbol{L}_\pm(z),
\end{equation}
where
$$\boldsymbol{L}_\pm(z)
= \mathrm{e}^{-\mathrm{i}\boldsymbol{\Theta}\left(-\frac{u_0^2}{z}\right)}
\boldsymbol{A}_\pm^{-1}\left(-\frac{u_0^2}{z}\right)
\boldsymbol{A}_\pm(z)
\mathrm{e}^{\mathrm{i}\boldsymbol{\Theta}(z)}
=
\begin{pmatrix}
0 & \boldsymbol{0}_{1\times(N-1)} & \dfrac{u_\pm}{\mathrm{i}z}\\[4pt]
\boldsymbol{0}_{(N-1)\times 1} & \boldsymbol{I}_{(N-1)\times(N-1)} & \boldsymbol{0}_{(N-1)\times 1}\\[4pt]
\dfrac{u_\pm^*}{\mathrm{i}z} & \boldsymbol{0}_{1\times(N-1)} & 0
\end{pmatrix}.$$
Using the analyticity of $\boldsymbol{\varphi}_\pm(z)$ and $\boldsymbol{\upsilon}_{\pm}(z)$, we can further conclude that their columns satisfy
\begin{equation}\label{eq49}
\begin{aligned}
&\boldsymbol{\varphi}_{\pm 1}(z) = \dfrac{u_\pm^*}{\mathrm{i}z}\boldsymbol{\varphi}_{\pm 3}\left(-\dfrac{u_0^2}{z}\right), &
&\boldsymbol{\upsilon}_{\pm 1}(z) = \dfrac{u_\pm^*}{\mathrm{i}z}\boldsymbol{\upsilon}_{\pm 3}\left(-\dfrac{u_0^2}{z}\right), & z &\in \Xi_\mp,\\
&\boldsymbol{\varphi}_{\pm 2}\left(-\dfrac{u_0^2}{z}\right) = \boldsymbol{\varphi}_{\pm 2}\left(z\right), &
&\boldsymbol{\upsilon}_{\pm 2}\left(-\dfrac{u_0^2}{z}\right) = \boldsymbol{\upsilon}_{\pm 2}(z), & z &\in \Sigma.
\end{aligned}
\end{equation}
Applying these symmetry relations to the tensors defined in Eqs. (\ref{eq24}) yields
\begin{equation}\label{eq50}
\begin{array}{l}
\boldsymbol{P}_{n - }^*\left( {{z^*}} \right) = {\left( { - 1} \right)^{N - 2}}\dfrac{{\mathrm{i}{u_ + }}}{z}\boldsymbol{P}_{n + }^*\left( { - \dfrac{{u_0^2}}{{{z^*}}}} \right),\quad z \in {\Xi _ - },\\
\boldsymbol{P}_{n + }^*\left( {{z^*}} \right) = {\left( { - 1} \right)^{N - 2}}\dfrac{{\mathrm{i}u_ + ^*}}{z}\boldsymbol{P}_{n - }^*\left( { - \dfrac{{u_0^2}}{{{z^*}}}} \right),\quad z \in {\Xi _ + }.
\end{array}
\end{equation}
Combining these results with Eqs. (\ref{eq34}) and (\ref{eq35}), and using the identity $\alpha \left( { - u_0^2/z} \right) = \frac{u_ + }{u_ - }\alpha \left( z \right)$, we get the third symmetry relations for ${\boldsymbol{\psi} _ \pm }(z)$ and ${\boldsymbol{\mu} _ \pm }(z)$ as follows:
\begin{equation}\label{eq51}
\begin{array}{l}
{\boldsymbol{\psi} _ + }(z) = \dfrac{{{u_ + }}}{{{u_ - }}}{\boldsymbol{\psi} _ - }\left( { - \dfrac{{u_0^2}}{z}} \right),\quad{\boldsymbol{\mu} _ + }(z) = \dfrac{{{u_ + }}}{{{u_ - }}}{\boldsymbol{\mu} _ - }\left( { - \dfrac{{u_0^2}}{z}} \right),\quad z \in {\Xi _ + },\\
{\boldsymbol{\psi} _ - }(z) = \dfrac{{{u_ - }}}{{{u_ + }}}{\boldsymbol{\psi} _ + }\left( { - \dfrac{{u_0^2}}{z}} \right),\quad {\boldsymbol{\mu} _ - }(z) = \dfrac{{{u_ - }}}{{{u_ + }}}{\boldsymbol{\mu} _ + }\left( { - \dfrac{{u_0^2}}{z}} \right),\quad z \in {\Xi _ - }.
\end{array}
\end{equation}
Finally, substituting Eqs. (\ref{eq48}) into the scattering relations (\ref{eq21}) and using the invertibility of $\boldsymbol{L}_{\pm}(z)$, we find that the symmetry relations for $\boldsymbol{G}$ and $\boldsymbol{B}$ are
\begin{equation}\label{eq52}
\boldsymbol{B}\left( { - \frac{{u_0^2}}{z}} \right) = {\boldsymbol{L}_ - }(z)\boldsymbol{B}(z)\boldsymbol{L}_ + ^{ - 1}(z),\quad \boldsymbol{G}\left( { - \frac{{u_0^2}}{z}} \right) = {\boldsymbol{L}_ + }(z)\boldsymbol{G}(z)\boldsymbol{L}_ - ^{ - 1}(z),\quad z \in \Sigma .
\end{equation}
Comparing the corresponding matrix entries yields
\begin{equation}\label{eq53}
\begin{aligned}
b_{12}(z) &= \dfrac{\mathrm{i}z}{u_-^*} b_{32}\left(-\dfrac{u_0^2}{z}\right), &
b_{13}(z) &= \dfrac{u_+}{u_-^*} b_{31}\left(-\dfrac{u_0^2}{z}\right), &
b_{21}(z) &= -\dfrac{\mathrm{i}u_+^*}{z} b_{23}\left(-\dfrac{u_0^2}{z}\right),\\
b_{22}(z) &= b_{22}\left(-\dfrac{u_0^2}{z}\right), &
b_{23}(z) &= -\dfrac{\mathrm{i}u_+}{z} b_{21}\left(-\dfrac{u_0^2}{z}\right), &
b_{31}(z) &= \dfrac{u_+^*}{u_-} b_{13}\left(-\dfrac{u_0^2}{z}\right),\\
b_{32}(z) &= \dfrac{\mathrm{i}z}{u_-} b_{12}\left(-\dfrac{u_0^2}{z}\right), &
g_{12}(z) &= \dfrac{\mathrm{i}z}{u_+^*} g_{32}\left(-\dfrac{u_0^2}{z}\right), &
g_{13}(z) &= \dfrac{u_-}{u_+^*} g_{31}\left(-\dfrac{u_0^2}{z}\right),\\
g_{21}(z) &= -\dfrac{\mathrm{i}u_-^*}{z} g_{23}\left(-\dfrac{u_0^2}{z}\right), &
g_{22}(z) &= g_{22}\left(-\dfrac{u_0^2}{z}\right), &
g_{23}(z) &= -\dfrac{\mathrm{i}u_-}{z} g_{21}\left(-\dfrac{u_0^2}{z}\right),\\
g_{31}(z) &= \dfrac{u_-^*}{u_+} g_{13}\left(-\dfrac{u_0^2}{z}\right), &
g_{32}(z) &= \dfrac{\mathrm{i}z}{u_+} g_{12}\left(-\dfrac{u_0^2}{z}\right), & z &\in \Sigma,
\end{aligned}
\end{equation}
and
\begin{equation}\label{eq54}
\begin{array}{l}
{g_{11}}(z) = \dfrac{{u_ - ^*}}{{u_ + ^*}}{g_{33}}\left( { - \dfrac{{u_0^2}}{z}} \right),\quad{b_{33}}(z) = \dfrac{{{u_ + }}}{{{u_ - }}}{b_{11}}\left( { - \dfrac{{u_0^2}}{z}} \right),\quad z \in {\Xi _ + },\\
{g_{33}}(z) = \dfrac{{{u_ - }}}{{{u_ + }}}{g_{11}}\left( { - \dfrac{{u_0^2}}{z}} \right),\quad{b_{11}}(z) = \dfrac{{u_ + ^*}}{{u_ - ^*}}{b_{33}}\left( { - \dfrac{{u_0^2}}{z}} \right),\quad z \in {\Xi _ - }.
\end{array}
\end{equation}

\subsubsection{The combined symmetry}

The three symmetries established above allow us to derive further symmetry relations for ${\boldsymbol{\boldsymbol{\mu}} _ \pm }(z)$ and $\boldsymbol{\upsilon}_{\pm}\left(z\right)$. In particular, the symmetry relations (\ref{eq43}) and (\ref{eq49}) together yield
\begin{equation}\label{eq55}
\begin{aligned}
&{\boldsymbol{\upsilon} _{ \pm 1}}\left( {z} \right) = \dfrac{{\mathrm{i}u_ \pm ^*}}{z}{\boldsymbol{\sigma} _3}{\boldsymbol{\upsilon} _{ \pm 3}}\left( {\dfrac{{u_0^2}}{z}} \right), &z& \in \Xi_\mp ,\\
&{\boldsymbol{\upsilon} _{ \pm 2}}\left( {z} \right) =  - {\boldsymbol{\sigma} _3}{\boldsymbol{\upsilon} _{ \pm 2}}\left( {\dfrac{{u_0^2}}{z}} \right), &z& \in \Sigma ,\\
&{\boldsymbol{\upsilon} _{ \pm 3}}\left( {z} \right) =  - \dfrac{{\mathrm{i}{u_ \pm }}}{z}{\boldsymbol{\sigma} _3}{\boldsymbol{\upsilon} _{ \pm 1}}\left( {\dfrac{{u_0^2}}{z}} \right), &z& \in \Xi_\pm .
\end{aligned}
\end{equation}
The modified auxiliary Jost eigenfunctions satisfy
\begin{equation}\label{eq56}
\begin{aligned}
{\boldsymbol{\mu} _ + }\left( {z} \right) =  - \dfrac{{{u_ + }}}{{{u_ - }}}{\boldsymbol{\sigma} _3}{\boldsymbol{\mu} _ - }\left( {\dfrac{{u_0^2}}{z}} \right),\quad z \in \Xi_+ ,\\
{\boldsymbol{\mu} _ - }\left( {z} \right) =  - \dfrac{{{u_ - }}}{{{u_ + }}}{\boldsymbol{\sigma} _3}{\boldsymbol{\mu} _ + }\left( {\dfrac{{u_0^2}}{z}} \right),\quad z \in \Xi_- .
\end{aligned}
\end{equation}
Substituting Eqs. (\ref{eq55}) into the orthogonality relations (\ref{eq39}) yields
\begin{equation}\label{eq57}
\begin{aligned}
\boldsymbol{\upsilon} _{ \pm 3}^\dag \left( {z} \right){\boldsymbol{\sigma} _3}{\boldsymbol{\upsilon} _{ \pm 3}}\left( {\frac{{u_0^2}}{{{z^*}}}} \right) = 0,\quad z\in\Xi_\pm,\\
\boldsymbol{\upsilon} _{ \pm 1}^\dag \left( {z} \right){\boldsymbol{\sigma} _3}{\boldsymbol{\upsilon} _{ \pm 1}}\left( {\frac{{u_0^2}}{{{z^*}}}} \right) = 0,\quad z\in\Xi_\mp.
\end{aligned}
\end{equation}
Combining Eqs. (\ref{eq55}) with the first symmetry (\ref{eq40}) of the scattering coefficients, we obtain
\begin{equation}\label{eq58}
\begin{aligned}
&b_{11}^* ({z^*}) = {g_{11}}(z) =  - \frac{1}{{\alpha (z)}}\frac{{\mathrm{i}{u_ + }}}{z}\boldsymbol{\upsilon} _{ + 3}^\dag \left( {\frac{{u_0^2}}{{{z^*}}}} \right){\boldsymbol{\sigma} _3}{\boldsymbol{\upsilon} _{ - 1}}\left( {z} \right), &z&\in\Xi_+,\\
&b_{33}^* ({z^*}) = {g_{33}}(z) =   \frac{1}{{\alpha (z)}}\frac{{\mathrm{i}{u_ +^* }}}{z}\boldsymbol{\upsilon} _{ + 1}^\dag \left( {\frac{{u_0^2}}{{{z^*}}}} \right){\boldsymbol{\sigma} _3}{\boldsymbol{\upsilon} _{ - 3}}\left( {z} \right), &z&\in\Xi_-.
\end{aligned}
\end{equation}
Equations (\ref{eq36}) and (\ref{eq55}) yield the following relations:
\begin{equation}\label{eq59}
\begin{aligned}
\boldsymbol{\upsilon} _{ \pm 3}^\dag \left( {\frac{{u_0^2}}{{{z^*}}}} \right){\boldsymbol{\sigma} _3}{\boldsymbol{\mu} _ \pm }\left( {z} \right) = {\boldsymbol{0}_{1 \times (N - 1)}}, \quad \boldsymbol{\upsilon} _{ \mp 1}^\dag \left( {\frac{{u_0^2}}{{{z^*}}}} \right){\boldsymbol{\sigma} _3}{\boldsymbol{\mu} _ \pm }\left( {z} \right) = {\boldsymbol{0}_{1 \times (N - 1)}},\quad z\in\Xi_\pm.
\end{aligned}
\end{equation}

The reflection coefficients on the contour $\Sigma$ are defined as follows:
\begin{equation}\label{eq60}
\begin{aligned}
{\beta _1}(z) = \frac{{{g_{21}}(z)}}{{{g_{11}}(z)}} = \frac{{\alpha b_{12}^\dag ({z^*})}}{{b_{11}^*({z^*})}},\quad {\beta _2}(z) = \frac{{{g_{31}}(z)}}{{{g_{11}}(z)}} = \frac{{b_{13}^* ({z^*})}}{{b_{11}^*({z^*})}},\quad{\beta _3}(z) = \frac{{{g_{23}}(z)}}{{{g_{33}}(z)}} = \frac{{\alpha b_{32}^\dag ({z^*})}}{{b_{33}^*({z^*})}},
\end{aligned}
\end{equation}
which are used in the inverse problem. Using the symmetry relations (\ref{eq45}), (\ref{eq53}) and (\ref{eq54}), we have
\begin{equation}\label{eq61}
\begin{aligned}
&{\beta _1}( - z) =  - {\beta _1}(z),&{\beta _2}&( - z) =  - {\beta _2}(z),&{\beta _3}&( - z) = {\beta _3}(z),\\
&{\beta _1}\left( { - \frac{{u_0^2}}{z}} \right) = \frac{{\mathrm{i}z}}{{{u_ + }}}{\beta _3}(z),&{\beta _3}&\left( { - \frac{{u_0^2}}{z}} \right) = \frac{{\mathrm{i}z}}{{u_ + ^*}}{\beta _1}(z),&{\beta _2}&\left( { - \frac{{u_0^2}}{z}} \right) = \frac{{u_ + ^*}}{{{u_ + }}}\frac{{b_{31}^*({z^*})}}{{b_{33}^*({z^*})}}.
\end{aligned}
\end{equation}
The first symmetry (\ref{eq39a}) of the scattering matrices then implies
\begin{equation}\label{eq61a}
\begin{aligned}
{\beta _1}(z) = \alpha\frac{{\boldsymbol{\varphi} _{ + 2}^\dag ({z^*}){\boldsymbol{\varphi} _{ - 1}}(z)}}{{\boldsymbol{\varphi} _{ + 1}^\dag ({z^*}){\boldsymbol{\varphi} _{ - 1}}(z)}},\quad{\beta _2}(z) = \frac{{\boldsymbol{\varphi} _{ + 3}^\dag ({z^*}){\boldsymbol{\varphi} _{ - 1}}(z)}}{{\boldsymbol{\varphi} _{ + 1}^\dag ({z^*}){\boldsymbol{\varphi} _{ - 1}}(z)}},\quad{\beta _3}(z) = \alpha \frac{{\boldsymbol{\varphi} _{ + 2}^\dag ({z^*}){\boldsymbol{\varphi} _{ - 3}}(z)}}{{\boldsymbol{\varphi} _{ + 3}^\dag ({z^*}){\boldsymbol{\varphi} _{ - 3}}(z)}}.
\end{aligned}
\end{equation}

\subsection{Asymptotic behavior}
\label{subsec:2.6}

The asymptotic behavior of the modified Jost eigenfunctions and the scattering coefficients as $k\to\infty$ provides the basis for formulating and normalizing the Riemann-Hilbert problem. Under the uniformization transformation, the limit $k\to\infty$ corresponds to the two limits $z \to \infty$ and $z \to 0$. Using the classical Wentzel-Kramers-Brillouin asymptotic expansion method, we substitute the asymptotic expansions of $\boldsymbol{\upsilon} _ \pm(z)$ and the $\check{\boldsymbol{p}}_{n\pm}(z)$ into their respective differential equations (\ref{eq11}) and (\ref{eq29}). Matching the terms of the same order in $z$ and imposing Eq. (\ref{eq9}), we determine their asymptotic behavior. In the corresponding analytic regions, the modified Jost eigenfunctions have the following asymptotic behavior as $z \to \infty$:
\begin{equation}\label{eq62}
\begin{aligned}
{\boldsymbol{\upsilon} _ \pm } = \begin{pmatrix}
1&{ - \dfrac{{\mathrm{i}}}{z}{\boldsymbol{u}^{{T}}}}\\
{ - \dfrac{{\mathrm{i}}}{z}{\boldsymbol{u}^*}}&{{\boldsymbol{I}_{N \times N}}}
\end{pmatrix} + O\left( z^{-1} \right).
\end{aligned}
\end{equation}
Since $z\to 0$ corresponds to $-u_0^2/z\to\infty$, the symmetry relation (\ref{eq49}) yields
\begin{equation}\label{eq63}
\begin{aligned}
\boldsymbol{\upsilon}_{\pm 1} &=\begin{pmatrix}
\dfrac{\boldsymbol{u}^{T}\tilde{\boldsymbol{e}}_N}{u_\pm} \\[10pt]
-\dfrac{\mathrm{i}\boldsymbol{u}^*_\pm}{z}
\end{pmatrix}+ O(1),\quad \boldsymbol{\upsilon}_{\pm 3} =
\begin{pmatrix}
\dfrac{u_\pm}{\mathrm{i}z} \\[10pt]
\dfrac{\boldsymbol{u}^*}{u_\pm^*}
\end{pmatrix}+ O(1),\\
\boldsymbol{\upsilon}_{\pm 2} &=
\begin{pmatrix}
\dfrac{\mathrm{i}z}{u_0^2}\boldsymbol{u}^{T} \\[10pt]
\boldsymbol{I}_{N\times N}
\end{pmatrix}
\left(\tilde{\boldsymbol{e}}_1,\dots,\tilde{\boldsymbol{e}}_{N-1}\right)+ O(z).
\end{aligned}
\end{equation}
Here, ${\tilde{\boldsymbol{e}}_l}$ $(l=1,\dots,N)$ denote the standard basis vectors in $\mathbb{C}^N$.

The same method also determines the asymptotic behavior of the coordinate vectors in both limiting cases described above. For each $n = 2, \dots, N$, the coordinate vectors admit the following asymptotic expansions in the corresponding analytic regions as $z \to \infty$:
\begin{equation}\label{eq64}
\begin{aligned}
\check{\boldsymbol{p}}_{n+} = \check{\boldsymbol{e}}_{N+n-2}+O\left(z^{-1}\right),\quad
\check{\boldsymbol{p}}_{n-} = \check{\boldsymbol{e}}_{n-1}+O\left(z^{-1}\right).
\end{aligned}
\end{equation}
As $z \to 0$, their asymptotic behavior is
\begin{equation}\label{eq65}
\begin{aligned}
\check{\boldsymbol{p}}_{n+} = \check{\boldsymbol{e}}_{N+n-2}+O(z),\quad
\check{\boldsymbol{p}}_{n-} = \check{\boldsymbol{e}}_{n-1}+O(z).
\end{aligned}
\end{equation}
Substituting the asymptotic expansions (\ref{eq64}) and (\ref{eq65}) of the coordinate vectors into the definition (\ref{eq35}) of the modified auxiliary Jost eigenfunctions then yields the asymptotic behavior of $\boldsymbol{\mu} _{n \pm }(z)$ in their respective analytic regions. The resulting expansions as $z\to\infty$ are
\begin{equation}\label{eq66}
\begin{aligned}
{\boldsymbol{\mu} _{n + }} = {\boldsymbol{e}_n} + O\left(z^{-1}\right),\quad
{\boldsymbol{\mu} _{n - }} = {\boldsymbol{e}_n} + O\left(z^{-1}\right),
\end{aligned}
\end{equation}
and those as $z\to 0$ are
\begin{equation}\label{eq67}
\begin{aligned}
{\boldsymbol{\mu} _{n + }} = \dfrac{{{u_ + }}}{{{u_ - }}}{\boldsymbol{e}_n} + O(z),\quad
{\boldsymbol{\mu} _{n - }} = \dfrac{{{u_ - }}}{{{u_ + }}}{\boldsymbol{e}_n} + O(z).
\end{aligned}
\end{equation}
Finally, substituting the expressions (\ref{eq62}) and (\ref{eq63}) into Eqs. (\ref{eq40}) yields the asymptotic behavior of $g_{11}(z)$, $g_{33}(z)$, $b_{11}(z)$, and $b_{33}(z)$ in the appropriate regions:
\begin{equation}\label{eq68}
\begin{aligned}
{g_{11}} = 1 + O\left(z^{-1}\right),\quad{b_{33}} = 1 + O\left(z^{-1}\right),\quad
{g_{33}} = 1 + O\left(z^{-1}\right),\quad{b_{11}} = 1 + O\left(z^{-1}\right),
\end{aligned}
\end{equation}
and
\begin{equation}\label{eq69}
\begin{aligned}
{g_{11}} = \dfrac{{{u_ + }}}{{{u_ - }}} + O(z),\quad{b_{33}} = \dfrac{{{u_ + }}}{{{u_ - }}} + O(z),\quad
{g_{33}} = \dfrac{{{u_ - }}}{{{u_ + }}} + O(z),\quad{b_{11}} = \dfrac{{{u_ - }}}{{{u_ + }}} + O(z).
\end{aligned}
\end{equation}

\subsection{Behavior at the branch points}
\label{subsec:2.7}

In this subsection, we examine the behavior of the Jost eigenfunctions and the scattering coefficients at the branch points $z=\pm\mathrm{i}u_0$. Proposition \ref{Pro:1} and Eq. (\ref{eq8}) show that the first and last columns of the Jost eigenfunctions are well defined, continuous, and nonzero at $z=\pm\mathrm{i}u_0$. Combining these properties with symmetry (\ref{eq49}) and the boundary conditions for $\boldsymbol{\varphi}_{\pm1}(\pm\mathrm{i}u_0)$ and $\boldsymbol{\varphi}_{\pm3}(\pm\mathrm{i}u_0)$ as $x\to\pm\infty$, we obtain the following relations:
\begin{equation}\label{eq70}
\begin{aligned}
{\boldsymbol{\varphi} _{ - 1}}( \pm \mathrm{i}{u_0}) =  \mp {\mathrm{e}^{ - {\mathrm{i}}{\theta _ - }}}{\boldsymbol{\varphi} _{ - 3}}( \pm \mathrm{i}{u_0}),\quad{\boldsymbol{\varphi} _{ + 1}}( \pm \mathrm{i}{u_0}) =  \mp {\mathrm{e}^{ - {\mathrm{i}}{\theta _ + }}}{\boldsymbol{\varphi} _{ + 3}}( \pm \mathrm{i}{u_0}).
\end{aligned}
\end{equation}
The expressions (\ref{eq40}) show that $g_{11}(z)$, $g_{33}(z)$, $b_{11}(z)$, and $b_{33}(z)$ have poles at $z=\pm \mathrm{i}u_0$. Thus, as $z\to\pm\mathrm{i}u_0$, their asymptotic behavior is given by
\begin{equation}\label{eq71}
\begin{aligned}
(z \mp {\mathrm{i}u_0}){g_{11}} = O(1),\quad(z \mp {\mathrm{i}u_0}){g_{33}} = O(1),\\
(z \mp {\mathrm{i}u_0}){b_{11}} = O(1),\quad(z \mp {\mathrm{i}u_0}){b_{33}} = O(1).
\end{aligned}
\end{equation}
For the reflection coefficient $\beta_2(z)$, the second equation in Eqs. (\ref{eq61a}) and the relations (\ref{eq70}) yield $\mathop {\lim }\limits_{z \to  \pm \mathrm{i}{u_0}} {\beta _2}\left( z \right) =  \pm {\mathrm{e}^{ - \mathrm{i}{\theta _ + }}}$. The behavior of $\beta_1(z)$ at the branch points follows from the first equation in Eqs. (\ref{eq61a}) and the relations (\ref{eq33}):
$$\mathop {\lim }\limits_{z \to  \pm \mathrm{i}{u_0}} \frac{{\beta _1^\dag \left( {{z^*}} \right)}}{{\alpha (z)}} = \frac{{\boldsymbol{\varphi} _{ - 1}^\dag ( \mp \mathrm{i}{u_0}){\boldsymbol{\varphi} _{ + 2}}( \pm \mathrm{i}{u_0})}}{{\boldsymbol{\varphi} _{ - 1}^\dag ( \mp \mathrm{i}{u_0}){\boldsymbol{\varphi} _{ + 1}}( \pm \mathrm{i}{u_0})}} = {\mathrm{e}^{ - \mathrm{i}(N - 1)\left( { - \frac{1}{2}u_0^2x + \left( {2u_0^4 - \frac{3}{4}u_0^2} \right)t} \right)}}\frac{{\boldsymbol{\varphi} _{ - 1}^\dag ( \mp \mathrm{i}{u_0}){\boldsymbol{\varphi} _{ + 2}}( \pm \mathrm{i}{u_0})}}{{\det \left( {{\boldsymbol{\varphi} _{ + 1}},{\boldsymbol{\varphi} _{ - 2}},{\boldsymbol{\varphi} _{ - 3}}} \right)( \pm \mathrm{i}{u_0})}}.$$
These results indicate that the local behavior at $z=\pm\mathrm{i}u_0$ is closely related to the determinant of the associated Jost eigenfunctions. In particular, when $\det(\boldsymbol{\varphi}_{+1},\boldsymbol{\varphi}_{-2},\boldsymbol{\varphi}_{-3})(\pm \mathrm{i}u_0)=0$, the corresponding branch points are referred to as ``virtual levels'' \cite{A49}. However, in the general case where $\det(\boldsymbol{\varphi}_{+1},\boldsymbol{\varphi}_{-2},\boldsymbol{\varphi}_{-3})(\pm \mathrm{i}u_0)\ne0$ the jump matrix of the Riemann-Hilbert problem introduced in the next section remains finite at the branch points.

\subsection{Decomposition of the auxiliary Jost eigenfunctions}
\label{subsec:2.8}

To facilitate the derivation of the jump matrix and the residue conditions in the Riemann-Hilbert problem, we now express the auxiliary Jost eigenfunctions as combinations of the scattering coefficients and the Jost eigenfunctions.

We begin with the first symmetry (\ref{eq38}), which gives
\begin{equation}\label{eq72}
\begin{aligned}
&\boldsymbol{\varphi} _{ \pm ,j}^T(z)\boldsymbol{\varphi} _{ \pm ,l}^*({z^*}) = {\varsigma _{jl}},\quad &j&,l = 2, \dots ,N,\\
&\boldsymbol{\varphi} _{ \pm 1}^T(z)\boldsymbol{\varphi} _{ \pm 3}^*({z^*}) = 0,\quad&\boldsymbol{\varphi}& _{ - 1}^T(z)\boldsymbol{\varphi} _{ + 1}^*({z^*}) = \alpha b_{11}^*({z^*}),\\
&\boldsymbol{\varphi} _{ \pm 3}^T(z)\boldsymbol{\varphi} _{ \pm 1}^*({z^*}) = 0,\quad&\boldsymbol{\varphi}& _{ + 3}^T(z)\boldsymbol{\varphi} _{ - 1}^*({z^*}) = \alpha {b_{33}}(z),
\end{aligned}
\end{equation}
where $\varsigma _{jl}$ is defined by $\varsigma _{jl} = 1$ if $j= l$ and $\varsigma _{jl}=0$ otherwise. Combining the above relations with the definition (\ref{eq34}) of the auxiliary Jost eigenfunctions yields
\begin{equation}\label{eq73}
\begin{aligned}
\boldsymbol{\psi} _{j - }^\dag ({z^*}){\boldsymbol{\psi} _{l + }}(z) &= \frac{{{{( - 1)}^{j + l}}}}{{{\alpha ^2}(z)}}{\left( {\boldsymbol{\mathcal{H}}\left[ {{\boldsymbol{\varphi} _{ - ,1}},{\boldsymbol{\varphi} _{ + ,2}}, \dots ,\widehat{\boldsymbol\varphi}_{+,j}, \dots ,{\boldsymbol{\varphi} _{ + ,N + 1}}} \right](z)} \right)^T}\\
&\qquad \boldsymbol{\mathcal{H}}\left[ {\boldsymbol{\varphi} _{ + ,1}^*, \dots ,\widehat{\boldsymbol{\varphi}^*} _{ + ,l}, \dots ,\boldsymbol{\varphi} _{ + ,N}^*,\boldsymbol{\varphi} _{ - ,N + 1}^*} \right]({z^*})\\
 &= {( - 1)^{j + l}}b_{11}^*({z^*}){b_{33}}(z){\varsigma _{jl}}.
\end{aligned}
\end{equation}
An analogous calculation gives
\begin{equation}\label{eq74}
\begin{aligned}
\boldsymbol{\psi} _{j + }^\dag ({z^*}){\boldsymbol{\psi} _{l - }}(z) = {( - 1)^{j + l}}{b_{11}}(z)b_{33}^*({z^*}){\varsigma _{jl}}.
\end{aligned}
\end{equation}
These calculations use $\left({\mathcal{H}}\left[ {{\boldsymbol{a}_1}, \dots ,{\boldsymbol{a}_N}} \right]\right)^T \mathcal{H}\left[ {{\boldsymbol{s}_1}, \dots ,{\boldsymbol{s}_N}} \right] = \det \left( {\boldsymbol{a}_i^T{\boldsymbol{s}_j}} \right)$, where $\boldsymbol{a}_i, \boldsymbol{s}_i \in\mathbb{C}^{N+1}$. Equations (\ref{eq73}) and (\ref{eq74}) then imply
\begin{equation}\label{eq75}
\begin{aligned}
\boldsymbol{\psi} _ - ^\dag ({z^*}){\boldsymbol{\psi} _ + }(z) = b_{11}^*({z^*}){b_{33}}(z){\boldsymbol{I}_{(N - 2) \times (N - 2)}},\\
\boldsymbol{\psi} _ + ^\dag ({z^*}){\boldsymbol{\psi} _ - }(z) = {b_{11}}(z)b_{33}^*({z^*}){\boldsymbol{I}_{(N - 2) \times (N - 2)}}.
\end{aligned}
\end{equation}
In view of symmetry (\ref{eq56}), these relations are equivalent to
\begin{equation}\label{eq76}
\begin{aligned}
\boldsymbol{\mu} _ + ^\dag \left( {\frac{{u_0^2}}{{{z^*}}}} \right){\sigma _3}{\boldsymbol{\mu} _ + }\left( {z} \right) =  - \frac{{u_ + ^*}}{{u_ - ^*}}b_{11}^*({z^*}){b_{33}}(z){\boldsymbol{I}_{(N - 1) \times (N - 1)}},\\
\boldsymbol{\mu} _ - ^\dag \left( {\frac{{u_0^2}}{{{z^*}}}} \right){\sigma _3}{\boldsymbol{\mu} _ - }\left( {z} \right) =  - \frac{{u_ - ^*}}{{u_ + ^*}}{b_{11}}(z)b_{33}^*({z^*}){\boldsymbol{I}_{(N - 1) \times (N - 1)}}.
\end{aligned}
\end{equation}

Next, let $\boldsymbol{v}^{adj}$ denote the adjugate matrix of $\boldsymbol{v}=(\boldsymbol{v}_1,\dots,\boldsymbol{v}_{N+1})$, where $\boldsymbol{v}_i\in\mathbb{C}^{N+1}$. A direct calculation gives
\begin{equation}\label{eq77}
{ {{\boldsymbol{v}^{adj}}} } = \left(\boldsymbol{\mathcal{D}}\left[ {{\boldsymbol{v}_1}, \dots ,{\boldsymbol{v}_{N + 1}}} \right]\right)^T,
\end{equation}
whose $l$-th row is given by
\begin{equation}\label{eq77a}
\boldsymbol{v}^{adj}_l={\left( { - 1} \right)^{N + 1 - l}}\left(\boldsymbol{\mathcal{H}}\left[ {{\boldsymbol{v}_1}, \dots ,{\widehat{\boldsymbol{v}}_{l}}, \dots {\boldsymbol{v}_{N + 1}}} \right]\right)^T=\left(\boldsymbol{\mathcal{D}}_l\left[ {{\boldsymbol{v}_1}, \dots ,{\boldsymbol{v}_{N + 1}}} \right]\right)^T,
\end{equation}
where the operator $\boldsymbol{\mathcal{D}}$ is defined as
$$\boldsymbol{\mathcal{D}}\left[ {{\boldsymbol{v}_1}, \dots {\boldsymbol{v}_{N + 1}}} \right] =  - \sum\limits_{l = 1}^{N + 1} {\sum\limits_{j = 1}^{N + 1} {\det } } \begin{pmatrix}
\boldsymbol{v}&{{{\boldsymbol{e}}_j}}\\
{{\boldsymbol{e}}_l^T}&0
\end{pmatrix}{{\boldsymbol{e}}_j}{\boldsymbol{e}}_l^T.$$
Here, $\boldsymbol{\mathcal{D}}_l\left[ {{\boldsymbol{v}_1}, \dots ,{\boldsymbol{v}_{N + 1}}} \right]$ denotes the $l$-th column of $\boldsymbol{\mathcal{D}}\left[ {{\boldsymbol{v}_1}, \dots ,{\boldsymbol{v}_{N + 1}}} \right]$, that is, $\boldsymbol{\mathcal{D}}_l\left[ {{\boldsymbol{v}_1}, \dots ,{\boldsymbol{v}_{N + 1}}} \right]=\boldsymbol{\mathcal{D}}\left[ {{\boldsymbol{v}_1}, \dots {\boldsymbol{v}_{N + 1}}} \right]\boldsymbol{e}_l$.

The preceding identities and the definition of the auxiliary Jost eigenfunctions show that $\boldsymbol{\psi}_\pm(z)$ can be expressed as products of the scattering coefficients and the Jost eigenfunctions.

\begin{proposition}\label{Pro:4}
For $z\in\Sigma$, the auxiliary Jost eigenfunctions admit the following decompositions:
\begin{subequations}\label{eq78}
\begin{flalign}
{\boldsymbol{\psi} _ + }(z) = {\boldsymbol{\varphi} _{ + 2}}(z){b_{33}}(z) - {\boldsymbol{\varphi} _{ + 3}}(z){b_{32}}(z),\label{eq78a}\\
{\boldsymbol{\psi} _ - }(z) = {\boldsymbol{\varphi} _{ + 2}}(z){b_{11}}(z) - {\boldsymbol{\varphi} _{ + 1}}(z){b_{12}}(z).\label{eq78b}
\end{flalign}
\end{subequations}
\end{proposition}
\begin{proof}
Using Eqs. (\ref{eq77a}) and (\ref{eq33}), by direct calculation yields
$$\begin{aligned}
{\boldsymbol{\mathcal{D}}_l}\left[ {{\boldsymbol{\varphi} _{ + 1}^*,\boldsymbol{\varphi} _{ + 2}^*,\boldsymbol{\varphi} _{ + 3}^*} } \right]({z^*})
 &= {( - 1)^{N + 1 - l}}\boldsymbol{\mathcal{H}}\left[ {\boldsymbol{\varphi} _{ + ,1}^*, \dots ,\widehat{\boldsymbol{\varphi}^*} _{ + ,l}, \dots ,\boldsymbol{\varphi} _{ + ,N + 1}^*} \right]({z^*})\\
 &= {\alpha_l}{{\mathrm{e}}^{  {\mathrm{i}}(1-N ){\theta _2}(z)}}{\boldsymbol{\varphi} _{ + ,l}}(z),
\end{aligned}$$
which implies
\begin{equation}\label{eq79}
\begin{aligned}
\boldsymbol{\mathcal{D}}\left[ { {\boldsymbol{\varphi} _{ + 1}^*,\boldsymbol{\varphi} _{ + 2}^*,\boldsymbol{\varphi} _{ + 3}^*} } \right]({z^*}) = {{\mathrm{e}}^{  {\mathrm{i}}(1-N){\theta _2}(z)}}\left( {{\boldsymbol{\varphi} _{ + 1}},\alpha {\boldsymbol{\varphi} _{ + 2}},{\boldsymbol{\varphi} _{ + 3}}} \right)(z).
\end{aligned}
\end{equation}
Incorporating relations (\ref{eq77}) and (\ref{eq79}), along with Eqs. (\ref{eq40}), we further derive
$$\begin{aligned}
\boldsymbol{\mathcal{D}}\left[ {\boldsymbol{\varphi} _{ + 1}^*,\boldsymbol{\varphi} _{ + 2}^*,\boldsymbol{\varphi} _{ - 3}^*} \right]({z^*})
 &= \boldsymbol{\mathcal{D}}\left[ \left( {\left( {\boldsymbol{\varphi} _{ + 1}^*,\boldsymbol{\varphi} _{ + 2}^*,\boldsymbol{\varphi} _{ + 3}^*} \right)({z^*})} \right)\begin{pmatrix}
1&{{{\bf{0}}_{1 \times (N - 1)}}}&{g_{13}^*({z^*})}\\
{{{\bf{0}}_{(N - 1) \times 1}}}&{{{\bf{I}}_{(N - 1) \times (N - 1)}}}&{g_{23}^*({z^*})}\\
0&{{{\bf{0}}_{1 \times (N - 1)}}}&{g_{33}^*({z^*})}
\end{pmatrix} \right]\\
 &= {\left[ {{{\left( {\boldsymbol{\varphi} _{ + 1}^*,\boldsymbol{\varphi} _{ + 2}^*,\boldsymbol{\varphi} _{ + 3}^*} \right)}^{adj}}({z^*})} \right]^T}{\left[ {{{\begin{pmatrix}
1&{{{\bf{0}}_{1 \times (N - 1)}}}&{g_{13}^*({z^*})}\\
{{{\bf{0}}_{(N - 1) \times 1}}}&{{{\bf{I}}_{(N - 1) \times (N - 1)}}}&{g_{23}^*({z^*})}\\
0&{{{\bf{0}}_{1 \times (N - 1)}}}&{g_{33}^*({z^*})}
\end{pmatrix}}^{adj}}} \right]^T}\\
 &= \boldsymbol{\mathcal{D}}\left[ {\left( {\boldsymbol{\varphi} _{ + 1}^*,\boldsymbol{\varphi} _{ + 2}^*,\boldsymbol{\varphi} _{ + 3}^*} \right)({z^*})} \right]\begin{pmatrix}
{{b_{33}}(z)}&{{{\bf{0}}_{1 \times (N - 1)}}}&0\\
{{{\bf{0}}_{(N - 1) \times 1}}}&{{b_{33}}(z){{\bf{I}}_{(N - 1) \times (N - 1)}}}&{{{\bf{0}}_{(N - 1) \times 1}}}\\
{ - {b_{31}}(z)}&{ - \alpha {b_{32}}(z)}&1
\end{pmatrix}\\
 &= {{\mathrm{e}}^{  {\mathrm{i}}(1-N ){\theta _2}(z)}}\left( {{\boldsymbol{\varphi} _{ + 1}},\alpha {\boldsymbol{\varphi} _{ + 2}},{\boldsymbol{\varphi} _{ + 3}}} \right)(z)\begin{pmatrix}
{{b_{33}}(z)}&{{{\bf{0}}_{1 \times (N - 1)}}}&0\\
{{{\bf{0}}_{(N - 1) \times 1}}}&{{b_{33}}(z){{\bf{I}}_{(N - 1) \times (N - 1)}}}&{{{\bf{0}}_{(N - 1) \times 1}}}\\
{ - {b_{31}}(z)}&{ - \alpha {b_{32}}(z)}&1
\end{pmatrix}.
\end{aligned}$$
Therefore,
$$\begin{aligned}
\left( {{\boldsymbol{\mathcal{D}}_2}, \dots ,{\boldsymbol{\mathcal{D}}_N}} \right)\left[ {\left( {\boldsymbol{\varphi} _{ + 1}^*,\boldsymbol{\varphi} _{ + 2}^*,\boldsymbol{\varphi} _{ - 3}^*} \right)} \right]\left( {{z^*}} \right) = {{\mathrm{e}}^{ {\mathrm{i}}(1-N ){\theta _2}(z)}}\alpha \left( {b_{33}}(z){{\boldsymbol{\varphi} _{ + 2}}(z) -{\boldsymbol{\varphi} _{ + 3}}(z){b_{32}}(z)} \right).
\end{aligned}$$
Substituting Eq. (\ref{eq77a}) into the definition (\ref{eq34}) of the auxiliary Jost eigenfunctions gives
$$\begin{aligned}
\alpha {\boldsymbol{\psi} _{l + }}(z)
 &= {( - 1)^{N + 1 - l}}{{\mathrm{e}}^{{\mathrm{i}}(N - 1){\theta _2}(z)}}\boldsymbol{\mathcal{H}}\left[ {\boldsymbol{\varphi} _{ + ,1}^*, \dots ,\boldsymbol{\varphi} _{ + ,l - 1}^*,\boldsymbol{\varphi} _{ + ,l + 1}^*, \dots ,\boldsymbol{\varphi} _{ + ,N}^*,\boldsymbol{\varphi} _{ - ,N + 1}^*} \right]({z^*})\\
 &= {{\mathrm{e}}^{{\mathrm{i}}(N - 1){\theta _2}(z)}}{\boldsymbol{\mathcal{D}}_l}\left[ {\boldsymbol{\varphi} _{ + ,1}^*, \dots ,\boldsymbol{\varphi} _{ + ,N}^*,\boldsymbol{\varphi} _{ - ,N + 1}^*} \right]({z^*})\\
 &= {{\mathrm{e}}^{{\mathrm{i}}(N - 1){\theta _2}(z)}}{\boldsymbol{\mathcal{D}}_l}\left[ {\boldsymbol{\varphi} _{ + 1}^*,\boldsymbol{\varphi} _{ + 2}^*,\boldsymbol{\varphi} _{ - 3}^*} \right]({z^*}).
\end{aligned}$$
In summary, we derive
$$\begin{aligned}
{\boldsymbol{\psi} _ + }(z)
 &= \left( {{\boldsymbol{\psi} _{2 + }}, \dots ,{\boldsymbol{\psi} _{N + }}} \right)(z)\\
 &= \frac{1}{{\alpha }}{{\mathrm{e}}^{{\mathrm{i}}(N - 1){\theta _2}(z)}}\left( {{\boldsymbol{\mathcal{D}}_2}, \dots ,{\boldsymbol{\mathcal{D}}_N}} \right)\left[ {\boldsymbol{\varphi} _{ + 1}^*,\boldsymbol{\varphi} _{ + 2}^*,\boldsymbol{\varphi} _{ - 3}^*} \right]({z^*})\\
 &= {\boldsymbol{\varphi} _{ + 2}}(z){b_{33}}(z) - {\boldsymbol{\varphi} _{ + 3}}(z){b_{32}}(z).
\end{aligned}$$
These steps establish Eq. (\ref{eq78a}). The decomposition (\ref{eq78b}) of $\boldsymbol{\psi}_-(z)$ follows by an analogous argument.
\end{proof}

\section{Inverse problem}
\label{sec:3}

In this section, we formulate the inverse scattering problem as a matrix Riemann-Hilbert problem with specified analytic properties, jump condition, and normalization condition. This formulation requires a set of analytic eigenfunctions comprising the modified and auxiliary Jost eigenfunctions. Consequently, establishing the jump condition requires not only the scattering relations (\ref{eq21}) but also the decompositions of ${\boldsymbol{\psi} _ \pm }(z)$ in Eqs. (\ref{eq78}).

\subsection{Riemann-Hilbert problem}
\label{subsec:3.1}

Using the analyticity of the Jost eigenfunctions and the scattering coefficients, we define the following sectionally meromorphic matrices:
\begin{equation}\label{eq80}
\begin{aligned}
{\boldsymbol{M}_ + }(x,t,z) &=\left( {\frac{{{\boldsymbol{\upsilon} _{ - 1}}}}{{{g_{11}}}},\frac{{{\boldsymbol{\mu} _ + }}}{{{b_{33}}}},{\boldsymbol{\upsilon} _{ + 3}}} \right)(x,t,z),\quad z \in {\Xi _ + },\\
{\boldsymbol{M}_ - }(x,t,z) &= \left( {{\boldsymbol{\upsilon} _{ + 1}},\frac{{{\boldsymbol{\mu} _ - }}}{{{b_{11}}}},\frac{{{\boldsymbol{\upsilon} _{ - 3}}}}{{{g_{33}}}}} \right)(x,t,z),\quad z \in {\Xi _ - }.
\end{aligned}
\end{equation}
For $z\in\Xi_\pm$, the symmetries in Subsection \ref{subsec:2.5} imply that these matrices satisfy the following three fundamental symmetry relations:
\begin{equation}\label{eq81}
\begin{aligned}
\quad{\boldsymbol{M}}_ \mp ^\dag ({z^*}) = {\boldsymbol{Q}}(z){{\boldsymbol{M}}^{-1}_ \pm }(z),\quad{{\boldsymbol{M}}_ \pm }(z) = {{\boldsymbol{\sigma }}_3}{{\boldsymbol{M}}_ \pm }( - z){{\boldsymbol{\sigma }}_3},\quad
{{\boldsymbol{M}}_ \pm }(z) = {{\boldsymbol{M}}_ \mp }\left( { - \frac{{u_0^2}}{z}} \right){{\boldsymbol{L}}_ + }(z).
\end{aligned}
\end{equation}
A direct calculation using the scattering relations (\ref{eq21}) and Eqs. (\ref{eq78}) yields
\begin{equation}\label{eq81b}
\det {\boldsymbol{M}_ \pm }(z) = \alpha,\quad z\in\Xi_\pm\cup{\Sigma}.
\end{equation}
Combining this identity with Eq. (\ref{eq77}) gives
\begin{equation}\label{eq81a}
\boldsymbol{\mathcal{D}}\left[ {{\boldsymbol{M}_ \pm }(z)} \right] = \alpha\boldsymbol{M}_ \mp ^*({z^*}){\boldsymbol{Q}^{ - 1}}(z).
\end{equation}

Next, we derive the jump condition, suppressing the explicit dependence on $(x, t, z)$ for brevity. The scattering relation (\ref{eq21b}) yields
\begin{subequations}\label{eq82}
\begin{align}
\frac{{{\boldsymbol{\varphi} _{ - 1}}}}{{{g_{11}}}} &= {\boldsymbol{\varphi} _{ + 3}}\frac{{{g_{31}}}}{{{g_{11}}}}+ {\boldsymbol{\varphi} _{ + 2}}\frac{{{g_{21}}}}{{{g_{11}}}}+ {\boldsymbol{\varphi} _{ + 1}},\label{eq82a} \\
{\boldsymbol{\varphi} _{ + 3}} &= \frac{1}{{{g_{33}}}}{\boldsymbol{\varphi} _{ - 3}} - {\boldsymbol{\varphi} _{ + 1}}\frac{{{g_{13}}}}{{{g_{33}}}} - {\boldsymbol{\varphi} _{ + 2}}\frac{{{g_{23}}}}{{{g_{33}}}}.\label{eq82b}
\end{align}
\end{subequations}
From decomposition (\ref{eq78b}), we further obtain
\begin{equation}\label{eq83}
{\boldsymbol{\varphi} _{ + 2}} = \frac{{{\boldsymbol{\psi} _ - }}}{{{b_{11}}}} + {\boldsymbol{\varphi} _{ + 1}}\frac{{{b_{12}}}}{{{b_{11}}}}.
\end{equation}
With the aid of Eqs. (\ref{eq82b}) and (\ref{eq83}), Eq. (\ref{eq82a}) can be rewritten as
\begin{equation}\label{eq84}
\begin{aligned}
\frac{{{\boldsymbol{\varphi} _{ - 1}}}}{{{g_{11}}}}
&= {\boldsymbol{\varphi} _{ + 1}}\left( {1 + \frac{{{b_{12}}}}{{{b_{11}}}}\left( {\frac{{{g_{21}}}}{{{g_{11}}}} - \frac{{{g_{23}}}}{{{g_{33}}}}\frac{{{g_{31}}}}{{{g_{11}}}}} \right) - \frac{{{g_{13}}}}{{{g_{33}}}}\frac{{{g_{31}}}}{{{g_{11}}}}} \right) + \frac{{{\boldsymbol{\psi} _ - }}}{{{b_{11}}}}\left( {\frac{{{g_{21}}}}{{{g_{11}}}} - \frac{{{g_{23}}}}{{{g_{33}}}}\frac{{{g_{31}}}}{{{g_{11}}}}} \right) + \frac{{{\boldsymbol{\varphi} _{ - 3}}}}{{{g_{33}}}}\frac{{{g_{31}}}}{{{g_{11}}}}\\
&=\boldsymbol{\varphi} _{ + 1}\left({1 + \dfrac{1}{{\alpha }}\beta _1^\dag ({z^*}){\beta _1}(z) + \beta _2^*({z^*}){\beta _2}(z)}\right)+\frac{{{\boldsymbol{\psi} _ - }}}{{{b_{11}}}}\left( {{\beta _1}(z) - {\beta _3}(z){\beta _2}(z)}\right)+\frac{{{\boldsymbol{\varphi} _{ - 3}}}}{{{g_{33}}}}{{\beta _2}(z)}.
\end{aligned}
\end{equation}
Combining Eqs. (\ref{eq82b}) and (\ref{eq83}) with the decomposition (\ref{eq78a}) of $\boldsymbol{\psi}_+(z)$, we obtain
\begin{equation}\label{eq85}
\begin{aligned}
\frac{{{\boldsymbol{\psi} _ + }}}{{{b_{33}}}}
 &= {\boldsymbol{\varphi} _{ + 2}} - {\boldsymbol{\varphi} _{ + 3}}\frac{{{b_{32}}}}{{{b_{33}}}}\\
 &= {\boldsymbol{\varphi} _{ + 1}}\left( {\frac{{{b_{12}}}}{{{b_{11}}}} + \frac{{{b_{12}}}}{{{b_{11}}}}\frac{{{g_{23}}}}{{{g_{33}}}}\frac{{{b_{32}}}}{{{b_{33}}}} + \frac{{{g_{13}}}}{{{g_{33}}}}\frac{{{b_{32}}}}{{{b_{33}}}}} \right) + \frac{{{\boldsymbol{\psi} _ - }}}{{{b_{11}}}}\left( {{\boldsymbol{I}_{(N - 1) \times (N - 1)}} + \frac{{{g_{23}}}}{{{g_{33}}}}\frac{{{b_{32}}}}{{{b_{33}}}}} \right) - \frac{{{\boldsymbol{\varphi} _{ - 3}}}}{{{g_{33}}}}\frac{{{b_{32}}}}{{{b_{33}}}}\\
 &={\boldsymbol{\varphi} _{ + 1}}\left( {\dfrac{1}{{\alpha }}\left( {\beta _1^\dag ({z^*}) -\beta _2^*({z^*})\beta _3^\dag ({z^*})} \right)} \right)+\frac{{{\boldsymbol{\psi} _ - }}}{{{b_{11}}}}\left( {{\boldsymbol{I}_{(N - 1) \times (N - 1)}} + \dfrac{1}{{\alpha }}{\beta _3}(z)\beta _3^\dag ({z^*})} \right)+\frac{{{\boldsymbol{\varphi} _{ - 3}}}}{{{g_{33}}}}\left( { - \dfrac{1}{{\alpha }}\beta _3^\dag ({z^*})} \right).
\end{aligned}
\end{equation}
Finally, substituting Eq. (\ref{eq83}) into Eq. (\ref{eq82b}) gives the relation
\begin{equation}\label{eq86}
\begin{aligned}
{\boldsymbol{\varphi} _{ + 3}}
&= {\boldsymbol{\varphi} _{ + 1}}\left( { - \frac{{{b_{12}}}}{{{b_{11}}}}\frac{{{g_{23}}}}{{{g_{33}}}} - \frac{{{g_{13}}}}{{{g_{33}}}}} \right) + \frac{{{\boldsymbol{\psi} _ - }}}{{{b_{11}}}}\left( { - \frac{{{g_{23}}}}{{{g_{33}}}}} \right) + \frac{{{\boldsymbol{\varphi} _{ - 3}}}}{{{g_{33}}}}\\
&= {\boldsymbol{\varphi} _{ + 1}} {  \beta _2^*({z^*})}+\frac{{{\boldsymbol{\psi} _ - }}}{{{b_{11}}}}\left( { - {\beta _3}(z)} \right)+\frac{{{\boldsymbol{\varphi} _{ - 3}}}}{{{g_{33}}}}.
\end{aligned}
\end{equation}
Expressing the preceding three relations in matrix form yields
\begin{equation}\label{eq87}
{\boldsymbol{M}_ + }(z) = {\boldsymbol{M}_ - }(z){\mathrm{e}^{\mathrm{i}\boldsymbol{\Theta} (z)}}\boldsymbol{V}(z){\mathrm{e}^{ - \mathrm{i}\boldsymbol{\Theta} (z)}}, \quad z\in\Sigma,
\end{equation}
where
$$\begin{aligned}
\boldsymbol{V}(z) = \begin{pmatrix}
{1 + \dfrac{1}{{\alpha }}\beta _1^\dag ({z^*}){\beta _1}(z) + \beta _2^*({z^*}){\beta _2}(z)}&{\dfrac{1}{{\alpha }}\left( {\beta _1^\dag ({z^*}) - \beta _2^*({z^*})\beta _3^\dag ({z^*})} \right)}&{  \beta _2^*({z^*})}\\
{{\beta _1}(z) - {\beta _3}(z){\beta _2}(z)}&{{\boldsymbol{I}_{(N - 1) \times (N - 1)}} + \dfrac{1}{{\alpha }}{\beta _3}(z)\beta _3^\dag ({z^*})}&{ - {\beta _3}(z)}\\
{{\beta _2}(z)}&{ - \dfrac{1}{{\alpha }}\beta _3^\dag ({z^*})}&1
\end{pmatrix}.
\end{aligned}$$
Here, the meromorphic matrix-valued functions are understood in the sense of non-tangential limit values. By Proposition \ref{Pro:1} and Eqs. (\ref{eq71}), these limit values are well defined for $z\in\Sigma$. Following the same procedure used to derive $\boldsymbol{V}(z)$, we obtain $\boldsymbol{V}^{-1}(z)$ in the following form:
$$\begin{aligned}
{\boldsymbol{V}^{ - 1}}(z) = \begin{pmatrix}
1&{ - \dfrac{1}{{\alpha }}\beta _1^\dag ({z^*})}&{\dfrac{{{u_ + }}}{{u_ + ^*}}{\beta _2}\left( { - \dfrac{{u_0^2}}{z}} \right)}\\
{ - {\beta _1}(z)}&{{\boldsymbol{I}_{(N - 1) \times (N - 1)}} + \dfrac{1}{{\alpha }}{\beta _1}(z)\beta _1^\dag ({z^*})}&{{\beta _3}(z) - \dfrac{{{u_ + }}}{{u_ + ^*}}{\beta _1}(z){\beta _2}\left( { - \dfrac{{u_0^2}}{z}} \right)}\\
{\dfrac{{u_ + ^*}}{{{u_ + }}}\beta _2^*\left( { - \dfrac{{u_0^2}}{{{z^*}}}} \right)}&{\left( {\beta _3^\dag ({z^*}) - \dfrac{{u_ + ^*}}{{{u_ + }}}\beta _2^*\left( { - \dfrac{{u_0^2}}{{{z^*}}}} \right)\beta _1^\dag ({z^*})} \right)\dfrac{1}{{\alpha }}}&{1 + \dfrac{1}{{\alpha }}\beta _3^\dag ({z^*}){\beta _3}(z) + \beta _2^*\left( { - \dfrac{{u_0^2}}{{{z^*}}}} \right){\beta _2}\left( { - \dfrac{{u_0^2}}{z}} \right)}
\end{pmatrix}.
\end{aligned}$$
Furthermore, combining the symmetries (\ref{eq81}) of the sectionally meromorphic matrices with the jump condition (\ref{eq87}) implies that $\boldsymbol{V}$ satisfies
$$\begin{aligned}
\boldsymbol{V}(z) = {\boldsymbol{Q}^{ - 1}}(z){\boldsymbol{V}^\dag }({z^*})\boldsymbol{Q}(z),\quad \boldsymbol{V}(z) = \boldsymbol{L}_ + ^{ - 1}(z){\boldsymbol{V}^{ - 1}}\left( { - \frac{{u_0^2}}{z}} \right){\boldsymbol{L}_ + }(z),\quad\boldsymbol{V}(z) = {\boldsymbol{\sigma} _3}\boldsymbol{V}( - z){\boldsymbol{\sigma} _3}.
\end{aligned}$$

The jump condition (\ref{eq87}) is supplemented by a suitable normalization condition to guarantee uniqueness of the Riemann-Hilbert problem. This condition is determined by both the leading asymptotic behavior of the sectionally meromorphic matrices ${\boldsymbol{M}_ \pm }(z)$ as $z \to \infty$ and the contribution from the pole at $z = 0$. We substitute the asymptotic behavior given in Subsection \ref{subsec:2.6} into the definition (\ref{eq80}) of the matrices ${\boldsymbol{M}_ \pm }(z)$ to obtain the following normalization condition:
\begin{equation}\label{eq88}
\begin{aligned}
{\boldsymbol{M}_ \pm }(z) &= {\boldsymbol{I}_{(N + 1) \times (N + 1)}} + O\left( {{z^{ - 1}}} \right),\\
{\boldsymbol{M}_ \pm }(z) &= \frac{1}{z}{\boldsymbol{M}_0} + O\left( 1 \right),
\end{aligned}
\end{equation}
where
$$\begin{aligned}
{\boldsymbol{M}_0} = \begin{pmatrix}
0&{{\boldsymbol{0}_{1 \times (N - 1)}}}&{ - \mathrm{i}{u_ + }}\\
{{\boldsymbol{0}_{(N - 1) \times 1}}}&{{\boldsymbol{0}_{(N - 1) \times (N - 1)}}}&{{\boldsymbol{0}_{(N - 1) \times 1}}}\\
{ - \mathrm{i}u_ + ^*}&{{\boldsymbol{0}_{1 \times (N - 1)}}}&0
\end{pmatrix},
\end{aligned}$$
and ${\boldsymbol{I}_{(N + 1) \times (N + 1)}}+\frac{1}{z}{\boldsymbol{M}_0}=\boldsymbol{A}_+$.

\subsection{Residue conditions }
\label{subsec:3.2}
We next examine the discrete spectrum and the distribution of discrete eigenvalues before deriving the corresponding residue conditions. The definition (\ref{eq80}) of the sectionally meromorphic matrices shows that ${\boldsymbol{M}_ \pm }(z)$ have singularities only at the zeros of $g_{11}, g_{33}, b_{11}$ and $b_{33}$. Assume that $g_{11}$ has $N_1$ simple zeros $z_m$ satisfying $\left| {{z_m}} \right| \ne {u_0}$ and $N_2$ simple zeros ${{\omega _l}}$ satisfying $\left| {{\omega_l}} \right| ={u_0}$. All these zeros lie in the region ${\Xi _ + } \cap \left\{ {{\mathop{\rm Im}\nolimits} z > 0} \right\}$. We further assume that no spectral singularities are present. From the three symmetries satisfied by the scattering coefficients, the following equivalence relations hold:
\begin{equation}\label{eq88b}
\begin{aligned}
&{b_{11}}(z_m^*) =0 \Leftrightarrow {g_{11}}({z_m}) = 0  \Leftrightarrow {g_{33}}\left( { - \dfrac{{u_0^2}}{{{z_m}}}} \right) = 0 \Leftrightarrow {b_{33}}\left( { - \dfrac{{u_0^2}}{{z_m^*}}} \right) = 0 \Leftrightarrow \\
&{b_{11}}( - z_m^*) = 0 \Leftrightarrow {g_{11}}( - {z_m}) = 0 \Leftrightarrow {g_{33}}\left( {\dfrac{{u_0^2}}{{{z_m}}}} \right) = 0 \Leftrightarrow {b_{33}}\left( {\dfrac{{u_0^2}}{{z_m^*}}} \right) = 0.
\end{aligned}
\end{equation}
Therefore, the discrete spectrum is
\begin{equation}\label{eq88a}
\left\{ { \pm {z_m}, \pm z_m^*, \pm \frac{{u_0^2}}{{{z_m}}}, \pm \frac{{u_0^2}}{{z_m^*}}} \right\}_{m = 1}^{{N_1}} \cup \left\{ { \pm {\omega _l}, \pm \omega _l^*} \right\}_{l = 1}^{{N_2}}.
\end{equation}

\begin{proposition}\label{Pro:5}
Let $\zeta$ be a simple pole of the sectionally meromorphic matrix in its corresponding domain of analyticity. Then
\begin{subequations}\label{eq89}
\begin{align}
\mathop {{\mathrm{Res}}}\limits_{z =  - u_0^2/\zeta } \boldsymbol{M}(z) = \frac{{u_0^2}}{{{\zeta ^2}}}\mathop {{\mathrm{Res}}}\limits_{z = \zeta } \boldsymbol{M}(z){\boldsymbol{L}_ + }\left( { - \frac{{u_0^2}}{\zeta }} \right),\label{eq89a} \\
\mathop {{\mathrm{Res}}}\limits_{z =  - \zeta } \boldsymbol{M}(z) =  - {\boldsymbol{\sigma} _3}\mathop {{\mathrm{Res}}}\limits_{z = \zeta } \boldsymbol{M}(z){\boldsymbol{\sigma} _3},\hspace{28pt}\label{eq89b}
\end{align}
\end{subequations}
where $\boldsymbol{M}(z) = {\boldsymbol{M}_ \pm }(z)$, $z \in {\Xi _ \pm }$. Suppose further that there exists a matrix $\boldsymbol{\Omega}$ with $\boldsymbol{\Omega}^2=\boldsymbol{0}_{(N+1)\times(N+1)}$ such that $\mathop {{\mathrm{Res}}}\limits_{z = \zeta } \boldsymbol{M}(z) = \mathop {\lim }\limits_{z \to \zeta } \boldsymbol{M}(z)\boldsymbol{\Omega} $. Then the following holds:
\begin{equation}\label{eq90}
\mathop {{\mathrm{Res}}}\limits_{z = {\zeta ^*}} \boldsymbol{M}(z) =  - \mathop {\lim }\limits_{z \to {\zeta ^*}} \boldsymbol{M}(z){\boldsymbol{Q}^{ - 1}}(z){\boldsymbol{\Omega} ^\dag}\boldsymbol{Q}(z).
\end{equation}
\end{proposition}
\begin{proof}
From the discrete spectrum (\ref{eq88a}), $-u_0^2/\zeta$ is also a simple pole of $\boldsymbol{M}(z)$. Applying the symmetries (\ref{eq81}), we obtain
$$\begin{aligned}
\mathop {{\mathrm{Res}}}\limits_{z =  - u_0^2/\zeta } \boldsymbol{M}(z)
 &= \mathop {\lim }\limits_{z \to  - u_0^2/\zeta } \boldsymbol{M}\left( { - \frac{{u_0^2}}{z}} \right){\boldsymbol{L}_ + }(z)\left( {z + \frac{{u_0^2}}{\zeta }} \right)
 = \mathop {\lim }\limits_{z \to  - u_0^2/\zeta }  - \frac{z}{\zeta }\boldsymbol{M}\left( { - \frac{{u_0^2}}{z}} \right){\boldsymbol{L}_ + }(z)\left( { - \frac{{u_0^2}}{z} - \zeta } \right)\\
 &= \frac{{u_0^2}}{{{\zeta ^2}}}\mathop {{\mathrm{Res}}}\limits_{z = \zeta } \boldsymbol{M}(z){\boldsymbol{L}_ + }\left( { - \frac{{u_0^2}}{\zeta }} \right).
\end{aligned}$$
Similarly, a direct calculation yields
$$\begin{aligned}
\mathop {{\mathrm{Res}}}\limits_{ z=- \zeta } \boldsymbol{M}(z)
 = \mathop {\lim }\limits_{z \to  - \zeta }{\boldsymbol{\sigma} _3}\boldsymbol{M}\left( { - z} \right){\boldsymbol{\sigma} _3} \left( {z + \zeta } \right)
 =  - {\boldsymbol{\sigma} _3}\mathop {{\mathrm{Res}}}\limits_{z=\zeta}  \boldsymbol{M}(z){\boldsymbol{\sigma} _3}.
\end{aligned}$$
Since $\boldsymbol{M}(z)$ has a simple pole at $z=\zeta$, its Laurent expansion near $\zeta$ takes the form
$$\boldsymbol{M}(z) = \frac{{{\boldsymbol{M}^{[ - 1]}}}}{{z - \zeta }} + {\boldsymbol{M}^{[0]}} + O(z - \zeta ),$$
which yields ${\boldsymbol{M}^{[ - 1]}} = \mathop {{\mathrm{Res}}}\limits_{z=\zeta}  \boldsymbol{M}(z)$. By Right-multiplying both sides of the above equation by matrix $\boldsymbol{\Omega}$, it follows that ${\boldsymbol{M}^{[ - 1]}}\boldsymbol{\Omega}  = {\boldsymbol{0}_{(N + 1) \times (N + 1)}}$ and ${\boldsymbol{M}^{[ - 1]}} = {\boldsymbol{M}^{[0]}}\boldsymbol{\Omega} $. Introducing the auxiliary matrix
\begin{equation}\label{eq91}
{\boldsymbol{\Phi}}(z) = \boldsymbol{M}(z)\left(   \frac{\boldsymbol{\Omega} }{{\zeta - z }}+ {\boldsymbol{I}_{(N + 1) \times (N + 1)}} \right),
\end{equation}
we obtain ${\boldsymbol{\Phi}}(z) = O(1),$ which implies that ${\boldsymbol{\Phi}}(z)$ admits an analytic continuation to $z = \zeta $. Solving Eq. (\ref{eq91}) for $\boldsymbol{M}$ gives
$$\boldsymbol{M}(z) = {\boldsymbol{\Phi}}(z)\left( {{\boldsymbol{I}_{(N + 1) \times (N + 1)}} + \frac{\boldsymbol{\Omega} }{{z - \zeta }}} \right),$$
and for $z\ne\zeta$, this yields $\det \boldsymbol{M} = \det {\boldsymbol{\Phi}} = \alpha $. Given the analyticity of ${\boldsymbol{\Phi}}(z)$ at $z=\zeta$, continuity yields $\det {\boldsymbol{\Phi}}(\zeta ) = \alpha (\zeta ) \ne 0$, which implies that the invertibility of ${\boldsymbol{\Phi}}(\zeta )$. Combining this invertibility with the first symmetry of the matrix $\boldsymbol{M}(z)$ in Eqs. (\ref{eq81}), we obtain
\begin{equation}\label{eq92}
{ {{{{\boldsymbol{\Phi}}}^\dag}({z^*})}}\boldsymbol{M}(z) = \boldsymbol{Q}(z) - \frac{{{\boldsymbol{\Omega} ^\dag}\boldsymbol{Q}(z)}}{{z - {\zeta ^*}}}.
\end{equation}
Taking the residue at $z=\zeta^*$ of this relation yields
\begin{equation}\label{eq93}
\mathop {{\mathrm{Res}}}\limits_{{z=\zeta ^*}} \boldsymbol{M}(z) =  - {\left( {{{{\boldsymbol{\Phi}}}^\dag}(\zeta )} \right)^{ - 1}}{\boldsymbol{\Omega} ^\dag}\boldsymbol{Q}({\zeta ^*}).
\end{equation}
On the other hand, Eq. (\ref{eq92}) gives $\boldsymbol{M}(z){\boldsymbol{Q}^{ - 1}}(z){\boldsymbol{\Omega} ^\dag}\boldsymbol{Q}(z) = {\left( {{{{\boldsymbol{\Phi}}}^\dag}({z^*})} \right)^{ - 1}}{\boldsymbol{\Omega} ^\dag}\boldsymbol{Q}(z)$. Since ${\boldsymbol{\Phi}}(z)$ extends analytically to $z = \zeta $ and is invertible there, ${{\boldsymbol{\Phi}}^\dag}({z^*})$ and its inverse are analytic in a neighborhood of $z = {\zeta ^*}$. Therefore, $z = {\zeta ^*}$ is a removable singularity of $\boldsymbol{M}(z){\boldsymbol{Q}^{ - 1}}(z){\boldsymbol{\Omega} ^\dag}\boldsymbol{Q}(z)$, and we obtain
 $$\mathop {\lim }\limits_{z \to {\zeta ^*}} \boldsymbol{M}(z){\boldsymbol{Q}^{ - 1}}(z){\boldsymbol{\Omega} ^\dag}\boldsymbol{Q}(z) = {\left( {{{{\boldsymbol{\Phi}}}^\dag}(\zeta )} \right)^{ - 1}}{\boldsymbol{\Omega} ^\dag}\boldsymbol{Q}({\zeta ^*}).$$
Finally, combining Eq. (\ref{eq93}) with the preceding expression yields Eq. (\ref{eq90}), which completes the proof.
\end{proof}

Using the residue transformation relations established in the preceding proposition, we derive explicit expressions for the residues at the discrete eigenvalues. These results are summarized in the following proposition. Here, $f'({z_m}) = {\left. {\frac{\mathrm{d}}{{\mathrm{d}z}}f(z)} \right|_{z = {z_m}}}$.

\begin{proposition}\label{Pro:6}
For discrete eigenvalues in the sets (\ref{eq88a}), the following residue conditions hold:
\begin{subequations}\label{eq94}
\begin{align}
\mathop {{\mathrm{Res}}}\limits_{z = {\omega _l}} {\boldsymbol{M}_ + }(z) = \left( {{d _l}{{\mathrm{e}}^{ -2 {\mathrm{i}}{\theta _1}({\omega _l})}}{\boldsymbol{M}_{ + 3}}({\omega _l}),{\boldsymbol{0}_{(N + 1) \times (N - 1)}},{\tilde{\boldsymbol{0}}}} \right),\hspace{24pt}\mathop {{\mathrm{Res}}}\limits_{z = \omega _l^*} {\boldsymbol{M}_ - }(z) = \left( {{\tilde{\boldsymbol{0}}},{\boldsymbol{0}_{(N + 1) \times (N - 1)}},d _l^*{{\mathrm{e}}^{{\mathrm{2i}}{\theta _1}(\omega _l^*)}}{\boldsymbol{M}_{ - 1}}(\omega _l^*)} \right),\hspace{36pt}\label{eq94a1}\\
\mathop {{\mathrm{Res}}}\limits_{z =  - {\omega _l}} {\boldsymbol{M}_ + }(z) =  - {\boldsymbol{\sigma} _3}\left( {{d _l}{{\mathrm{e}}^{ - {\mathrm{2i}}{\theta _1}({\omega _l})}}{\boldsymbol{M}_{ + 3}}({\omega _l}),{\boldsymbol{0}_{(N + 1) \times (N - 1)}},{\tilde{\boldsymbol{0}}}} \right),\mathop {{\mathrm{Res}}}\limits_{z =  - \omega _l^*} {\boldsymbol{M}_ - }(z) = {\boldsymbol{\sigma} _3}\left( {{\tilde{\boldsymbol{0}}},{\boldsymbol{0}_{(N + 1) \times (N - 1)}},d _l^*{{\mathrm{e}}^{{\mathrm{2i}}{\theta _1}(\omega _l^*)}}{\boldsymbol{M}_{ - 1}}(\omega _l^*)} \right),\hspace{22pt}\label{eq94b1}\\
\mathop {{\mathrm{Res}}}\limits_{z = {z_m}} {\boldsymbol{M}_ + }(z) = \left( { - {{\mathrm{e}}^{ - {\mathrm{i}}({\theta _1} - {\theta _2})({z_m})}}{\boldsymbol{M}_{ + 2}}({z_m}){\boldsymbol{W} _m},{\boldsymbol{0}_{(N + 1) \times (N - 1)}},{\tilde{\boldsymbol{0}}}} \right),\mathop {{\mathrm{Res}}}\limits_{z = u_0^2/z_m^*} {\boldsymbol{M}_ + }(z) =  - {\boldsymbol{\sigma} _3}\left( {{\tilde{\boldsymbol{0}}},\dfrac{{\mathrm{i}u_ + ^*}}{{z_m^*}}\dfrac{{{{\mathrm{e}}^{{\mathrm{i}}({\theta _1} - {\theta _2})(z_m^*)}}}}{{\alpha \left( { - \frac{{u_0^2}}{{z_m^*}}} \right)}}{\boldsymbol{M}_{ + 3}}\left( { - \dfrac{{u_0^2}}{{z_m^*}}} \right)\boldsymbol{W} _m^\dag,{\tilde{\boldsymbol{0}}}} \right),\hspace{8pt}\label{eq94a}\\
\mathop {{\mathrm{Res}}}\limits_{z =  - {z_m}} {\boldsymbol{M}_ + }(z) = {\boldsymbol{\sigma} _3}\left( {{{\mathrm{e}}^{ - {\mathrm{i}}({\theta _1} - {\theta _2})({z_m})}}{\boldsymbol{M}_{ + 2}}({z_m}){\boldsymbol{W} _m},{\boldsymbol{0}_{(N + 1) \times (N - 1)}},{\tilde{\boldsymbol{0}}}} \right),\mathop {{\mathrm{Res}}}\limits_{z =  - u_0^2/z_m^*} {\boldsymbol{M}_ + }(z) = \left( {{\tilde{\boldsymbol{0}}}, - \dfrac{{\mathrm{i}u_ + ^*}}{{z_m^*}}\dfrac{{{{\mathrm{e}}^{{\mathrm{i}}({\theta _1} - {\theta _2})(z_m^*)}}}}{{\alpha \left( { - \frac{{u_0^2}}{{z_m^*}}} \right)}}{\boldsymbol{M}_{ + 3}}\left( { - \dfrac{{u_0^2}}{{z_m^*}}} \right)\boldsymbol{W} _m^\dag,{\tilde{\boldsymbol{0}}}} \right),\hspace{9pt}\label{eq94b}\\
\mathop {{\mathrm{Res}}}\limits_{z = z_m^*} {\boldsymbol{M}_ - }(z) = \left( {{\tilde{\boldsymbol{0}}},\dfrac{{{{\mathrm{e}}^{{\mathrm{i}}({\theta _1} - {\theta _2})(z_m^*)}}}}{{\alpha (z_m^*)}}{\boldsymbol{M}_{ - 1}}(z_m^*)\boldsymbol{W} _m^\dag,{\tilde{\boldsymbol{0}}}} \right),\hspace{11pt}\mathop {{\mathrm{Res}}}\limits_{z =  - u_0^2/{z_m}} {\boldsymbol{M}_ - }(z) = \left( {{\tilde{\boldsymbol{0}}},{\boldsymbol{0}_{(N + 1) \times (N - 1)}}, - \dfrac{{\mathrm{i}{u_ + }}}{{{z_m}}}{{\mathrm{e}}^{{\mathrm{ - i}}({\theta _1} - {\theta _2})({z_m})}}{\boldsymbol{M}_{ - 2}}\left( { - \dfrac{{u_0^2}}{{{z_m}}}} \right){\boldsymbol{W} _m}} \right),\hspace{11pt}\label{eq94c}\\
\mathop {{\mathrm{Res}}}\limits_{z =  - z_m^*} {\boldsymbol{M}_ - }(z) = {\boldsymbol{\sigma} _3}\left( {{\tilde{\boldsymbol{0}}},\dfrac{{{{\mathrm{e}}^{{\mathrm{i}}({\theta _1} - {\theta _2})(z_m^*)}}}}{{\alpha (z_m^*)}}{\boldsymbol{M}_{ - 1}}(z_m^*)\boldsymbol{W} _m^\dag,{\tilde{\boldsymbol{0}}}} \right),\mathop {{\mathrm{Res}}}\limits_{z = u_0^2/{z_m}} {\boldsymbol{M}_ - }(z) = -{\boldsymbol{\sigma} _3}\left( {{\tilde{\boldsymbol{0}}},{\boldsymbol{0}_{(N + 1) \times (N - 1)}},\dfrac{{\mathrm{i}{u_ + }}}{{{z_m}}}{{\mathrm{e}}^{{\mathrm{ - i}}({\theta _1} - {\theta _2})({z_m})}}{\boldsymbol{M}_{ - 2}}\left( { - \dfrac{{u_0^2}}{{{z_m}}}} \right){\boldsymbol{W} _m}} \right).\label{eq94d}
\end{align}
\end{subequations}
Here $\tilde{\boldsymbol{0}}\in\mathbb{C}^{N+1}$, $\boldsymbol{W}^\dag _m = {\boldsymbol{\gamma} _m}/{{s}_{11}'}(z_m^*)$, $d_l=\rho_l/{{g}_{11}'}(\omega_l)$, and $\arg d _l^2 = \arg \omega _l^2 - 2{\theta _ + }$. The vector ${\boldsymbol{\gamma} _m}$ is a constant row vector of size $1\times(N-1)$, while $\rho_l$ is an arbitrary constant.
\end{proposition}
\begin{proof}
If ${\omega _l}\in{\Xi _ + } \cap \left\{ {{\mathop{\mathrm {Im}\hspace{2pt}}\nolimits} z > 0} \right\}$ is a simple zero of $g_{11}$ with $\left| {{\omega_l}} \right| = {u_0}$, then ${\omega _l} = u_0^2/\omega _l^*$. We define the matrix $\boldsymbol{\varpi} ({\omega _l}) = \left( {{\boldsymbol{\upsilon} _{ - 1}},{\boldsymbol{\mu} _ + },{\boldsymbol{\upsilon} _{ + 3}}} \right)({\omega _l})$. Equations (\ref{eq56}), (\ref{eq57}), (\ref{eq59}), and (\ref{eq76}) imply that ${\boldsymbol{\varpi} ^\dag}({\omega _l}){\boldsymbol{\sigma} _3}\boldsymbol{\varpi} ({\omega _l}) = {\boldsymbol{0}_{(N + 1) \times (N + 1)}}$. Let ${\Upsilon _l}$ be the column space of $\boldsymbol{\varpi} ({\omega _l})$, and consider the Hermitian form ${\left\langle {{\boldsymbol{v}_1},{\boldsymbol{v}_2}} \right\rangle _{{\boldsymbol{\sigma} _3}}} = \boldsymbol{v}_1^\dag{\boldsymbol{\sigma} _3}{\boldsymbol{v}_2}$. For any ${\boldsymbol{v}_1},{\boldsymbol{v}_2} \in {\Upsilon _l}$, we have ${\left\langle {{\boldsymbol{v}_1},{\boldsymbol{v}_2}} \right\rangle _{{\boldsymbol{\sigma} _3}}} = 0$. Thus, ${\Upsilon _l}$ is a totally isotropic subspace with respect to this Hermitian form. Since ${\boldsymbol{\sigma} _3}$ has inertia $(1,N)$, the dimension of any totally isotropic subspace is at most $ \min \{ 1,N\}  = 1$. It follows that ${\mathrm{rank}}\hspace{2pt}\boldsymbol{\varpi} ({\omega _l}) = \dim {\Upsilon _l} \le 1$. Moreover, since both ${\boldsymbol{\upsilon} _{ - 1}}({\omega _l})$ and ${\boldsymbol{\upsilon} _{ + 3}}({\omega _l})$ are nonzero column vectors, $\boldsymbol{\varpi} ({\omega _l}) \ne \boldsymbol{0}_{(N+1)\times(N+1)}$. Therefore, ${\mathrm{rank}}\hspace{2pt}\boldsymbol{\varpi} ({\omega _l}) = 1$, so the columns of $\boldsymbol{\varpi} ({\omega _l})$ are linearly dependent. Combining this fact with the definition (\ref{eq35}) of ${\boldsymbol{\mu} _ + }(z)$, we obtain ${\boldsymbol{\mu} _ + }({\omega _l}) = {\boldsymbol{0}_{(N + 1) \times (N - 1)}}$. By Proposition \ref{Pro:3}, there exists a constant ${\rho _l}$ satisfying ${\boldsymbol{\upsilon} _{ - 1}}({\omega _l}) = {\rho _l}{\mathrm{e}^{ - 2\mathrm{i}{\theta _1}({\omega _l})}}{\boldsymbol{\upsilon} _{ + 3}}({\omega _l})$. We calculated that
$$\mathop {{\mathop{\rm Res}\nolimits}}\limits_{z = {\omega _l}} \frac{{{\boldsymbol{\upsilon} _{ - 1}}(z)}}{{{g_{11}}(z)}} = \frac{{{\boldsymbol{\upsilon} _{ - 1}}({\omega _l})}}{{{{g'}_{11}}({\omega _l})}}.$$
This yields the residue condition for the matrix $\boldsymbol{M}_+(z)$ at $z=\omega_l$, as given by the first expression in Eqs. (\ref{eq94a1}).

We next examine the residue conditions associated with discrete eigenvalues lying off the circle $\boldsymbol{C}_0$. Suppose that $z_m\in{\Xi _ + } \cap \left\{ {{\mathop{\mathrm {Im}\hspace{2pt}}\nolimits} z > 0} \right\}$ is a simple zero of $g_{11}$ with $\left| {{z_m}} \right| \ne {u_0}$. For $0 < \left| {z - z_m^*} \right| < \varepsilon $, Eqs. (\ref{eq77}) and (\ref{eq81a}) imply that
$${\boldsymbol{\mathcal D}}\left( {{\boldsymbol{\upsilon}^* _{ - 1}},\frac{{{\boldsymbol{\mu}^* _ + }}}{{{s^*_{33}}}},{\boldsymbol{\upsilon}^* _{ + 3}}} \right)(z^*) = \left( {{\boldsymbol{\upsilon} _{ + 1}},\alpha {\boldsymbol{\mu} _ - },\frac{{b_{11}}}{{{g_{33}}}}{\boldsymbol{\upsilon} _{ - 3}}} \right)(z).$$
Taking the limit as $z \to z_m^*$ in this relation yields
$${\boldsymbol{\mathcal D}}\left( {{\boldsymbol{\upsilon}^* _{ - 1}},\frac{{{\boldsymbol{\mu}^* _ + }}}{{{s^*_{33}}(z_m)}},{\boldsymbol{\upsilon}^* _{ + 3}}} \right)(z_m) = \left( {{\boldsymbol{\upsilon} _{ + 1}},\alpha {\boldsymbol{\mu} _ - },\tilde{\boldsymbol{0}}} \right)(z_m^*).$$
Equation (\ref{eq81b}) gives
$$\det \left( {\boldsymbol{\upsilon} _{ - 1}^*,\frac{{\boldsymbol{\mu} _ + ^*}}{{b_{33}^*}},{\boldsymbol{\upsilon}^* _{ + 3}}} \right)({z^*}) = g_{11}^*({z^*})\alpha (z).$$
Since $z = z_m^*$ is a simple zero of the determinant in the preceding relation, we obtain
$$\mathrm{rank}\left[ {{\boldsymbol{\mathcal D}}\left( {\boldsymbol{\upsilon} _{ - 1}^*,\frac{{\boldsymbol{\mu} _ + ^*}}{{b_{33}^*}},{\boldsymbol{\upsilon}^* _{ + 3}}} \right)}({z_m}) \right] = 1,$$
which in turn implies $\mathrm{rank}\left( {{\boldsymbol{\upsilon} _{ + 1}},\alpha {\boldsymbol{\mu} _ - },{\tilde{\boldsymbol{0}}}} \right)(z_m^*) = 1$. Since ${\boldsymbol{\upsilon} _{ + 1}}(z_m^*) \ne \tilde{\boldsymbol{0}}$, Proposition \ref{Pro:3}, together with the preceding result, implies the existence of a $1\times(N-1)$ constant vector $\boldsymbol{\gamma}_m$ such that
$${\boldsymbol{\mu} _ - }(z_m^*)\alpha (z_m^*) = {\boldsymbol{\upsilon} _{ + 1}}(z_m^*){\boldsymbol{\gamma} _m}{\mathrm{e}^{\mathrm{i}\left( {{\theta _1} - {\theta _2}} \right)(z_m^*)}}.$$
The equivalence relations (\ref{eq88b}) show that $z=z^*_m$ is a simple zero of $b_{11}$. A direct calculation of the residue then yields
$$\mathop {{\mathop{\rm Re}\nolimits} {\mathrm{s}}}\limits_{z = z_m^*} \frac{{{\boldsymbol{\mu} _ - }(z)}}{{{b_{11}}(z)}} = \frac{{{\boldsymbol{\mu} _ - }(z_m^*)}}{{{{b}_{11}'}(z_m^*)}}.$$
We thus obtain the residue condition for $\boldsymbol{M}_-(z)$ at $z=z^*_m$, as given by the first expression in Eqs. (\ref{eq94c}). The remaining residue conditions in this proposition can be derived from the residue transformation relations established in Proposition \ref{Pro:5}.
\end{proof}

\subsection{Trace formula}
\label{subsec:3.3}
The objective of this subsection is to reconstruct the analytic scattering coefficients from discrete eigenvalues and reflection coefficients. Recall that $g_{11}$ has simple zeros at the set $\left\{ { \pm {z_m}} \right\}_{m = 1}^{{N_1}}$ off $\boldsymbol{C}_0$ and $\left\{ { \pm {\omega _l}} \right\}_{l = 1}^{{N_2}}$ on $\boldsymbol{C}_0$. Accordingly, we define
\begin{equation}\label{eq95}
\begin{aligned}
{J_ + }(z) &= \frac{1}{{{g_{11}}(z)}}\prod\limits_{m = 1}^{{N_1}} {\frac{{(z - {z_m})(z + {z_m})}}{{(z - z_m^*)(z + z_m^*)}}} \prod\limits_{l = 1}^{{N_2}} {\frac{{(z - {\omega _l})(z + {\omega _l})}}{{(z - \omega _l^*)(z + \omega _l^*)}}},\quad &z&\in\Xi_+, \\
{J_ - }(z) &= {b_{11}}(z)\prod\limits_{m = 1}^{{N_1}} {\frac{{(z - {z_m})(z + {z_m})}}{{(z - z_m^*)(z + z_m^*)}}} \prod\limits_{l = 1}^{{N_2}} {\frac{{(z - {\omega _l})(z + {\omega _l})}}{{(z - \omega _l^*)(z + \omega _l^*)}}},\quad &z&\in\Xi_-,
\end{aligned}
\end{equation}
where ${J_ + }(z)$ is analytic and zero-free in $\Xi_+$ and satisfies ${J_ + }(z)\to 1$ as $z\to\infty$. The corresponding properties hold for ${J_ - }(z)$ in $\Xi_-$. The identity $\boldsymbol{G}=\boldsymbol{B}^{-1}$, together with Eqs. (\ref{eq60}), yields
$$\log {J_ + }(z) - \log {J_ - }(z) = \log \left( \beta _2^*({z^*}){\beta _2}(z)+1 + \frac{1}{{\alpha }}\beta _1^\dag({z^*}){\beta _1}(z) \right),\quad z\in\Sigma.$$
The Plemelj  formula then gives
$$J = \exp \left( {\frac{1}{{2\pi \mathrm{i}}}\int_\Sigma  {\log \left( \beta _2^*({\xi ^*}){\beta _2}(\xi ) + \frac{1}{{\alpha (\xi )}}\beta _1^\dag({\xi ^*}){\beta _1}(\xi ) + 1 \right)\frac{{\mathrm{d}\xi }}{{\xi  - z}}} } \right),$$
where $J(z)=J_\pm(z)$, $z\in\Xi_\pm$. Substituting this expression into Eqs. (\ref{eq95}) yields the trace formula
$${g_{11}}(z) = \prod\limits_{m = 1}^{{N_1}} {\frac{{(z - {z_m})(z + {z_m})}}{{(z - z_m^*)(z + z_m^*)}}\prod\limits_{l = 1}^{{N_2}} {\frac{{(z - {\omega _l})(z + {\omega _l})}}{{(z - \omega _l^*)(z + \omega _l^*)}}} } \exp \left\{ { - \frac{1}{{2\pi {\mathrm{i}}}}\int_\Sigma  {\log } \left( \beta _2^*({\xi ^*}){\beta _2}(\xi ) + \frac{1}{{\alpha (\xi )}}\beta _1^\dag({\xi ^*}){\beta _1}(\xi ) + 1 \right)\frac{{{\mathrm{d}}\xi }}{{\xi  - z}}} \right\}.$$
The other expressions for $g_{33}$, $b_{11}$, and $b_{33}$ follow from the corresponding symmetries. Finally, the asymptotic phase difference follows from the $z\to0$ behavior of the trace formula and the first expression in Eqs. (\ref{eq69}):
$$\Delta \theta=\theta_+-\theta_-  = 4\sum\limits_{m = 1}^{{N_1}} {\arg } ({z_m}) + 4\sum\limits_{l = 1}^{{N_2}} {\arg } ({\omega _l}) + \frac{1}{{2\pi }}\int_\Sigma  {\log } \left( {\beta _2^*({\xi ^*}){\beta _2}(\xi ) + \frac{1}{{\alpha (\xi )}}\beta _1^\dag({\xi ^*}){\beta _1}(\xi ) + 1} \right)\frac{{{\mathrm{d}}\xi }}{\xi }.$$

\subsection{Restructuring the potential }
\label{subsec:3.4}

To simplify the notation in the reconstruction formula for the potential, we rewrite the matrix $\boldsymbol{M}$ in the following block structure:
$$\boldsymbol{M}=\left(\boldsymbol{M}_1,\boldsymbol{M}_2,\boldsymbol{M}_3\right)=\begin{pmatrix}
\boldsymbol{M}_{11} & \boldsymbol{M}_{12} & \boldsymbol{M}_{13}\\
\boldsymbol{M}_{21} & \boldsymbol{M}_{22} & \boldsymbol{M}_{23}
\end{pmatrix},$$
where $\boldsymbol{M}_{11}$ and $\boldsymbol{M}_{13}$ are scalars, $\boldsymbol{M}_{12}$ is a $1\times(N-1)$ row vector, $\boldsymbol{M}_{21}$ and $\boldsymbol{M}_{23}$ are $N\times 1$ column vectors, and $\boldsymbol{M}_{22}$ is an $N\times(N-1)$ matrix. The reconstruction formula for the potential $\boldsymbol{u}(x,t)$ can be derived from the asymptotic behavior in Eqs. (\ref{eq62}) and (\ref{eq68}), as stated in the following theorem.

\newtheorem{theorem}{Theorem}
\begin{theorem}\label{th1}
The reconstruction formula for the $N$-component GI equation with NZBCs is given by
\begin{equation}\label{eq96}
\boldsymbol{u}^*(x,t) =\mathrm{i}\lim_{z\to\infty} z \boldsymbol{M}_{21}(x,t,z).
\end{equation}
\end{theorem}

To solve the Riemann-Hilbert problem, we first regularize it by subtracting the leading asymptotic terms and the pole contributions. The Plemelj formula transforms the Riemann-Hilbert problem into the following algebraic integral equation:
$$\begin{aligned}
\boldsymbol{M} &= \boldsymbol{A}_+ + \sum_{m=1}^{N_1} \left(
\frac{\mathop{\mathrm{Res}}\limits_{z=z_m} \boldsymbol{M}_+}{z-z_m}
+ \frac{\mathop{\mathrm{Res}}\limits_{z=-z_m} \boldsymbol{M}_+}{z+z_m}
+ \frac{\mathop{\mathrm{Res}}\limits_{z=-z_m^*} \boldsymbol{M}_-}{z+z_m^*}
+ \frac{\mathop{\mathrm{Res}}\limits_{z=z_m^*} \boldsymbol{M}_-}{z-z_m^*}
+ \frac{\mathop{\mathrm{Res}}\limits_{z=-u_0^2/z_m} \boldsymbol{M}_-}{z+u_0^2/z_m}
+ \frac{\mathop{\mathrm{Res}}\limits_{z=u_0^2/z_m} \boldsymbol{M}_-}{z-u_0^2/z_m}
+ \frac{\mathop{\mathrm{Res}}\limits_{z=-u_0^2/z_m^*} \boldsymbol{M}_+}{z+u_0^2/z_m^*}
+ \frac{\mathop{\mathrm{Res}}\limits_{z=u_0^2/z_m^*} \boldsymbol{M}_+}{z-u_0^2/z_m^*}
\right) \\
&\quad
+ \sum_{l=1}^{N_2}\left(
\frac{\mathop{\mathrm{Res}}\limits_{z=\omega_l} \boldsymbol{M}_+}{z-\omega_l}
+\frac{\mathop{\mathrm{Res}}\limits_{z=-\omega_l} \boldsymbol{M}_+}{z+\omega_l}
+\frac{\mathop{\mathrm{Res}}\limits_{z=\omega^*_l} \boldsymbol{M}_+}{z-\omega^*_l}
+\frac{\mathop{\mathrm{Res}}\limits_{z=-\omega^*_l} \boldsymbol{M}_+}{z+\omega^*_l}
 \right)
+ \frac{\mathrm{i}}{2\pi} \int_{\Sigma} \frac{\boldsymbol{M}_-(\xi)\left(\mathrm{e}^{\mathrm{i}\boldsymbol{\Theta}(\xi)}\boldsymbol{V}(\xi)\mathrm{e}^{-\mathrm{i}\boldsymbol{\Theta}(\xi)}-\boldsymbol{I}_{(N+1)\times(N+1)}\right)}{z-\xi} \mathrm{d}\xi.
\end{aligned}$$
This equation is equivalent to
$$\begin{aligned}
\boldsymbol{M}_1 &= \boldsymbol{A}_{+1} - \sum_{m=1}^{N_1} \left( \frac{1}{z-z_m} - \frac{\boldsymbol{\sigma}_3}{z+z_m} \right)  \boldsymbol{M}_{+2}(z_m) \tilde{\boldsymbol{W}}_m
+\sum_{l=1}^{N_2} \left( \frac{1}{z-\omega_l} - \frac{\boldsymbol{\sigma}_3}{z+\omega_l} \right) \tilde{d}_l \boldsymbol{M}_{+3}(\omega_l)\\
&\quad + \frac{\mathrm{i}}{2\pi} \int_{\Sigma} \frac{\boldsymbol{M}_-(\xi)\left(\mathrm{e}^{\mathrm{i}\boldsymbol{\Theta}(\xi)}\boldsymbol{V}(\xi)\mathrm{e}^{-\mathrm{i}\boldsymbol{\Theta}(\xi)}-\boldsymbol{I}_{(N+1)\times(N+1)}\right)_1}{z-\xi} \mathrm{d}\xi,\\
\boldsymbol{M}_2&= \boldsymbol{A}_{+2} + \sum_{m=1}^{N_1} \left(
 \left( \frac{1}{z+\frac{u_0^2}{z_m^*}} + \frac{\boldsymbol{\sigma}_3}{z-\frac{u_0^2}{z_m^*}} \right)\dfrac{ u_+^*}{\mathrm{i}z^*_m} \frac{ \boldsymbol{M}_{+3}\left(-\frac{u_0^2}{z_m^*}\right) \tilde{\boldsymbol{W}}^\dag_m}{\alpha\left(-\frac{u_0^2}{z_m^*}\right)}+\left( \frac{1}{z-z_m^*} + \frac{\boldsymbol{\sigma}_3}{z+z_m^*} \right) \frac{ \boldsymbol{M}_{-1}(z_m^*) \tilde{\boldsymbol{W}}^\dag_m}{\alpha(z_m^*)} \right) \\
&\quad + \frac{\mathrm{i}}{2\pi} \int_{\Sigma} \frac{\boldsymbol{M}_-(\xi)\left(\mathrm{e}^{\mathrm{i}\boldsymbol{\Theta}(\xi)}\boldsymbol{V}(\xi)\mathrm{e}^{-\mathrm{i}\boldsymbol{\Theta}(\xi)}-\boldsymbol{I}_{(N+1)\times(N+1)}\right)_2}{z-\xi} \mathrm{d}\xi,\\
\boldsymbol{M}_{3}&= \boldsymbol{A}_{+3} - \sum_{m=1}^{N_1} \left(
\frac{1}{z+\frac{u_0^2}{z_m}} + \frac{\boldsymbol{\sigma}_3}{z-\frac{u_0^2}{z_m}}
\right)
\frac{\mathrm{i}u_+}{z_m}  \boldsymbol{M}_{-2}\left(-\frac{u_0^2}{z_m}\right) \tilde{\boldsymbol{W}}_m
+ \sum_{l=1}^{N_2} \left(\frac{1}{z-\omega^*_l} + \frac{\boldsymbol{\sigma}_3}{z+\omega^*_l}\right)
\tilde{d}^*_l \boldsymbol{M}_{-1}\left(\omega^*_l\right)\\
&\quad + \frac{\mathrm{i}}{2\pi} \int_{\Sigma} \frac{\boldsymbol{M}_-(\xi)\left(\mathrm{e}^{\mathrm{i}\boldsymbol{\Theta}(\xi)}\boldsymbol{V}(\xi)\mathrm{e}^{-\mathrm{i}\boldsymbol{\Theta}(\xi)}-\boldsymbol{I}_{(N+1)\times(N+1)}\right)_3}{z-\xi} \mathrm{d}\xi,
\end{aligned}$$
where $\tilde{\boldsymbol{W}}_m=\mathrm{e}^{-\mathrm{i}(\theta_1-\theta_2)(z_m)}\boldsymbol{W}_m$ and $\tilde{d}_l=\mathrm{e}^{-2\mathrm{i}\theta_1(\omega_l)}d_l$. Using the above expressions, the reconstruction formula (\ref{eq96}) can be rewritten as
$$\begin{aligned}
\boldsymbol{u}^* &= \boldsymbol{u}_+^* - 2\mathrm{i}\sum_{m=1}^{N_1} \boldsymbol{M}_{+22}(z_m) \tilde{\boldsymbol{W}}_m\\
 &\quad +2\mathrm{i}\sum_{l=1}^{N_2} \tilde{d}_l \boldsymbol{M}_{+23}(\omega_l)- \frac{1}{2\pi} \int_{\Sigma} \left(\boldsymbol{M}_-(\xi)\left(\mathrm{e}^{\mathrm{i}\boldsymbol{\Theta}(\xi)}\boldsymbol{V}(\xi)\mathrm{e}^{-\mathrm{i}\boldsymbol{\Theta}(\xi)}-\boldsymbol{I}_{(N+1)\times(N+1)}\right)\right)_{21} \mathrm{d}\xi.
 \end{aligned}$$

\subsection{Reflectionless potentials and exact solutions }
\label{subsec:3.5}

To get an explicit solution formula of the $N$-component GI equation with NZBCs, we consider the case where ${\beta _1}(z) = {\boldsymbol{0}_{(N - 1) \times 1}}$ and ${\beta _2}(z) = {0}$ for $z\in\Sigma$. In this case, $\boldsymbol{M}_+(z)$ and $\boldsymbol{M}_-(z)$ have no jump cross the contour $\Sigma$. Therefore, we have
\begin{subequations}\label{eq95a}
\begin{align}
\boldsymbol{M}_1 &= \boldsymbol{A}_{+1} + \sum_{l=1}^{N_2} \left( \frac{1}{z-\omega_l} - \frac{\boldsymbol{\sigma}_3}{z+\omega_l} \right) \tilde{d}_l \boldsymbol{M}_{+3}(\omega_l)- \sum_{m=1}^{N_1} \left( \frac{1}{z-z_m} - \frac{\boldsymbol{\sigma}_3}{z+z_m} \right) \boldsymbol{M}_{+2}(z_m)\tilde{\boldsymbol{W}}_m,\label{eq95a1} \\
\boldsymbol{M}_2 &= \boldsymbol{A}_{+2} + \sum_{m=1}^{N_1} \left(\frac{\boldsymbol{\sigma}_3}{z+z_m^*-\frac{u_0^2}{z}-\frac{u_0^2}{z_m^*}}+\frac{1}{z-z_m^*-\frac{u_0^2}{z}+\frac{u_0^2}{z_m^*}} \right) \boldsymbol{M}_{-1}(z_m^*)\tilde{\boldsymbol{W}}^\dag_m,\label{eq95a2}\\
\boldsymbol{M}_3 &= \boldsymbol{A}_{+3} - \sum_{m=1}^{N_1} \left( +\frac{1}{z+\frac{u_0^2}{z_m}}  \right) \frac{\mathrm{i}u_+}{z_m} \boldsymbol{M}_{+2}(z_m)\tilde{\boldsymbol{W}}_m
+ \sum_{l=1}^{N_2} \left( \frac{1}{z-\omega_l^*} + \frac{\boldsymbol{\sigma}_3}{z+\omega_l^*} \right) \tilde{d}_l^* \boldsymbol{M}_{-1}(\omega_l^*).\label{eq95a3}
\end{align}
\end{subequations}
The reconstruction formula then reduces to
\begin{equation}\label{eq97}
\boldsymbol{u}^* = \boldsymbol{u}_+^* - 2\mathrm{i}\sum_{m=1}^{N_1} \boldsymbol{M}_{+22}(z_m)\tilde{\boldsymbol{W}}_m + 2\mathrm{i}\sum_{l=1}^{N_2} \tilde{d}_l \boldsymbol{M}_{+23}(\omega_l).
\end{equation}
Similarly, the asymptotic phase difference and the trace formula reduce to
\begin{equation}\label{eq97a}
\Delta \theta  = 4\sum\limits_{m = 1}^{{N_1}} {\arg } ({z_m}) + 4\sum\limits_{l = 1}^{{N_2}} {\arg } ({\omega _l}),\quad {g_{11}}(z) = \prod\limits_{m = 1}^{{N_1}} {\frac{{(z - {z_m})(z + {z_m})}}{{(z - z_m^*)(z + z_m^*)}}\prod\limits_{l = 1}^{{N_2}} {\frac{{(z - {\omega _l})(z + {\omega _l})}}{{(z - \omega _l^*)(z + \omega _l^*)}}} }.
\end{equation}
Combining the preceding algebraic system with the reconstruction formula, we obtain an explicit solution formula in the reflectionless case, as stated in the following theorem.
\begin{theorem}\label{th2}
In the reflectionless case, the solution of the $N$-component GI equation with NZBCs is given by
\begin{equation}\label{eq98}
{\boldsymbol{u}^*}(x,t) = \boldsymbol{u}_ + ^* - 2\mathrm{i}\boldsymbol{F}{\left( {{\boldsymbol{I}_{\mathcal{N} \times \mathcal{N}}} - \boldsymbol{H}} \right)^{ - 1}}{\boldsymbol{1}_\mathcal N},
\end{equation}
where $\mathcal N =N_1+N_2$, $\boldsymbol{F}=(\boldsymbol{F}_1,\dots,\boldsymbol{F}_{\mathcal N})$ is an $N\times\mathcal N$ matrix-valued function, $\boldsymbol{H}=({\eta}_{mn})_{\mathcal N \times \mathcal N}$ is an $\mathcal N \times \mathcal N$ matrix-valued function, and ${\boldsymbol{1}_{\cal N}} = {(\underbrace {1, \dots ,1}_{{N_1}},\underbrace { - 1, \dots , - 1}_{{N_2}})^T}$. The detailed expressions for $\boldsymbol F_j$ and $\eta_{mn}$ are
$$\begin{aligned}{\boldsymbol{F}_j} = \left\{ {\begin{array}{*{20}{l}}
{{\boldsymbol{A}_{ + 22}}({z_j}){{\tilde {\boldsymbol{W}} }_j} + \displaystyle \sum\limits_{m = 1}^{{N_1}} {\left( {\frac{1}{{{z_j} - z_m^* - \frac{{u_0^2}}{{{z_j}}} + \frac{{u_0^2}}{{z_m^*}}}} - \frac{1}{{{z_j} + z_m^* - \frac{{u_0^2}}{{{z_j}}} - \frac{{u_0^2}}{{z_m^*}}}}} \right)} {\boldsymbol{A}_{ + 21}}(z_m^*)\tilde {\boldsymbol{W}} _m^\dag {{\tilde {\boldsymbol{W}} }_j},}&{1 \le j \le {N_1},}\\
{{{\tilde d }_{j - {N_1}}}{\boldsymbol{A}_{ + 23}}({\omega _{j - {N_1}}}) + \displaystyle \sum\limits_{\iota = 1}^{{N_2}} {\left( {\frac{1}{{{\omega _{j - {N_1}}} - \omega _\iota^*}} - \frac{1}{{{\omega _{j - {N_1}}} + \omega _\iota^*}}} \right)} {{\tilde d }_{j - {N_1}}}\tilde d _l^*{\boldsymbol{A}_{ + 21}}(\omega _\iota^*),}&{{N_1} < j \le \mathcal N,}
\end{array}} \right.
\end{aligned}$$
$$\begin{aligned}
{{\eta} _{mn}} = \left\{ \begin{array}{*{10}{l}}
\displaystyle - \sum\limits_{\iota = 1}^{{N_1}} {\left( {\frac{1}{{\left( {{z_n} - z_\iota^* - \frac{{u_0^2}}{{{z_n}}} + \frac{{u_0^2}}{{z_\iota^*}}} \right)\left( {z_\iota^* - {z_m}} \right)}} + \frac{1}{{\left( {{z_n} + z_\iota^* - \frac{{u_0^2}}{{{z_n}}} - \frac{{u_0^2}}{{z_\iota^*}}} \right)\left( {z_\iota^* + {z_m}} \right)}}} \right.} \\
\displaystyle \left. { + \frac{1}{{\left( {{z_n} + z_\iota^* - \frac{{u_0^2}}{{{z_n}}} - \frac{{u_0^2}}{{z_\iota^*}}} \right)\left( {z_\iota^* - {z_m}} \right)}} + \frac{1}{{\left( {{z_n} - z_\iota^* - \frac{{u_0^2}}{{{z_n}}} + \frac{{u_0^2}}{{z_\iota^*}}} \right)\left( {z_\iota^* + {z_m}} \right)}}} \right)\tilde {\boldsymbol{W}} _\iota^\dag {{\tilde {\boldsymbol{W}} }_n},&{n, m=1, \dots, {N_1},}\\
\displaystyle \sum\limits_{\iota = 1}^{{N_1}} {\left( {\frac{1}{{\left( {{z_n} - z_\iota^* - \frac{{u_0^2}}{{{z_n}}} + \frac{{u_0^2}}{{z_\iota^*}}} \right)\left( {z_\iota^* - {\omega _{m - {N_1}}}} \right)}} + \frac{1}{{\left( {{z_n} + z_\iota^* - \frac{{u_0^2}}{{{z_n}}} - \frac{{u_0^2}}{{z_\iota^*}}} \right)\left( {z_\iota^* + {\omega _{m - {N_1}}}} \right)}}} \right.} \\
\displaystyle \left. { + \frac{1}{{\left( {{z_n} + z_\iota^* - \frac{{u_0^2}}{{{z_n}}} - \frac{{u_0^2}}{{z_\iota^*}}} \right)\left( {z_\iota^* - {\omega _{m - {N_1}}}} \right)}} + \frac{1}{{\left( {{z_n} - z_\iota^* - \frac{{u_0^2}}{{{z_n}}} + \frac{{u_0^2}}{{z_\iota^*}}} \right)\left( {z_\iota^* + {\omega _{m - {N_1}}}} \right)}}} \right)\tilde {\boldsymbol{W}} _\iota^\dag {{\tilde {\boldsymbol{W}} }_n},&{n=1,\dots,{N_1},  m=N_1\dots \mathcal N,}\\
\displaystyle - \left( {\frac{1}{{{\omega_{n - {N_1}}} + \frac{{u_0^2}}{{{z_m}}}}} - \frac{1}{{{\omega_{n - {N_1}}} - \frac{{u_0^2}}{{{z_m}}}}}} \right)\frac{{{\mathrm{i}}{u_ + }{{\tilde d }_{n - {N_1}}}}}{{{z_m}}} - \sum\limits_{\iota = 1}^{{N_2}} {\left( {\frac{1}{{({\omega_{n - {N_1}}} - \omega_\iota^*)(\omega_\iota^* - {z_m})}} } \right.} \\
\displaystyle \left. { + \frac{1}{{({\omega_{n - {N_1}}} + \omega_\iota^*)(\omega_\iota^* + {z_m})}} + \frac{1}{{({\omega_{n - {N_1}}} + \omega_\iota^*)(\omega_\iota^* - {z_m})}} + \frac{1}{{({\omega_{n - {N_1}}} - \omega_\iota^*)(\omega_\iota^* + {z_m})}}} \right){{\tilde d }_j}\tilde d _\iota^*,&{m=1,\dots,{N_1}, n=N_1,\dots, \mathcal N,}\\
\displaystyle \sum\limits_{\iota = 1}^{{N_2}} {\left( {\frac{1}{{({\omega_{n - {N_1}}} - \omega_\iota^*)(\omega_\iota^* - {\omega_{m - {N_1}}})}} + \frac{1}{{({\omega_{n - {N_1}}} + \omega_\iota^*)(\omega_\iota^* + {\omega_{m - {N_1}}})}}} \right.} \\
\displaystyle \left. {  + \frac{1}{{({\omega_{n - {N_1}}} + \omega_\iota^*)(\omega_\iota^* - {\omega_{m - {N_1}}})}}+ \frac{1}{{({\omega_{n - {N_1}}} - \omega_\iota^*)(\omega_\iota^* + {\omega_{m - {N_1}}})}}} \right){{\tilde d }_j}\tilde d_\iota^*,&n,m={{N_1},\dots, \mathcal N.}
\end{array} \right.
\end{aligned}$$
\end{theorem}
\begin{proof}
First, we introduce an $N\times\mathcal N$ matrix-valued function $\boldsymbol{\chi}=(\boldsymbol{\chi}_1,\dots,\boldsymbol{\chi}_{\mathcal N})$, where
$$\begin{aligned}
{\boldsymbol{\chi} _j} = \left\{ \begin{array}{*{10}{l}}
{\boldsymbol{M}_{ + 22}}({z_j}){{\tilde {\boldsymbol{W}}}_j},&{1 \le j \le {N_1},}\\
{{\tilde d }_{j - {N_1}}}{\boldsymbol{M}_{ + 23}}({\omega _{j - {N_1}}}),&{{N_1} \le j \le \mathcal N.}
\end{array} \right.
\end{aligned}$$
In terms of this function, the reconstruction formula (\ref{eq97}) can be rewritten as
\begin{equation}\label{eq99}
{\boldsymbol{u}^*} = \boldsymbol{u}_ + ^* - 2{\mathrm{i}}\boldsymbol{\chi} {\boldsymbol{1}_\mathcal{N}}.
\end{equation}
Substituting Eq. (\ref{eq95a1}) into Eqs. (\ref{eq95a2}) and (\ref{eq95a3}) yields the system
$$\begin{aligned}
{\boldsymbol{M}_{ + 22}}({z_n}){{\tilde {\boldsymbol{W}}}_n} &= {\boldsymbol{A}_{ + 22}}({z_n}){{\boldsymbol{\tilde W}}_n} + \sum\limits_{m = 1}^{{N_1}} {\left( {\frac{1}{{{z_n} - z_m^* - \frac{{u_0^2}}{{{z_n}}} + \frac{{u_0^2}}{{z_m^*}}}} - \frac{1}{{{z_n} + z_m^* - \frac{{u_0^2}}{{{z_n}}} - \frac{{u_0^2}}{{z_m^*}}}}} \right){\boldsymbol{A}_{ + 21}}(z_m^*)\boldsymbol{\tilde W}_m^\dag {{\boldsymbol{\tilde W}}_n}}
- \sum\limits_{m = 1}^{{N_1}} {\sum\limits_{j = 1}^{{N_1}} {\left( {\frac{1}{{\left( {{z_n} - z_m^* - \frac{{u_0^2}}{{{z_n}}} + \frac{{u_0^2}}{{z_m^*}}} \right)\left( {z_m^* - {z_j}} \right)}}}   \right.} } \\
&\left. { - \frac{1}{{\left( {{z_n} + z_m^* - \frac{{u_0^2}}{{{z_n}}} - \frac{{u_0^2}}{{z_m^*}}} \right)\left( {z_m^* + {z_j}} \right)}}- \frac{1}{{\left( {{z_n} + z_m^* - \frac{{u_0^2}}{{{z_n}}} - \frac{{u_0^2}}{{z_m^*}}} \right)\left( {z_m^* - {z_j}} \right)}} + \frac{1}{{\left( {{z_n} - z_m^* - \frac{{u_0^2}}{{{z_n}}} + \frac{{u_0^2}}{{z_m^*}}} \right)\left( {z_m^* + {z_j}} \right)}}} \right)\left( {\boldsymbol{\tilde W}_m^\dag {{\boldsymbol{\tilde W}}_n}} \right){\boldsymbol{M}_{ + 22}}({z_j}){{\boldsymbol{\tilde W}}_j}\\
& + \sum\limits_{j = 1}^{{N_1}} {\sum\limits_{\iota = 1}^{{N_2}} {\left( {\frac{1}{{\left( {{z_n} - z_j^* - \frac{{u_0^2}}{{{z_n}}} + \frac{{u_0^2}}{{z_j^*}}} \right)\left( {z_j^* - {\omega_\iota}} \right)}}}- \frac{1}{{\left( {{z_n} + z_j^* - \frac{{u_0^2}}{{{z_n}}} - \frac{{u_0^2}}{{z_j^*}}} \right)\left( {z_j^* + {\omega_\iota}} \right)}}-\frac{1}{{\left( {{z_n} + z_j^* - \frac{{u_0^2}}{{{z_n}}} - \frac{{u_0^2}}{{z_j^*}}} \right)\left( {z_j^* - {\omega_\iota}} \right)}}   \right.} }\\
&\quad \left. { + \frac{1}{{\left( {{z_n} - z_j^* - \frac{{u_0^2}}{{{z_n}}} + \frac{{u_0^2}}{{z_j^*}}} \right)\left( {z_j^* + {\omega_\iota}} \right)}}} \right)\left( {\boldsymbol{\tilde W}_j^\dag {{\boldsymbol{\tilde W}}_n}} \right){{\tilde d }_\iota}{\boldsymbol{M}_{ + 23}}({\omega _\iota}),\\
{{\tilde d }_j}{\boldsymbol{M}_{ + 23}}({\omega_j}) &= {{\tilde d}_j}{\boldsymbol{A}_{ + 23}}({\omega_j}) - \sum\limits_{m = 1}^{{N_1}} {\left( {\frac{1}{{{\omega_j} + \frac{{u_0^2}}{{{z_m}}}}} - \frac{1}{{{\omega_j} - \frac{{u_0^2}}{{{z_m}}}}}} \right)} \frac{{{\mathrm{i}}{u_ + }}}{{{z_m}}}{{\tilde d}_j}{\boldsymbol{M}_{ + 22}}({z_m}){{\boldsymbol{\tilde W}}_m} + \sum\limits_{\iota = 1}^{{N_2}} {\left( {\frac{1}{{{\omega_j} - \omega_\iota^*}} - \frac{1}{{{\omega_j} + \omega_\iota^*}}} \right){{\tilde d}_j}\tilde d_\iota^*{\boldsymbol{A}_{ + 21}}(\omega_\iota^*)} \\
 &+ \sum\limits_{\iota = 1}^{{N_2}} {\sum\limits_{h = 1}^{{N_2}} {\left( {\frac{1}{{\left( {{\omega_j} - \omega_\iota^*} \right)\left( {\omega_\iota^* - {\omega_h}} \right)}}}- \frac{1}{{\left( {{\omega_j} + \omega_\iota^*} \right)\left( {\omega_\iota^* + {\omega_h}} \right)}} - \frac{1}{{\left( {{\omega_j} + \omega_\iota^*} \right)\left( {\omega_\iota^* - {\omega_h}} \right)}}  \right.} }
\left. { + \frac{1}{{\left( {{\omega_j} - \omega_\iota^*} \right)\left( {\omega_\iota^* + {\omega_h}} \right)}}} \right){{\tilde d}_j}\tilde d_\iota^*{{\tilde d}_h}{\boldsymbol{M}_{ + 23}}({\omega_h})\\
& - \sum\limits_{\iota = 1}^{{N_2}} {\sum\limits_{m = 1}^{{N_1}} {\left( {\frac{1}{{\left( {{\omega_j} - \omega_\iota^*} \right)\left( {\omega_\iota^* - {z_m}} \right)}}} \right.} }
 \left. { - \frac{1}{{\left( {{\omega_j} + \omega_\iota^*} \right)\left( {\omega_\iota^* + {z_m}} \right)}} - \frac{1}{{\left( {{\omega_j} + \omega_\iota^*} \right)\left( {\omega_\iota^* - {z_m}} \right)}} + \frac{1}{{\left( {{\omega_j} - \omega_\iota^*} \right)\left( {\omega_\iota^* + {z_m}} \right)}}} \right){{\tilde d}_j}\tilde d_\iota^*{\boldsymbol{M}_{ + 22}}({z_m}){{\boldsymbol{\tilde W}}_m}.
\end{aligned}$$
Equivalently, this system can be expressed as $\boldsymbol{\chi}=\boldsymbol{F}+\boldsymbol{\chi} \boldsymbol{H}$, and hence $\boldsymbol{\chi}=\boldsymbol{F}(\boldsymbol{I}_{\mathcal N \times \mathcal N}-\boldsymbol{H})^{-1}$. Inserting this expression into Eq. (\ref{eq99}) recovers the solution (\ref{eq98}) stated in Theorem \ref{th2}.
\end{proof}

\subsection{Soliton solutions}
\label{subsec:3.6}

In this subsection, we take the four-component GI equation as an example to investigate the structural features of different types of soliton solutions to Eq. (\ref{eq2}). As shown in Subsection \ref{subsec:3.2}, the norming constant $d_l$ associated with each discrete eigenvalue on the circle $\boldsymbol{C}_0$ satisfies the constraint stated in Proposition \ref{Pro:6}. This constraint is equivalent to ${\rho _l} = {\kappa _l}{\omega _l}{g'_{11}}({\omega _l}){\mathrm{e}^{ - \mathrm{i}{\theta _ + }}}$, where $\kappa_l$ is an arbitrary real number.

When only discrete eigenvalues lying on the $\boldsymbol{C}_0$ are present, all but the last row of matrix $\boldsymbol{F}$ vanish. Consequently, the first three components of the potential $\boldsymbol{u}$ vanish, whereas the fourth component $u_4$ corresponds dark soliton solutions. If only one set of discrete eigenvalues is present, whether on or off the circle, the resulting solutions are one-soliton solutions, in agreement with the corresponding result for the two-component GI equation \cite{A36}. Given the relative simplicity of soliton structures in these cases, we omit them for brevity.

\begin{figure}
  \centering
\subfigure[]{\includegraphics[height=1.3in,width=1.69in]{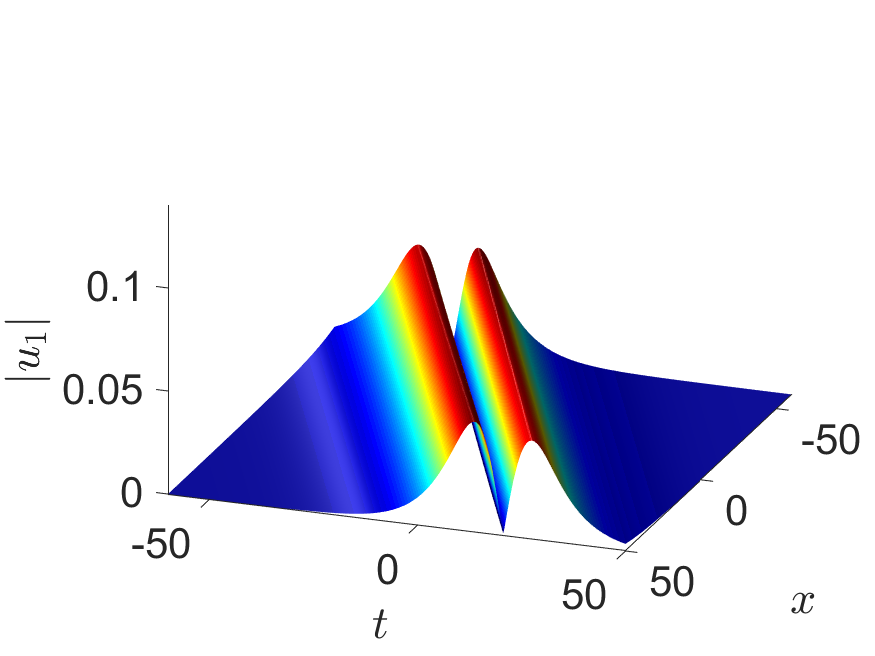}}\hspace{0.1cm}
\subfigure[]{\includegraphics[height=1.3in,width=1.69in]{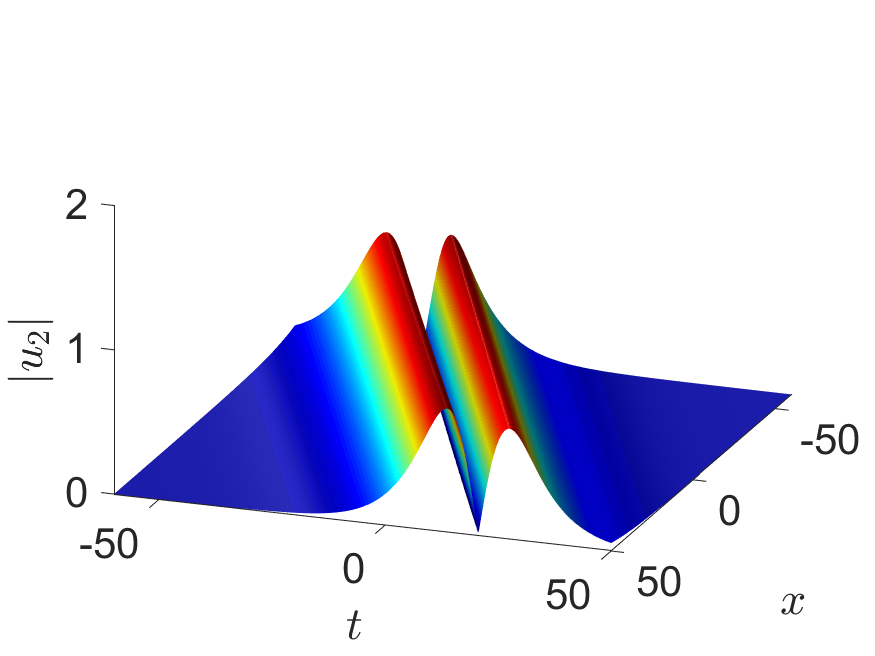}}\hspace{0.1cm}
\subfigure[]{\includegraphics[height=1.3in,width=1.69in]{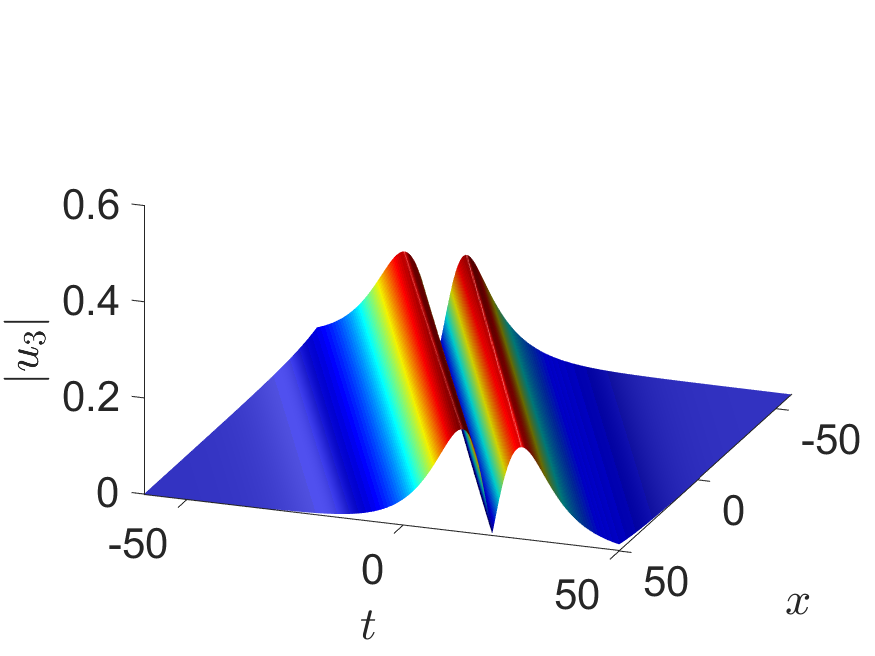}}\hspace{0.1cm}
\subfigure[]{\includegraphics[height=1.3in,width=1.69in]{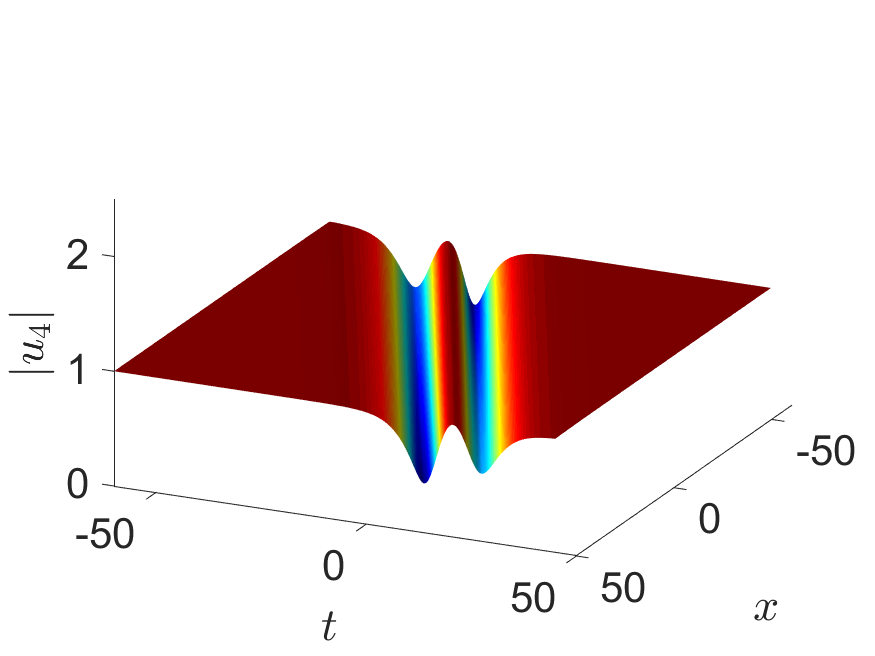}}\hspace{0.1cm}
  \caption{The two parallel solitons for $N_1=2$, $N_2=0$, $u_+=1$, $z_1={1}/{3}+\mathrm{i}/{2}$, $z_2={1}/{5}+{\mathrm{i}}/{3}$, $\boldsymbol{\gamma}_1=({1}/{4},4,1)$, and $\boldsymbol{\gamma}_2=(-{1}/{4},-4,-1)$.  }\label{fig:1}
\end{figure}

\begin{figure}
  \centering
\subfigure[]{\includegraphics[height=1.3in,width=1.69in]{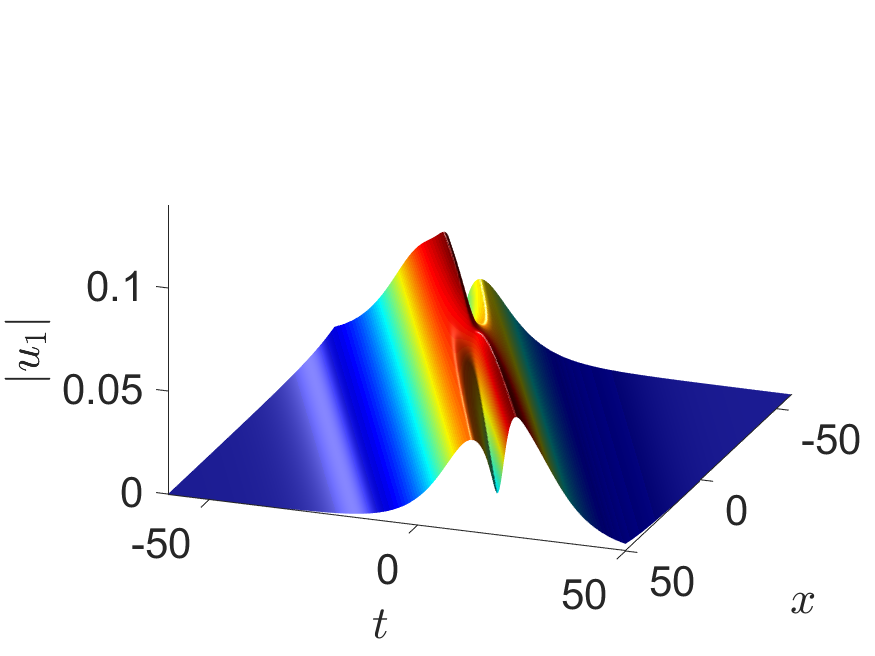}}\hspace{0.1cm}
\subfigure[]{\includegraphics[height=1.3in,width=1.69in]{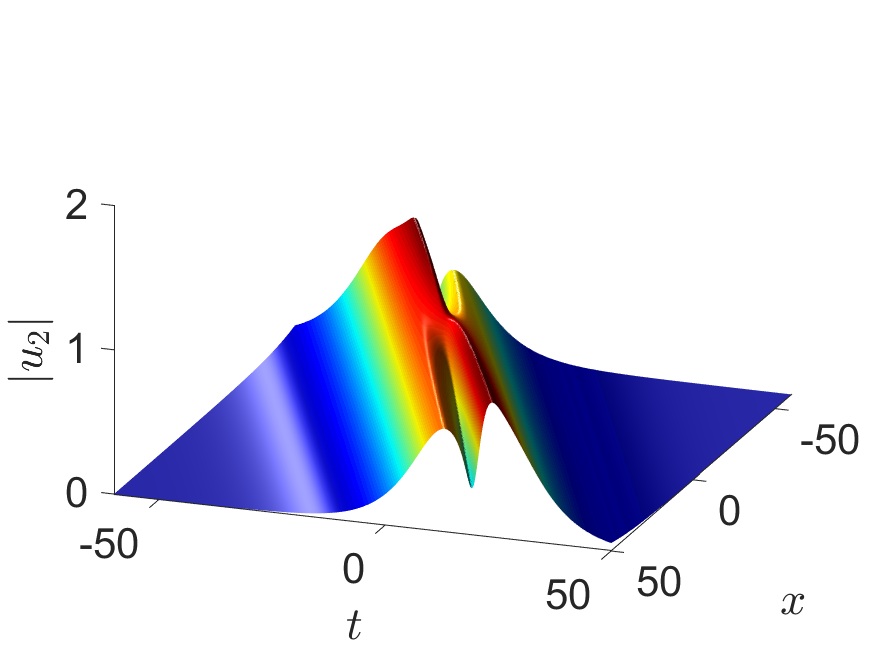}}\hspace{0.1cm}
\subfigure[]{\includegraphics[height=1.3in,width=1.69in]{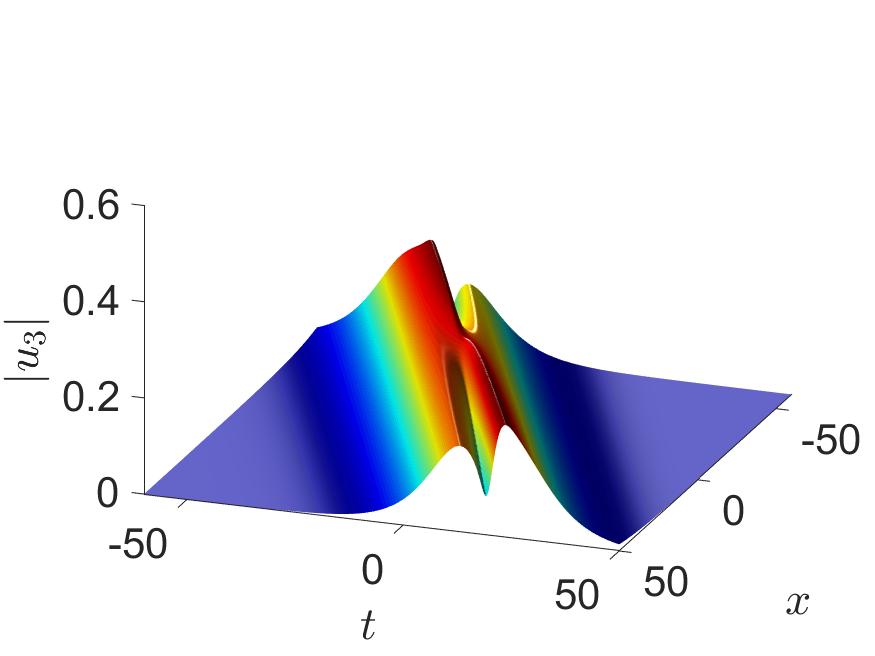}}\hspace{0.1cm}
\subfigure[]{\includegraphics[height=1.3in,width=1.69in]{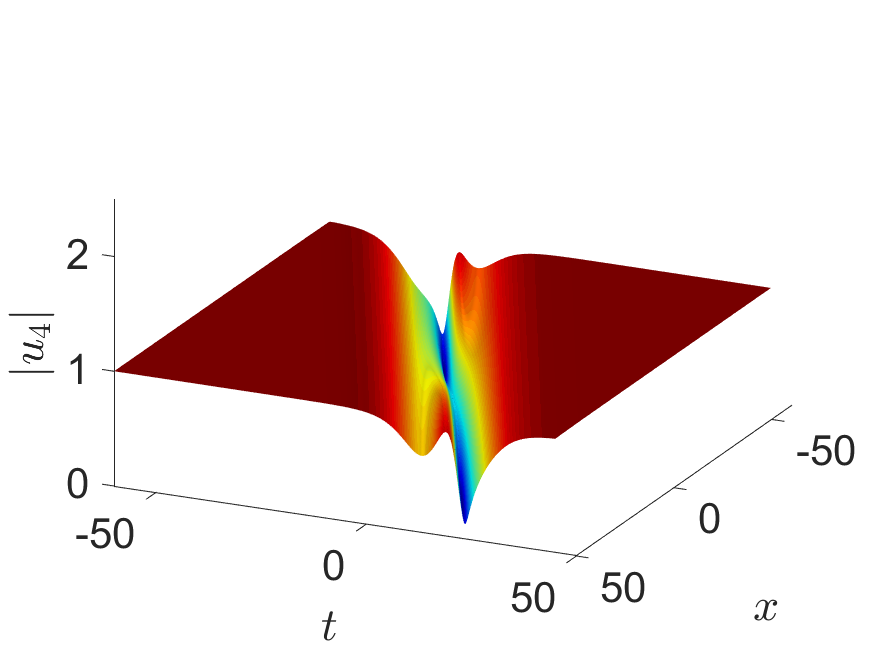}}\hspace{0.1cm}
  \caption{The two interacting solitons for $N_1=2$, $N_2=0$, $u_+=1$, $z_1={1}/{3}+\mathrm{i}/{2}$, $z_2={1}/{5}+{\mathrm{i}}/{3}$, $\boldsymbol{\gamma}_1=({1}/{4},4,1)$, and $\boldsymbol{\gamma}_2=({1}/{4},4,1)$.  }\label{fig:2}
\end{figure}

\begin{figure}
  \centering
\subfigure[]{\includegraphics[height=1.3in,width=1.69in]{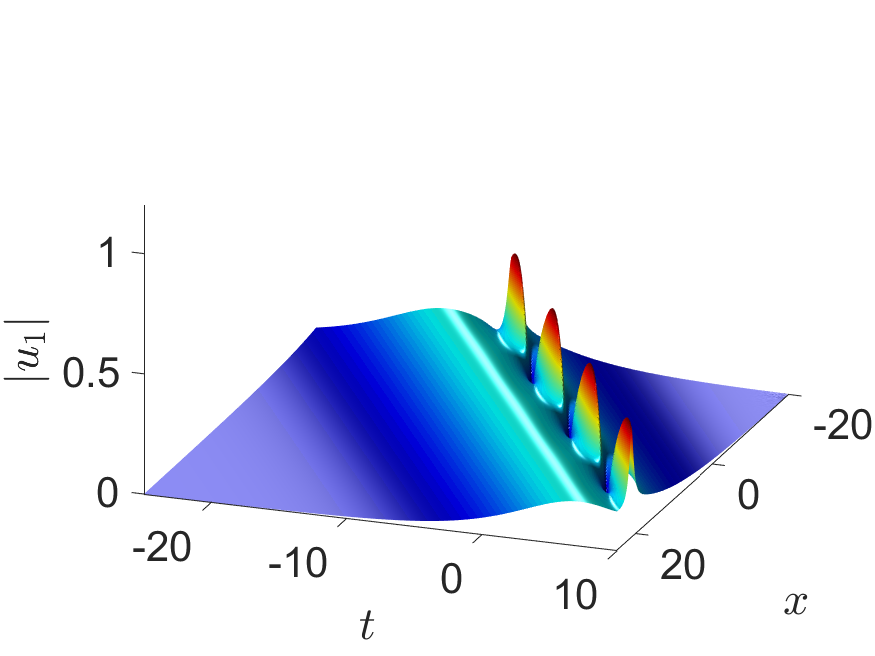}}\hspace{0.1cm}
\subfigure[]{\includegraphics[height=1.3in,width=1.69in]{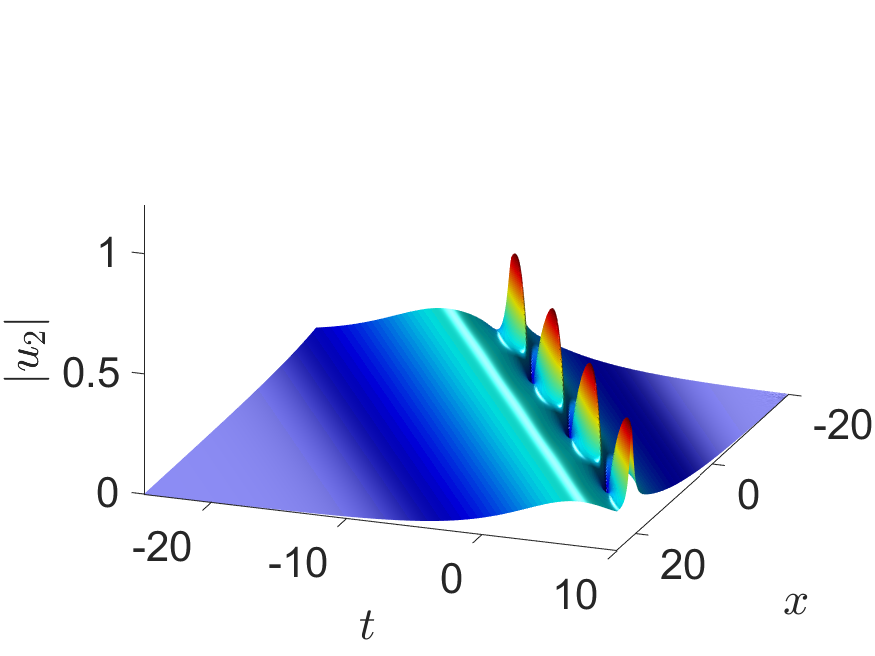}}\hspace{0.1cm}
\subfigure[]{\includegraphics[height=1.3in,width=1.69in]{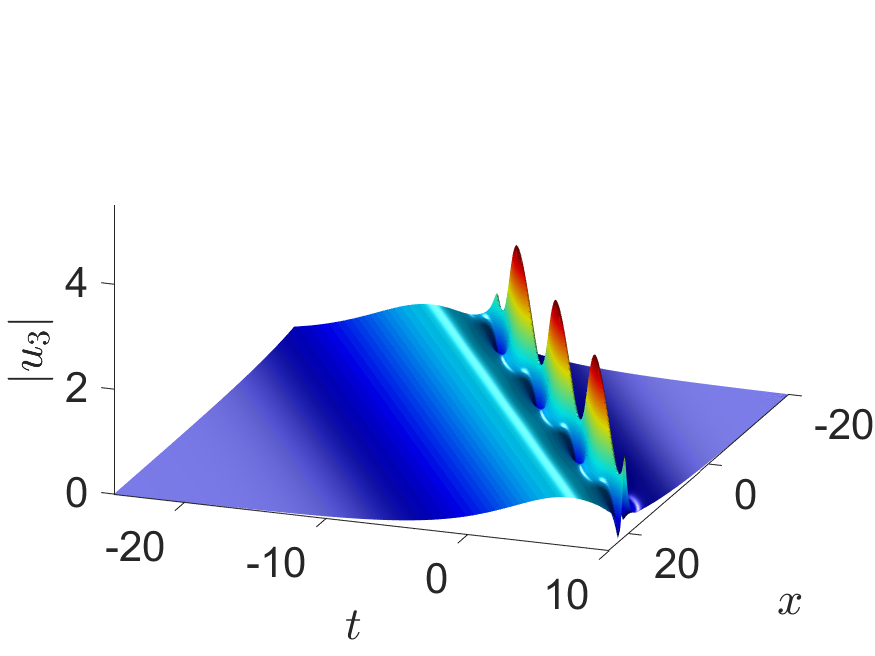}}\hspace{0.1cm}
\subfigure[]{\includegraphics[height=1.3in,width=1.69in]{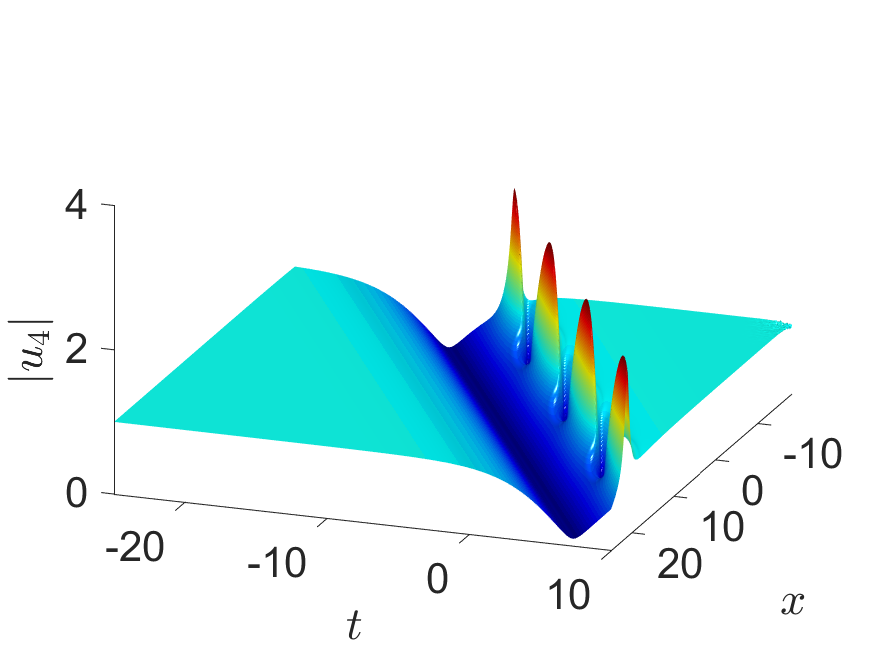}}\hspace{0.1cm}
  \caption{ A soliton and a breather for $N_1=2$, $N_2=0$, $u_+=1$, $z_1={1}+\mathrm{i}$, $z_2={1}/{3}+{\mathrm{i}}/{3}$, $\boldsymbol{\gamma}_1=({1},1,1)$, and $\boldsymbol{\gamma}_2=({1},1,5)$. }\label{fig:3}
\end{figure}

We consider two sets of discrete eigenvalues lying off the circle (i.e., $N_1 = 2$). Depending on the parameter values, three types of solutions arise. For $|z_m| < {u_0}$, the solutions are solitons. When the norming constants $\boldsymbol{\gamma}_1$ and $\boldsymbol{\gamma}_2$ have the same sign, the solution consists of two parallel solitons. Each of the first three components of the potential $\boldsymbol{u}$ exhibits two parallel bright solitons, whereas the fourth component $u_4$ exhibits two parallel dark solitons, as shown in Fig. \ref{fig:1}. By contrast, when these constants have opposite signs, the two solitons interact. This interaction manifests as two bright solitons in each of the first three components and two dark solitons in $u_4$, as illustrated in Fig. \ref{fig:2}. For $|z_m| > {u_0}$, breather solutions arise. Each of the first three components contains a breather and a bright soliton, whereas $u_4$ contains a dark soliton and a breather, as depicted in Fig. \ref{fig:3}.

\begin{figure}
  \centering
\subfigure[]{\includegraphics[height=1.3in,width=1.69in]{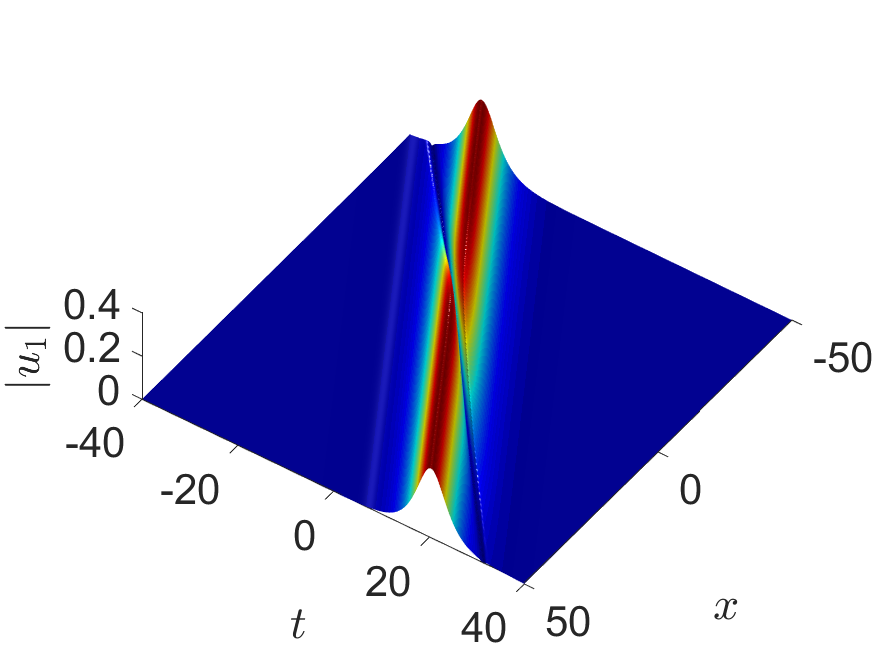}}\hspace{0.1cm}
\subfigure[]{\includegraphics[height=1.3in,width=1.69in]{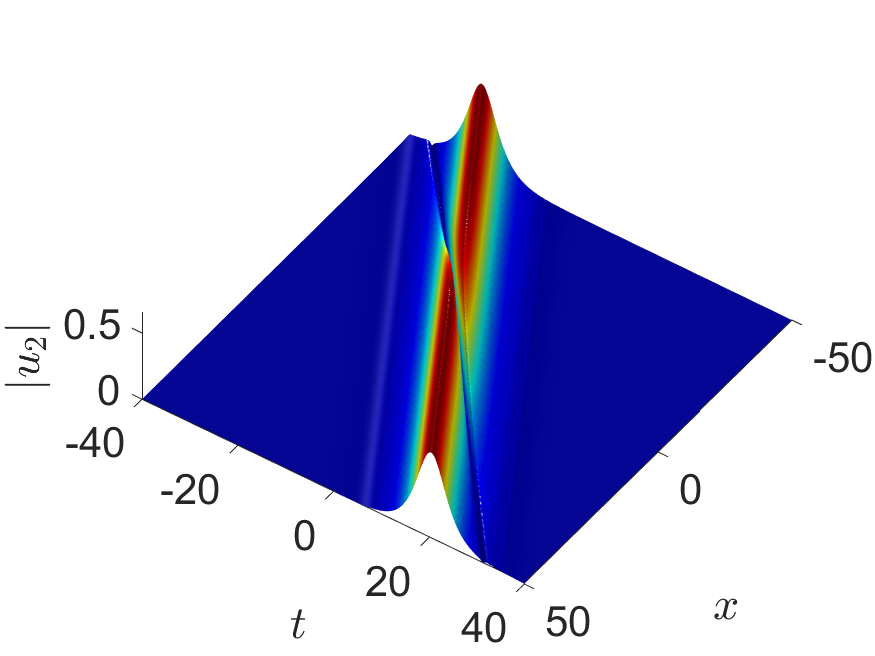}}\hspace{0.1cm}
\subfigure[]{\includegraphics[height=1.3in,width=1.69in]{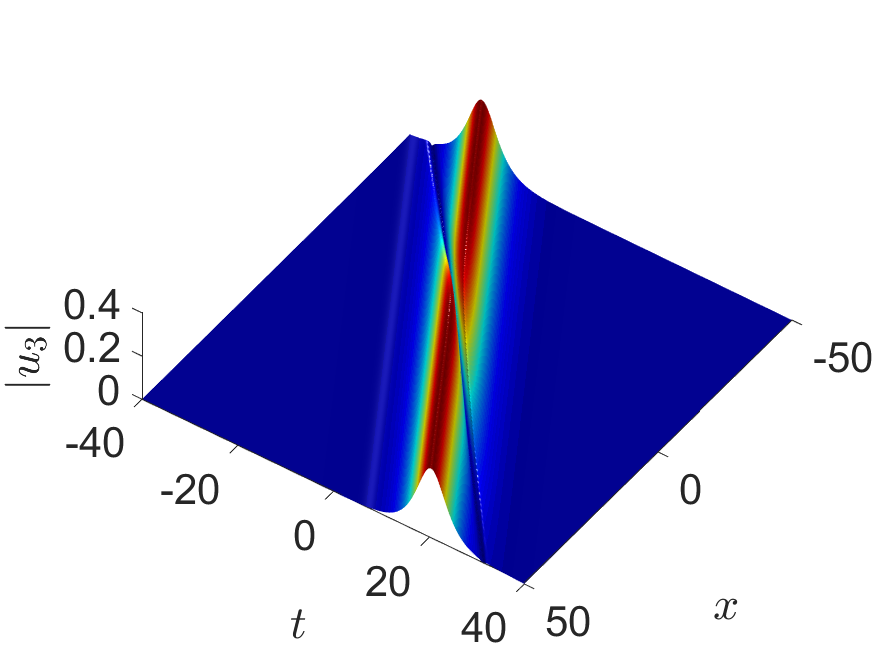}}\hspace{0.1cm}
\subfigure[]{\includegraphics[height=1.3in,width=1.69in]{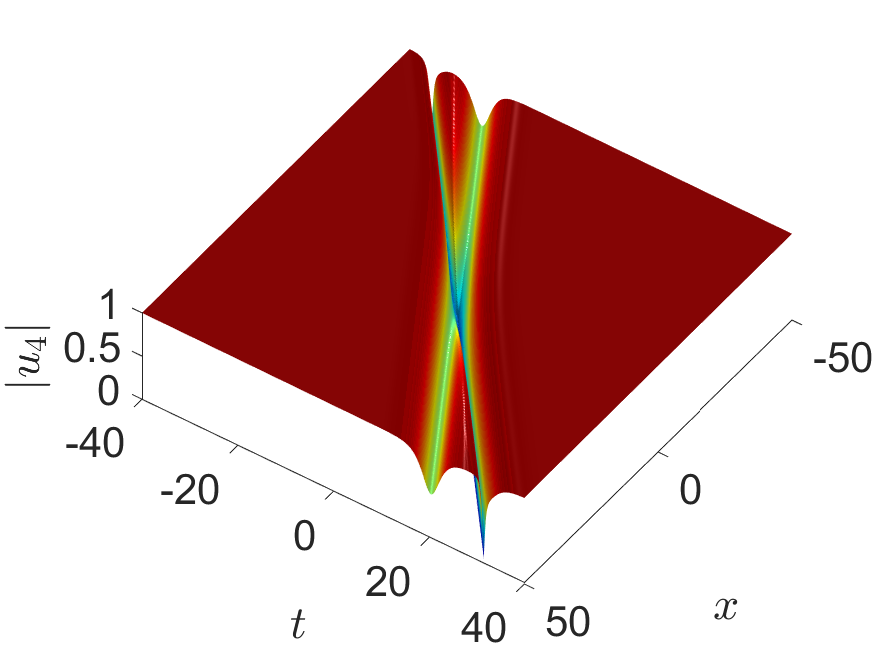}}\hspace{0.1cm}
  \caption{ The two interacting solitons for $N_1=1$, $N_2=1$, $u_+=1$, $z_1={1/3}+\mathrm{i}/2$, $\omega_1=\mathrm{e}^{\mathrm{i}\pi/6}$, $\boldsymbol{\gamma}_1=({-1},2\mathrm{i},-\mathrm{i})$, and $\kappa_1=5$.  }\label{fig:4}
\end{figure}

\begin{figure}
  \centering
\subfigure[]{\includegraphics[height=1.3in,width=1.69in]{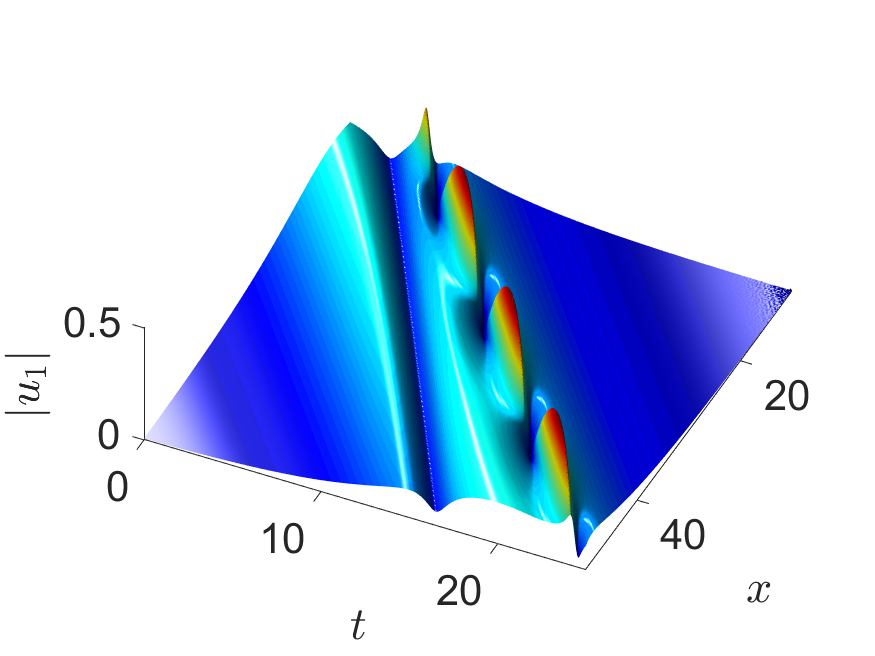}}\hspace{0.1cm}
\subfigure[]{\includegraphics[height=1.3in,width=1.69in]{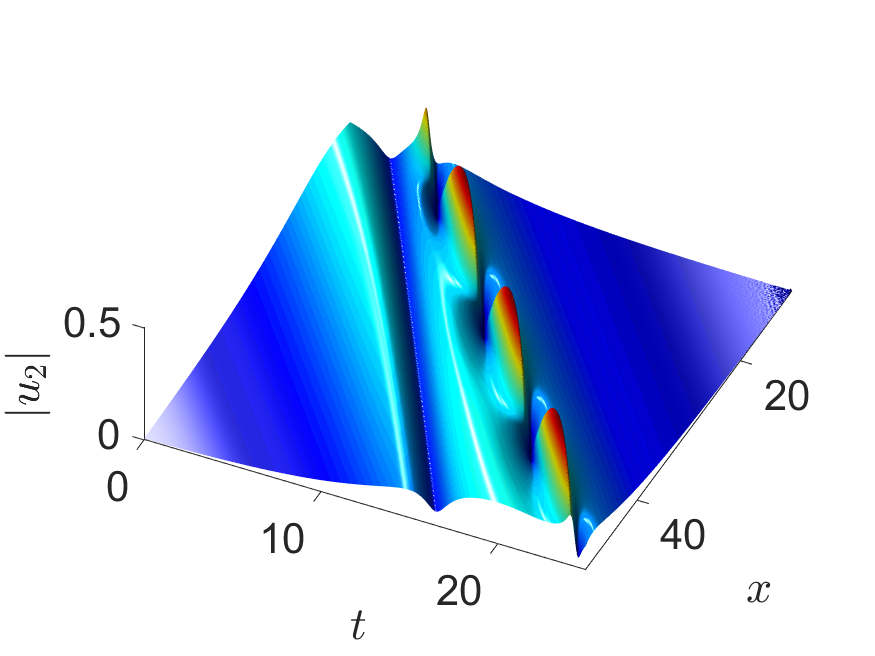}}\hspace{0.1cm}
\subfigure[]{\includegraphics[height=1.3in,width=1.69in]{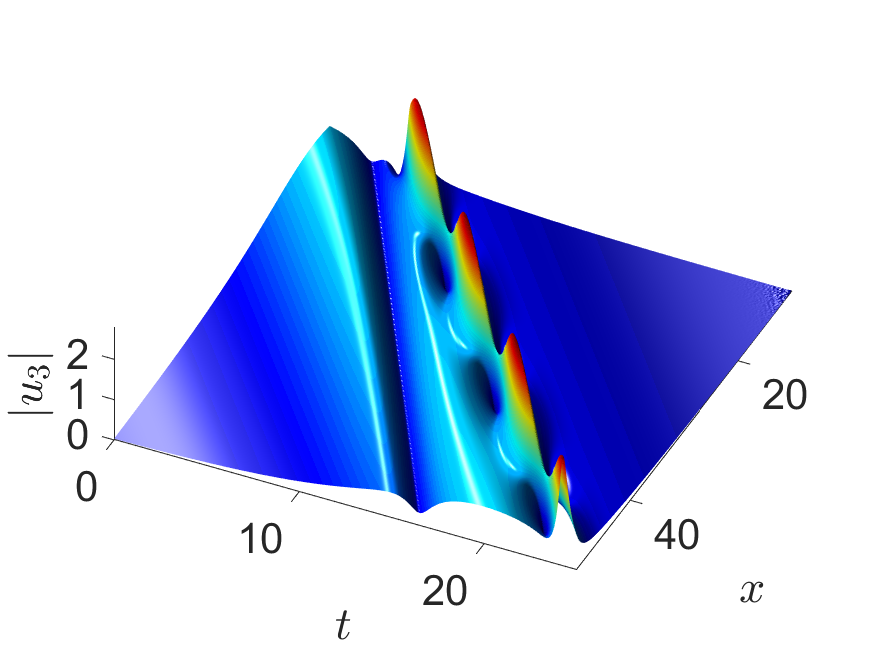}}\hspace{0.1cm}
\subfigure[]{\includegraphics[height=1.3in,width=1.69in]{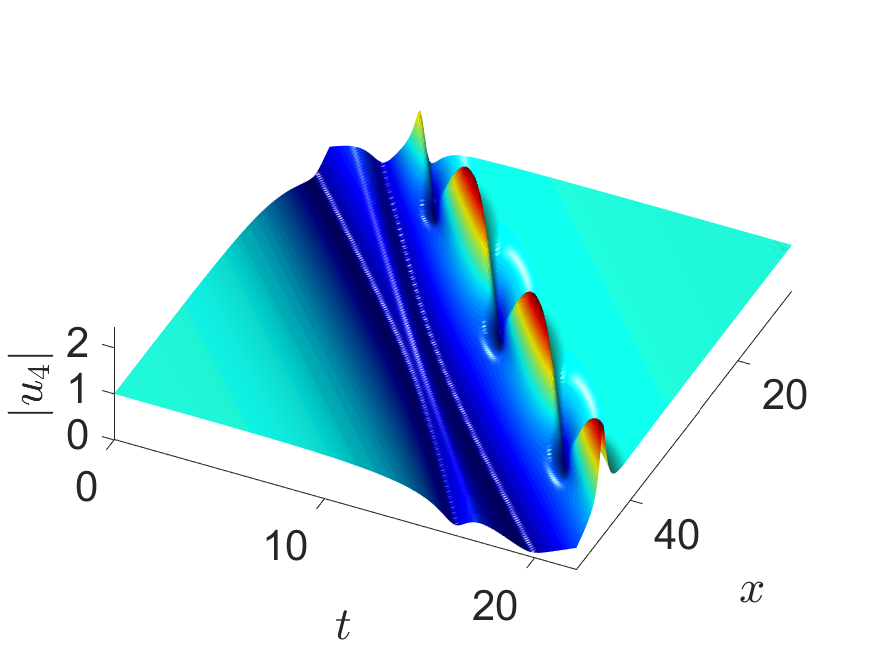}}\hspace{0.1cm}
  \caption{ The interaction of two solitons with a breather for $N_1=2$, $N_2=1$, $u_+=1$, $z_1={1}+\mathrm{i}$, $z_2={1}/3+\mathrm{i}/3$, $\omega_1=\mathrm{e}^{2\mathrm{i}\pi/5}$, $\boldsymbol{\gamma}_1=({1},1,1)$, $\boldsymbol{\gamma}_2=({1},1,5)$, and $\kappa_1=5$. }\label{fig:5}
\end{figure}

When one set of discrete eigenvalues lies on the circle and another lies off the circle, i.e., $N_1=N_2=1$, the resulting solution exhibits a two-soliton interaction. Specifically, a bright soliton interacts with a dark soliton in each of the first three components of the potential, whereas two dark solitons interact in the fourth component $u_4$, as shown in Fig. \ref{fig:4}. We next consider two sets of discrete eigenvalues off the circle and one set on the circle, i.e., $N_1 = 2$ and $N_2 = 1$. In this configuration, a breather, a bright soliton, and a dark soliton interact in each of the first three components, whereas two dark solitons and a breather interact in $u_4$, as illustrated in Fig. \ref{fig:5}.

\section{Summary and discussion}
\label{sec:4}

As discussed above, we develop the IST of the $N$-component GI equation with NZBCs. In contrast to the scalar and two-component cases with NZBCs, the $N$-component case poses greater complexity in constructing the auxiliary Jost eigenfunctions and formulating the inverse problem. The IST established here is therefore not a straightforward generalization of the corresponding scalar and two-component results. To overcome these difficulties, we employ tensor products and generalized cross products to construct the auxiliary Jost eigenfunctions, thereby obtaining a set of analytic eigenfunctions required for the inverse problem. We then derive the residue conditions for two classes of discrete eigenvalues, and formulate the inverse problem as a matrix Riemann-Hilbert problem. We note that the unitary invariance of Eq. (\ref{eq2a}) allows us to assume without loss of generality that only the last component of the background vector $\boldsymbol{u}_0$ is nonzero, thereby simplifying the analysis. Indeed, by constructing a suitable constant unitary matrix and applying the resulting transformation to the background vector and the corresponding blocks of the eigenvector matrix, explicit solutions under general parallel NZBCs can be obtained from the solution given in Theorem \ref{th2}. This transformation preserves the integrable structure of the equation and allows different soliton structures to be redistributed among the components. Consequently, each component can exhibit bright solitons, dark solitons, or breathers, yielding multi-component solutions with rich soliton structures and interaction patterns. Since these solutions follow directly from the present results via unitary transformations, we neither analyze their dynamics nor present the corresponding figures.

It is worth emphasizing that the IST framework presented in this work is restricted to NZBCs for which $\boldsymbol{u}_-$ and $\boldsymbol{u}_+$ are parallel. For the more general case of nonparallel asymptotic polarization vectors (i.e., $\boldsymbol{u}_+$ and $\boldsymbol{u}_-$ are not collinear), the IST for the Manakov system with such boundary conditions has been investigated in \cite{A52}. However, it remains an open problem to extend these techniques to the $N$-component equation with nonparallel NZBCs. Higher-order pole solutions of the $N$-component mKdV equation with ZBC were investigated in \cite{A53}, while \cite{A43} provided a systematic study of double-pole solutions for the Manakov system with NZBCs. Motivated by these results, another direction is to construct double-pole and higher-order pole solutions of Eq. (\ref{eq2}) and analyze their associated soliton structures and dynamics. These two directions constitute the main focus of our future work.

\begin{flushleft}
\textbf{Acknowledgments}
\end{flushleft}

This work is supported by the National Natural Science Foundation of China (No. 11371326 and No. 12271488).

\begin{flushleft}
\textbf{Data Availability Statements}
\end{flushleft}

No datasets were generated or analysed during the current study.

\begin{flushleft}
\textbf{Conflict of Interest}
\end{flushleft}

The authors have no conflicts to disclose.

\begin{flushleft}
 \textbf{References}
\end{flushleft}

\small{

\end{document}
\begin{thebibliography}{99}

\bibitem{A1} E.A. Kuznetsov, A.M. Rubenchik, V.E. Zakharov, Soliton stability in plasmas and hydrodynamics, Physics Reports, 1986, 142: 103--165.
\bibitem{A2} A. Mucci, P. Suret, F. Copie, S. Randoux, R. Mullyadzhanov, A. Gelash, Manipulation of strongly interacting solitons in optical fiber experiments, Physical Review Letters, 2025, 134: 193804.
\bibitem{A3} H. Huang, A.V. Ivlev, V. Nosenko, W. Yang, C.R. Du, Dissipative solitary waves in a two-dimensional complex plasma: Amorphous versus crystalline, Physical Review E, 2023, 107: 045205.
\bibitem{A4} P.D. Lax, Integrals of nonlinear equations of evolution and solitary waves, Communications on Pure and Applied Mathematics, 1968, 21: 467--490.
\bibitem{A5} M.J. Ablowitz, P.A. Clarkson, Solitons, nonlinear evolution equations and inverse scattering, Cambridge University Press, Cambridge, 1991.
\bibitem{A8} M.J. Ablowitz, D.J. Kaup, A.C. Newell, H. Segur, The inverse scattering transform-Fourier analysis for nonlinear problems, Studies in Applied Mathematics, 1974, 53: 249--315.
\bibitem{A7} V.B. Matveev, M.A. Salle, Darboux transformations and solitons, Springer-Verlag, Berlin, 1991.
\bibitem{A6} R. Hirota, Exact solution of the Korteweg-de Vries equation for multiple collisions of solitons, Physical Review Letters, 1971, 27: 1192--1194.
\bibitem{A9} C.S. Gardner, J.M. Greene, M.D. Kruskal, R.M. Miura, Method for solving the Korteweg-de Vries equation, Physical Review Letters, 1967, 19: 1095--1097.
\bibitem{A10} V.E. Zakharov, A.B. Shabat, Exact theory of two-dimensional self-focusing and one-dimensional self-modulation of waves in nonlinear media, Soviet Physics JETP, 1972, 34: 62--69.
\bibitem{A11} C. Sulem, P.L. Sulem, The nonlinear Schr\"odinger equation: Self-focusing and wave collapse, Springer, New York, 1999.
\bibitem{A12} L.M. Ling, W. Tang, On the $N$-elliptic localized wave solutions to the derivative nonlinear Schr\"odinger equation and their asymptotic analysis, Studies in Applied Mathematics, 2026, 156: e70176.
\bibitem{A13} V.S. Gerdjikov, M.I. Ivanov, A quadratic pencil of general type and nonlinear evolution equations. II. Hierarchies of Hamiltonian structures, Bulgarian Journal of Physics, 1983, 10: 130--143.
\bibitem{A14} D.J. Kaup, A.C. Newell, An exact solution for a derivative nonlinear Schr\"odinger equation, Journal of Mathematical Physics, 1978, 19: 798--801.
\bibitem{A15} H.H. Chen, Y.C. Lee, C.S. Liu, Integrability of nonlinear Hamiltonian systems by inverse scattering method, Physica Scripta, 1979, 20: 490--492.
\bibitem{A16} C.C. Ding, Y.T. Gao, L.Q. Li, Breathers and rogue waves on the periodic background for the Gerdjikov-Ivanov equation for the Alfv\'en waves in an astrophysical plasma, Chaos, Solitons \& Fractals, 2019, 120: 259--265.
\bibitem{A17} E.G. Fan, Integrable evolution systems based on Gerdjikov-Ivanov equations, bi-Hamiltonian structure, finite-dimensional integrable systems and $N$-fold Darboux transformation, Journal of Mathematical Physics, 2000, 41: 7769--7782.
\bibitem{A18} S.W. Xu, J.S. He, The rogue wave and breather solution of the Gerdjikov-Ivanov equation, Journal of Mathematical Physics, 2012, 53: 063507.
\bibitem{A19} S.S. Zhang, T. Xu, M. Li, X.F. Zhang, Higher-order algebraic soliton solutions of the Gerdjikov-Ivanov equation: Asymptotic analysis and emergence of rogue waves, Physica D, 2022, 432: 133128.
\bibitem{A20} B.L. Guo, N. Liu, The Gerdjikov-Ivanov-type derivative nonlinear Schr\"odinger equation: Long-time dynamics of nonzero boundary conditions, Mathematical Methods in the Applied Sciences, 2019, 42: 4839--4861.
\bibitem{A21} J.W. Hu, H.H. Dong, N. Zhang, Long-time asymptotics for the nonlocal Gerdjikov-Ivanov equation with decaying initial value, Chaos, Solitons \& Fractals, 2026, 208: 118276.
\bibitem{A22} J. Ieda, T. Miyakawa, M. Wadati, Exact analysis of soliton dynamics in spinor Bose-Einstein condensates, Physical Review Letters, 2004, 93: 194102.
\bibitem{A23} S.M. Mossman, G.C. Katsimiga, S.I. Mistakidis, A. Romero-Ros, T.M. Bersano, P. Schmelcher, P.G. Kevrekidis, P. Engels, Observation of dense collisional soliton complexes in a two-component Bose-Einstein condensate, Communications Physics, 2024, 7: 163.
\bibitem{A24} G.Q. Zhang, L.M. Ling, Z.Y. Yan, Multi-component nonlinear Schr\"odinger equations with nonzero boundary conditions: Higher-order vector Peregrine solitons and asymptotic estimates, Journal of Nonlinear Science, 2021, 31: 81.
\bibitem{A25} Z.J. Lin, Z.Y. Yan, Interactions and asymptotic analysis of $N$-soliton solutions for the $n$-component generalized higher-order Sasa-Satsuma equations, Chaos, 2024, 34: 123145.
\bibitem{A26} F. Zhang, K. Zhu, P.F. Han, Y. Zhang, Inverse scattering transform for the real matrix modified Korteweg-de Vries equation with nonzero boundary conditions, Physica D, 2026, 496: 135322.
\bibitem{A27} T. Tsuchida, M. Wadati, Complete integrability of derivative nonlinear Schr\"odinger-type equations, Inverse Problems, 1999, 15: 1363--1373.
\bibitem{A28} V.S. Gerdjikov, G.G. Grahovski, R.I. Ivanov, On integrable wave interactions and Lax pairs on symmetric spaces, Wave Motion, 2017, 71: 53--70.
\bibitem{A29} Y.Y. Zhai, T. Ji, X.G. Geng, Coupled derivative nonlinear Schr\"odinger III equation: Darboux transformation and higher-order rogue waves in a two-mode nonlinear fiber, Applied Mathematics and Computation, 2021, 411: 126551.
\bibitem{A30} Y. Zhang, H.H. Dong, Multi-component Gerdjikov-Ivanov system and its Riemann-Hilbert problem under zero boundary conditions, Nonlinear Analysis: Real World Applications, 2021, 60: 103279.
\bibitem{A31} T.S. Liu, T.C. Xia, Multi-component generalized Gerdjikov-Ivanov integrable hierarchy and its Riemann-Hilbert problem, Nonlinear Analysis: Real World Applications, 2022, 68: 103667.
\bibitem{A32} T.H. Luo, Z.Y. Yan, G.Q. Zhang, Multi-soliton interactions and asymptotic analysis for the integrable $n$-component higher-order Gerdjikov-Ivanov system, Wave Motion, 2027, 148: 103857.
\bibitem{A33} Y. Zhang, H.H. Dong, $N$-soliton solutions to the multi-component nonlocal Gerdjikov-Ivanov equation via Riemann-Hilbert problem with zero boundary conditions, Applied Mathematics Letters, 2022, 125: 107770.
\bibitem{A34} T. Ji, Y.Y. Zhai, Soliton, breather and rogue wave solutions of the coupled Gerdjikov-Ivanov equation via Darboux transformation, Nonlinear Dynamics, 2020, 101: 619--631.
\bibitem{A35} J.B. Zhang, Y.Y. Gongye, S.T. Chen, Soliton solutions to the coupled Gerdjikov-Ivanov equation with rogue-wave-like phenomena, Chinese Physics Letters, 2017, 34: 090201.
\bibitem{A36} X.X. Ma, J.Y. Zhu, Inverse scattering transform for the two-component Gerdjikov-Ivanov equation with nonzero boundary conditions: Dark-dark solitons, Studies in Applied Mathematics, 2023, 151: 676--715.
\bibitem{A37} Y. Li, S.F. Tian, J.J. Yang, Riemann-Hilbert problem and interactions of solitons in the $n$-component nonlinear Schr\"odinger equations, Studies in Applied Mathematics, 2022, 148: 577--605.
\bibitem{A38} Z.J. Lin, Z.Y. Yan, Long-time asymptotics of solutions for any $n$-component focusing nonlinear Schr\"odinger equations in spacetime solitonic regions, Studies in Applied Mathematics, 2025, 155: e70093.
\bibitem{A39} F. Demontis, B. Prinari, C. van der Mee, F. Vitale, The inverse scattering transform for the defocusing nonlinear Schr\"odinger equations with nonzero boundary conditions, Studies in Applied Mathematics, 2013, 131: 1--40.
\bibitem{A40} G. Biondini, G. Kova\v{c}i\v{c}, Inverse scattering transform for the focusing nonlinear Schr\"odinger equation with nonzero boundary conditions, Journal of Mathematical Physics, 2014, 55: 031506.
\bibitem{A41} Z.C. Zhang, E.G. Fan, Inverse scattering transform for the Gerdjikov-Ivanov equation with nonzero boundary conditions, Zeitschrift f\"ur angewandte Mathematik und Physik, 2020, 71: 149.
\bibitem{A42} B. Prinari, M.J. Ablowitz, G. Biondini, Inverse scattering transform for the vector nonlinear Schr\"odinger equation with nonvanishing boundary conditions, Journal of Mathematical Physics, 2006, 47: 063508.
\bibitem{A43} G. Biondini, D. Kraus, Inverse scattering transform for the defocusing Manakov system with nonzero boundary conditions, SIAM Journal on Mathematical Analysis, 2015, 47: 706--757.
\bibitem{A44} D. Kraus, G. Biondini, G. Kova\v{c}i\v{c}, The focusing Manakov system with nonzero boundary conditions, Nonlinearity, 2015, 28: 3101--3151.
\bibitem{A45} Y. Xiao, E.G. Fan, P. Liu, Inverse scattering transform for the coupled modified Korteweg-de Vries equation with nonzero boundary conditions, Journal of Mathematical Analysis and Applications, 2021, 504: 125567.
\bibitem{A46} P.F. Han, W.X. Ma, R.S. Ye, Y. Zhang, Inverse scattering transform for the defocusing-defocusing coupled Hirota equations with non-zero boundary conditions: Multiple double-pole solutions, Physica D, 2025, 471: 134434.
\bibitem{A47} F. Zhang, P.F. Han, Y. Zhang, Inverse scattering transform for the focusing two-component Hirota equation with nonzero boundary conditions, Journal of Mathematical Analysis and Applications, 2026, 554: 129941.
\bibitem{A48} B. Prinari, G. Biondini, A.D. Trubatch, Inverse scattering transform for the multi-component nonlinear Schr\"odinger equation with nonzero boundary conditions, Studies in Applied Mathematics, 2011, 126: 245--302.
\bibitem{A49} G. Biondini, D.K. Kraus, B. Prinari, The three-component defocusing nonlinear Schr\"odinger equation with nonzero boundary conditions, Communications in Mathematical Physics, 2016, 348: 475--533.
\bibitem{A50} H. Liu, J. Shen, X.G. Geng, Inverse scattering transformation for the $N$-component focusing nonlinear Schr\"odinger equation with nonzero boundary conditions, Letters in Mathematical Physics, 2023, 113: 23.
\bibitem{A51} H. Liu, J. Shen, X.G. Geng, Inverse scattering transformation for defocusing $N$-component NLS with nonzero boundaries, Bulletin of the Malaysian Mathematical Sciences Society, 2025, 48: 103.
\bibitem{A52} A. Abeya, G. Biondini, B. Prinari, Inverse scattering transform for the defocusing Manakov system with non-parallel boundary conditions at infinity, East Asian Journal on Applied Mathematics, 2022, 12: 715--760.
\bibitem{A53} Z.Z. Yang, H. Liu, J. Shen, Real-valued vector modified Korteweg-de Vries equation: Solitons featuring multiple poles, Physica D, 2025, 484: 135009.






\end{thebibliography}
